\documentclass[11pt]{article}
\usepackage{epsf,mathptm,times,amsmath,amssymb,fullpage}
\advance\parskip 6pt
\advance\footskip 0pt

\newdimen\figwidth 
\newdimen\smallerfigwidth
\newtheorem{theorem}{Theorem}[section]
\newtheorem{lemma}[theorem]{Lemma}
\newtheorem{proposition}[theorem]{Proposition}
\newtheorem{definition}[theorem]{Definition}
\newtheorem{corollary}[theorem]{Corollary}
\newtheorem{conjecture}[theorem]{Conjecture}
\newenvironment{proof}{\par\noindent{\bf Proof.} }{\proofbox\par\medskip}
\newenvironment{remark}{\par\smallskip\noindent\refstepcounter{theorem}{\bf Remark~\arabic{section}.\arabic{theorem}.} }{\par\medskip}
\def\proofbox{\hfill{\ensuremath\Box}}

\makeatletter
\@addtoreset{equation}{section}
\renewcommand\theequation{\arabic{section}.\arabic{equation}}
\renewcommand\paragraph{\@startsection{paragraph}{4}{\z@}%
   {1.4ex \@plus 0.4ex \@minus 0.2ex}{-1em}{\normalfont\normalsize\bfseries}}
\def\journal#1&#2,#3(#4){\begingroup \let\journal=\d@mmyjournal
  {\frenchspacing\sl #1\unskip}\/ {\bf\ignorespaces #2}\rm, #3
  (\afterassignment\y@ar \count255=#4)\endgroup}
\def\y@ar{\ifnum\count255<100 \advance\count255 by 1900 \fi\number\count255 }
\def\d@mmyjournal{\errmessage{Reference foul up: nested \journal macros}}
\let\@fullcite=\@cite
\def\fullcite#1{\begingroup\def\@cite##1##2{[{##1\if@tempswa , ##2\fi}]}\cite{#1}\endgroup}
\def\@cite#1#2{\raise0.8ex\hbox{\rm\scriptsize[{#1\if@tempswa , #2\fi}]}}
\long\def\@makecaption#1#2{\vskip\abovecaptionskip
  \sbox\@tempboxa{\small #1: #2}%
  \ifdim \wd\@tempboxa >\hsize\small #1: #2\par
  \else \global \@minipagefalse \hb@xt@\hsize{\hfil\box\@tempboxa\hfil}\fi
  \vskip\belowcaptionskip}
\def\@trivlist{%
  \if@noskipsec \leavevmode \fi
  \@topsepadd \topsep
  \ifvmode \advance\@topsepadd \partopsep
  \else \unskip \par \fi
  \if@inlabel \@noparitemtrue \@noparlisttrue
  \else \if@newlist \@noitemerr \fi \@noparlistfalse \@topsep 0.5\@topsepadd \fi
  \advance\@topsep \parskip
  \leftskip \z@skip
  \rightskip \@rightskip
  \parfillskip \@flushglue
  \par@deathcycles \z@
  \@setpar{\if@newlist \advance\par@deathcycles \@ne
             \ifnum \par@deathcycles >\@m \@noitemerr {\@@par}\fi
           \else {\@@par}\fi}%
  \global \@newlisttrue
  \@outerparskip \parskip}
\def\@endtheorem{\if@noparlist \else
  \if@inlabel\leavevmode\global \@inlabelfalse\fi
  \if@newlist\@noitemerr\global \@newlistfalse\fi
  \ifhmode\unskip \par\fi
    \ifdim\lastskip >\z@ \@tempskipa\lastskip \vskip -\lastskip
      \advance\@tempskipa\parskip \advance\@tempskipa -\@outerparskip
      \vskip\@tempskipa\fi
    \addpenalty\@endparpenalty\@endpetrue%
  \fi}
\newcommand\partialderiv[3][]{\frac{\partial^{#1}#2}{\partial {#3}^{#1}}}
\def\overl@ss#1#2{\vcenter{\offinterlineskip
        \ialign{$\m@th#1\hfil##\hfil$\crcr#2\crcr\raise0.6ex\hbox{$<$}\crcr } }}
\def\overgr@ater#1#2{\vcenter{\offinterlineskip
        \ialign{$\m@th#1\hfil##\hfil$\crcr#2\crcr\raise0.6ex\hbox{$>$}\crcr } }}
\def\gl{\mathrel{\mathpalette\overl@ss>}}
\def\lg{\mathrel{\mathpalette\overgr@ater<}}
\def\Real{\mathbb{R}}
\def\Complex{\mathbb{C}}
\def\max{\mathrm{max}}
\def\min{\mathrm{min}}
\def\tot{\mathrm{tot}}
\def\txtfrac#1#2{{\textstyle\frac{#1}{#2}}}
\let\<=\langle
\let\>=\rangle
\def\nl{\mathrm{nl}}
\def\d{{\bar d}}
\def\r{r}
\def\1{^{(1)}}
\def\2{^{(2)}}
\def\o{^0}
\def\v{{\mathbf v}}
\numberwithin{figure}{section}

\begin{document}
\title{On the Whitham equations for the defocusing nonlinear\\
Schr\"odinger equation with step initial data}
\author{Gino Biondini$^*$ and Yuji Kodama$^\dag$\\[0.4ex]
\small\it $^*$: State University of New York at Buffalo, Department of Mathematics, Buffalo, NY 14260\\
\small\it $^\dag$: Ohio State University, Department of Mathematics, Columbus, OH 43210}
\date{\small\today}
\maketitle
\begin{abstract}\noindent
The behavior of solutions of the finite-genus Whitham equations for
the weak dispersion limit of the defocusing nonlinear Schr\"odinger 
equation
is investigated analytically and numerically for piecewise-constant
initial data.
In particular, the dynamics of constant-amplitude initial conditions with
one or more frequency jumps (i.e., piecewise linear phase) 
are considered.
It is shown analytically and numerically that, 
for finite times, regions of arbitrarily high genus can be produced;
asymptotically with time, however, 
the solution can be divided into expanding regions which are either of 
genus-zero, genus-one or genus-two type, their precise arrangement
depending on the specifics of the initial datum given.
This behavior should be compared to that of the Korteweg-de~Vries equation, 
where the solution is devided into the regions which are either genus-zero or genus-one asymptotically.
Finally, the potential application of these results to the 
generation of short optical pulses is discussed:
the method proposed takes advantage of nonlinear compression via 
appropriate frequency modulation, and allows control of both the
pulse amplitude and its width, as well as the distance along the 
fiber at which the pulse is produced.
\end{abstract}

\kern\medskipamount
\noindent
The weak dispersion limit of the defocusing nonlinear Schr\"odinger (NLS)
equation has been extensively studied in recent years
(see e.g., Refs.~\fullcite{OTCMCC1986,CPAM52p613,FAA22p37,TMP71p584,CPAM52p655}),
and in some sense it is well-characterized mathematically.
Not as much is known, however, about the detailed behavior of
the solutions for specific choices
of initial datum\cite{PLA1995v203p77,PHYSD1995v87p186,JNLSCI7p43,SJAM59p2162}.
The purpose of this work is to present analytical and numerical results
regarding the behavior of a special class of solutions
of the NLS equation in the weak dispersion limit.
More precisely, we consider the initial value problem for
the NLS equation in the weak dispersion limit with the
initial data having a constant-amplitude and piecewise constant frequency.
We believe that, on one hand, this behavior is interesting 
mathematically, 
and that, on the other hand, it could have potential applications 
in the generation of intense, ultra-short optical pulses.
As such, it is worthy of further study.

The structure of this document is as follows:
in section~\ref{s:nlssmalldisp} we introduce the problem,
and in section~\ref{s:nlswhitham} 
we review some well-known results regarding the weak dispersion limit 
of the NLS equation.
In section~\ref{s:singlejump} we discuss the behavior of 
solutions corresponding to ``single-jump'' initial conditions, 
which are the starting point for our investigation.
Then, in sections~\ref{s:multiphase} and~\ref{s:arbitraryjumps}
we present the analytical calculations which are the main results of this work.
In section~\ref{s:numerics} we demonstrate these results 
through numerical simulations of the NLS equation
and obtain further information about the solution behavior,
and in section~\ref{s:opticalpulses} we discuss the application
of our results to the generation of intense, ultra-short optical pulses.
Appendix~\ref{s:units} describes our
nondimensionalizations and our choice of units,
Appendices~\ref{a:elliptic} and~\ref{a:NLS-Whitham}
review some known results regarding genus-one (i.e., periodic) solutions 
of the NLS equation, the Whitham averaging method and the NLS-Whitham equations,
and Appendix~\ref{a:characteristicspeeds} gives the details of 
some calculations whose results are presented in 
sections \ref{s:singlejump} and~\ref{s:multiphase}.

\section{The NLS equation with small dispersion}
\label{s:nlssmalldisp}

In this section we recall some basic results regarding 
the behavior of solutions of the nonlinear Schr\"odinger (NLS) equation 
with small dispersion.
This will establish the background and the notation necessary
to extend these results in the following sections.

\paragraph{The semiclassical limit of the NLS equation.}

We start from the defocusing NLS equation 
(that is, the NLS in the normal dispersion regime in optical fibers)
in dimensionless form:
\begin{equation}
i\epsilon\partialderiv qt - \frac12\epsilon^2\partialderiv[2]qx + |q|^2q= 0\,,
\label{e:NLS}
\end{equation}
where we assume $0<\epsilon\ll1$.
In the context of optical fibers, $t$ represents the dimensionless
propagation distance and $x$~is the dimensionless retarded time 
(cf.\ Appendix~\ref{s:units}).
To study the weak dispersion limit of the NLS Eq.~\eqref{e:NLS}, 
we first express the field 
$q(x,t)$ in a WKB form as
\begin{equation}
q(x,t)= \sqrt{\rho(x,t)}\,\,\exp[i\,\varphi(x,t)/\epsilon]\,,
\label{e:modphase}
\end{equation}
where $\rho(x,t)=|q(x,t)|^2$ and 
$\varphi(x,t)=\epsilon\,\arg\,q(x,t)=(i\epsilon/2)\ln[q^*(x,t)/q(x,t)]$ 
represent respectively 
the local intensity and the normalized local phase of $q(x,t)$.
We then introduce the normalized phase gradient
\begin{equation}
u(x,t)=\,\, \partial\varphi(x,t)/\partial x\,,
\end{equation}
that is, $u(x,t)=(i\epsilon/2)\,(q^*_x(x,t)q(x,t)-q^*(x,t)q^{}_x(x,t))/|q(x,t)|^2$. 
Throughout this work we refer to $x$ and $t$ as the space and time variables, 
respectively.
By analogy with the fiber optics context, however,
we will refer to $u(x,t)$ as the local \textit{frequency} 
of the solution.
With this decomposition, 
the NLS Eq.~\eqref{e:NLS} can be written in the form of a conservation law:
\begin{subequations}
\label{e:NLSmodphase}
\begin{align}
&\partialderiv \rho {t}= \partialderiv {(\rho u)}x\,, 
\\
&\partialderiv {(\rho u)}{t}= \partialderiv { }x \bigg(\,
  \rho u^2 + \frac12\rho^2 
  - \frac14\epsilon^2\rho\partialderiv[2]{}x\ln\rho \,\bigg)\,.
\end{align}
\end{subequations}
The weak dispersion limit of the defocusing NLS equation
is defined as the problem of studying the solutions of Eq.~\eqref{e:NLS}
as $\epsilon\to0^+$ 
with initial condition expressed in terms of Eq.~\eqref{e:modphase}.
The limit is singular, and it should be considered in the weak sense.

There are two main situations where the weak dispersion limit of the
NLS equation is relevant: nonlinear fiber optics
and Bose-Einstein Condensation (BEC).
In the context of nonlinear fiber optics, the weak dispersion limit
is relevant for the long-distance transmission of non-return-to-zero (NRZ)
pulses, as discussed in Ref.~\fullcite{SJAM59p2162}, 
or for the generation of intense short optical pulses,
as proposed in this work.
In the context of BEC, the semiclassical limit applies due to the 
very small value of Planck's constant $\hbar$ relative to quantities
associated with macroscopic objects, i.e., $\hbar=\epsilon$.
Normalizations appropriate for long distance optical fiber communications
were discussed in Ref.~\fullcite{JNLSCI7p43},
whereas in appendix~\ref{s:units}, we discuss scalings and 
nondimensionalizations relevant for the generation of intense 
short optical pulses.
We emphasize however that the results presented in this work
apply equally well to Bose-Einstein condensates and to other
physical context where the NLS equation is relevant, such as
for example ferromagnetics and water waves.

\paragraph{Hydrodynamic analogy and dam-breaking problem.}

If both $\rho$ and $u$ are smooth and $\rho> 0$, and if $\epsilon\ll1$,
Eqs.~\eqref{e:NLSmodphase} are approximated to leading order by
the following reduced hydrodynamical system\cite{OL20p2291,Whitham}
\begin{equation}
\partialderiv{ }{t} \left(\begin{matrix}\rho\\u\end{matrix}\right)
 = \left(\begin{matrix}u &\rho\\ 1 &u\end{matrix}\right)
  \partialderiv{ }x \left(\begin{matrix}\rho\\u\end{matrix}\right)\,.
\label{e:dam}
\end{equation}
Equation~\eqref{e:dam} is called the dispersionless NLS equaton, and is used to
describe a surface wave motion in shallow water.
In the hydrodynamical setting, $\rho$ and $-u$ represent respectively the depth 
and velocity of water, and $x$ and $t$ are dimensionless space and time.
For $\rho>0$, the eigenvalues $u\pm\sqrt{\rho}$ of the 
coefficient matrix are real,
and the system~\eqref{e:dam} is strictly hyperbolic.
This system, which is known as the shallow water wave equations
and has been intensively studied (see e.g. Ref.~\fullcite{Whitham}),
can be rewritten in Riemann invariant (i.e., diagonal) form
\begin{equation}
\partialderiv{r_k}t=s_k\partialderiv{r_k}x\,,\qquad k=1,2\,,
 \label{e:damRiemann}
\end{equation}
where the Riemann invariants $\r_{1,2}(x,t)$ are given by
\begin{equation}
\r_1= u-2\sqrt{\rho}\,,\qquad
\r_2= u+2\sqrt{\rho}\,,
\label{e:g0invariants}
\end{equation}
and the characteristic speeds $s_{1,2}(x,t)$ are
\[
s_1=\frac{1}{4}(3r_1+r_2)=u-\sqrt{\rho}\,,\quad
s_2=\frac{1}{4}(r_1+3r_2)=u+\sqrt{\rho}\,.
\]
Note that $s_k>0$ implies a left-moving wave (i.e., $dx/dt=-s_k$), 
and that Eqs.~\eqref{e:g0invariants} are equivalent to
\begin{equation}
\rho= \frac1{16}(r_2-r_1)^2\,,\qquad
u= \frac12(r_1+r_2)\,.
\label{e:g0amplitude}
\end{equation}
Since the system of PDEs described by Eq.~\eqref{e:damRiemann} is strictly hyperbolic,
it is possible to show that, for ``rarefaction'' initial data,
namely, when $\r_{1,2}(x,0)$ are both monotonically decreasing 
functions of~$x$, 
a global solution exists for all $t>0$.
In many cases of interest, however, the initial data do not satisfy
this property and as a consequence they develop a shock, as we shall see 
in the following.

\begin{figure}[b!]
\kern\medskipamount
\centerline{\epsfxsize0.495\textwidth\epsfbox{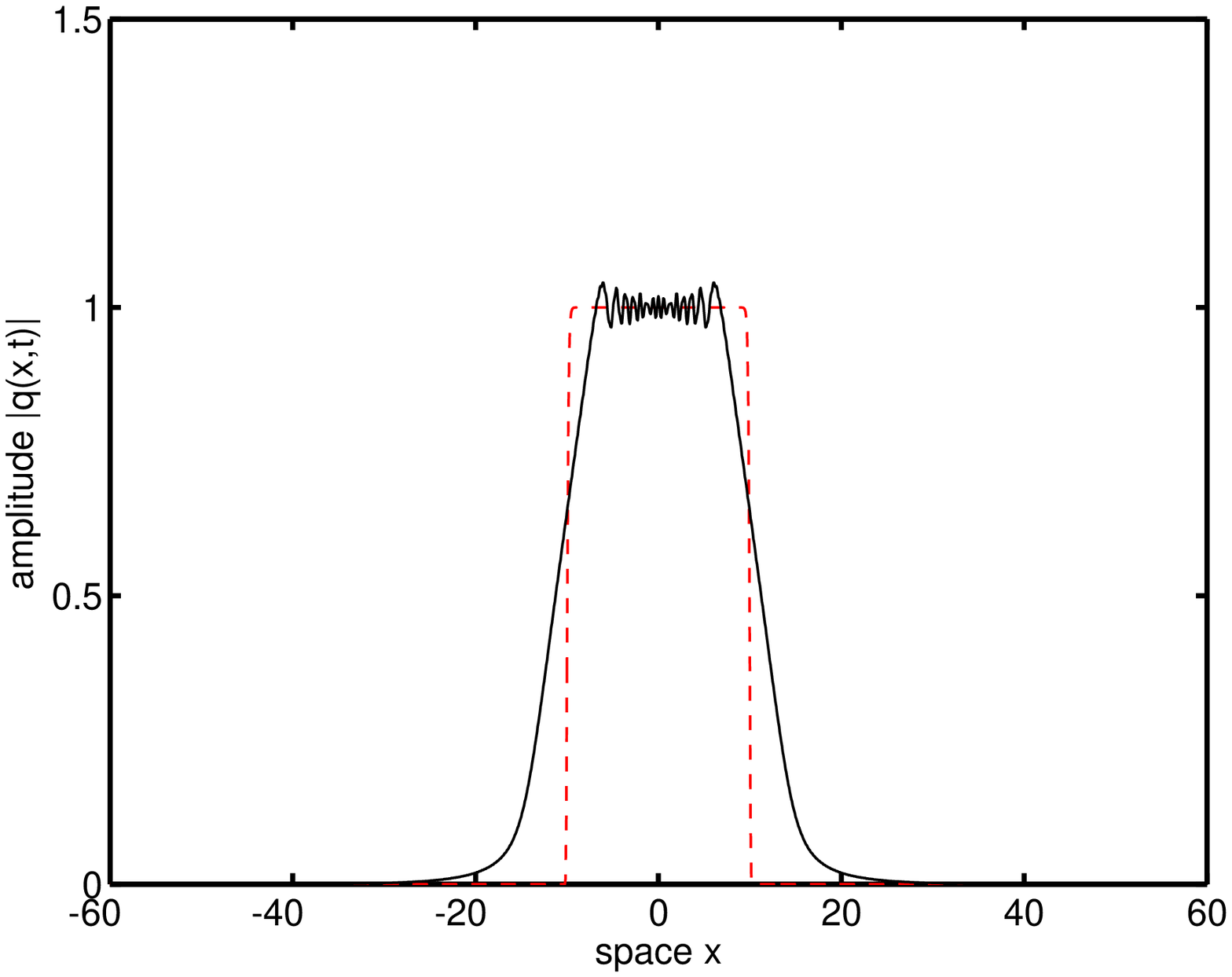}\hss
\epsfxsize0.495\textwidth\epsfbox{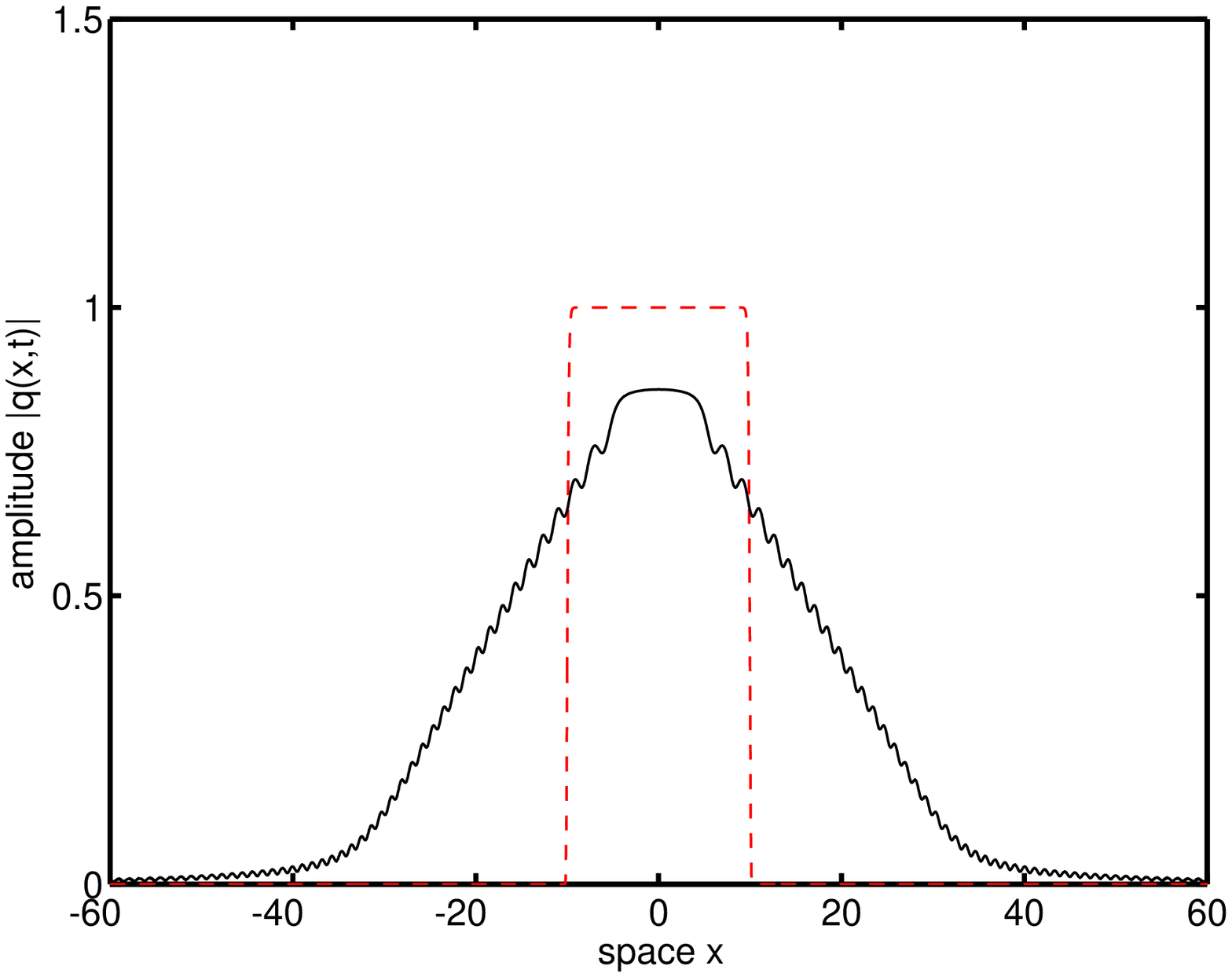}}
\caption{Deformation of an NRZ pulse (ie., the ``dam-breaking'' problem)
defined by the initial conditions in Eqs.~\eqref{e:dambreaking}, 
with $q_0=1$, $L=20$ and $\epsilon^2=0.1$:
(a, left)~$t'=10$;
(b, right)~$t'=40$, where $t'=t/\epsilon$.
The critical time here is $t_0=L/q_0=20$; 
hence Fig.~\ref{f:NRZnojump}a and Fig.~\ref{f:NRZnojump}b 
show the solution respectively before and after the time~$t_0$. 
The dotted lines show the initial condition, while the solid lines
show the result of numerical simulations of the NLS equation~\eqref{e:NLS},
which are performed in terms of the fast time scale $t'$,
as discussed in section~\ref{s:numerics}.}
\label{f:NRZnojump}
\kern-\medskipamount
\end{figure}

\paragraph{Square-wave initial conditions.}
Consider first the initial datum given by the following rectangular pulse
of width $2L$:
\begin{subequations}
\label{e:dambreaking}
\begin{align}
&\rho(x,0)= \bigg\{\!\!
  \begin{array}{ll}q_0^2 &\quad|x|<L\\ 0 &\quad|x|>L\,,\end{array}
\label{e:rho0}
\\
&u(x,0)=0\,,
\label{e:u0}
\end{align}
\end{subequations}
with $q_0>0$, corresponding in the optics framework to an NRZ pulse.
The system of equations~\eqref{e:dam} 
with initial conditions~\eqref{e:dambreaking}
is known in the literature as the ``dam-breaking'' problem.
(In the hydrodynamic analogy, the problem describes the behavior 
of a mass of water which is initially confined in a uniform, spatially 
localized state by two dams located at $x=\pm L$ and both of which are 
removed at $t=0$.)
The Riemann invariants for this situation are shown as the dashed lines in
Fig.~\ref{f:riemann0}a.
Note that the initial data for the Riemann invariants~$r_1$, $r_2$ 
corresponding to the initial condition~\eqref{e:dambreaking} is not
of rarefaction type.
It is possible however to obtain initial data of rarefaction-type 
by properly redefining the initial value of the invariants $r_1$ and $r_2$
for $|x|>L$, as shown in Fig.~\ref{f:riemann0}b later.
(This procedure is a special case of the process known as 
\textit{regularization}, see next section.)
Then the system has the following solution up to the time $t_0=L/q_0$:
for $0<x<L+2q_0\,t$,  it is
\begin{subequations}
\label{e:damsolution}
\begin{align}
&\rho(x,t)= 
  \min\big\{q_0^2,\,\,\txtfrac19\big[2q_0-(x-L)/t\big]^2\big\}\,,
\\
&u(x,t)= 
  \min\big\{0,\,\,-\txtfrac23\big[q_0+(x-L)/t\big]\big\}\,,
\end{align}
\end{subequations}
while for $x>L+2q_0\,t$, $\rho(x,t)=u(x,t)=0$, 
with $\rho(-x,t)=\rho(x,t)$ and $u(-x,t)=-u(x,t)$
(cf.\ Ref.~\fullcite{SJAM59p2162}).  
The full solution of the NLS equation~\eqref{e:NLS}
with initial condition~\eqref{e:dambreaking},
as obtained from numerical simulations
(described in section~\ref{s:numerics})
is depicted in Fig.~\ref{f:NRZnojump}a.
(This kind of solution is usually called a \textit{fan}
in the context of hydrodynamics.)
The speeds of the boundaries of the top ($\rho(x,t)=q_0^2$) 
and bottom ($\rho(x,t)=0$) regions
are easily obtained from Eqs.~\eqref{e:damsolution}
(see also~Fig.~\ref{f:riemann0}b later);
these two speeds are respectively $s_2^-=q_0$ and~$s_2^+=-2q_0$.

Equations~\eqref{e:damsolution} cease to be valid beyond the time 
$t_0=L/q_0$ when the boundaries of the top region meet at $x=0$ 
(i.e., the time at which the left-moving characteristic emanating 
from $x=L$ meets with the right-moving characteristic from $x=-L$).
An analytical expression for $q(x,t)$ when $t>L/q_0$ however
can be obtained using the dispersionless limit of the scattering
transform for the NLS equation\cite{CPAM52p655}.
The corresponding behavior of the full solution of the NLS equation
is shown in Fig.~\ref{f:NRZnojump}b.
The appearance of small oscillations in the numerical solution
in Figs.~\ref{f:NRZnojump}a,b
was discussed in Ref.~\fullcite{JNLSCI7p43}, 
and is the consequence of approximating the discontinuous 
initial data~\eqref{e:dambreaking} with a continuous initial datum
in the numerical simulations (as described in section~\ref{s:numerics}). 
It should also be noted that some care must be taken 
regarding the regularization of the discontinuous 
initial datum~\eqref{e:dambreaking}
for Eq.~\eqref{e:dam}, since a weak solution with a discontinuity 
is not unique. In fact, one can construct a different solution of (\ref{e:NLSmodphase})
with a discontinuity. 
In our case, however, 
the regularization described in section~\ref{s:nlswhitham}
enforces the continuity of~$\rho(x,t)$, 
thus removing the ambiguity and producing a unique solution.

\paragraph{Frequency jumps and high-frequency oscillations.}

In terms of optical pulses, the above results imply that the
initial condition in Eq.~\eqref{e:dambreaking}
rapidly spreads out, as shown in Figs.~\ref{f:NRZnojump}a,b.
This behavior can be partly prevented (or, alternatively, reinforced) 
by employing initial conditions with nontrivial phase.  
For example, consider the following:
\begin{equation}
u(x,0)= 
\bigg\{\!\!  \begin{array}{ll}-u_0 &\quad x<0\,,\\ u_0 &\quad x>0\,,\end{array}
\tag{\ref{e:u0}$'$}
\label{e:g0u}
\end{equation}
with $\rho(x,0)$ still given by Eq.~\eqref{e:rho0}.
Hereafter, we will use $r_1\o(x)$ and $r_2\o(x)$ to refer to the value
of the Riemann invariants~\eqref{e:g0invariants} at $t=0$.
If $\rho(x,0)$ is given by Eq.~\eqref{e:rho0} and $u(x,0)$ by Eq.~\eqref{e:g0u},
for $|x|<L$ we have
\begin{subequations}
\label{e:riemann0onejump}
\begin{gather}
r_1\o(x)= 
\bigg\{\!\!  \begin{array}{ll}-u_0-2q_0&\quad x<0\,,\\ u_0-2q_0&\quad x>0\,,\end{array}
\\
r_2\o(x)= 
\bigg\{\!\!  \begin{array}{ll}-u_0+2q_0 &\quad x<0\,,\\ u_0+2q_0&\quad x>0\,,\end{array}
\end{gather}
\end{subequations}
as shown by dashed lines in Figs.~\ref{f:riemann1jump}a,b.
Note that the value of the invariants can be redefined for $|x|>L$, 
since $\rho(x,0)=0$ there.

In terms of optical pulses, Eq.~\eqref{e:g0u} amounts to imposing a 
\textit{frequency jump} at the center of the pulse ($x=0$).
If $u_0>0$, the right half of the pulse acquires a positive frequency
and the left half a negative frequency.
Thus, owing to the normal dispersion, the two halves of the pulse
will tend to move towards each other.\cite{OL20p2291}
In terms of the hydrodynamical problem, this corresponds to assigning
an inward initial velocity to the mass of water, as if two pistons 
were acting on each side of it.
(For this reason, this case is often referred to as the ``piston'' problem.)
Note that if $u_0>0$ the initial data $r_{1,2}\o(x)$ are increasing.
Thus, another consequence of the initial frequency jump is that
if $u_0>0$
a shock develops at $x=0$, and the solution develops high-frequency 
oscillations, as shown in Fig.~\ref{f:damshock}a. 
(This type of shock is called \textit{collisionless}, or dispersive, to
distinguish it from the usual type of shock, which is dissipative;
e.g., see Ref.~\fullcite{Whitham}.) 
The characteristic frequency of these oscillations is $O(1/\epsilon)$
(i.e., one period of the oscillation shown in Fig.~\ref{f:damshock}a 
is of order $\epsilon$).
If $u_0<0$ instead, the two halves of the pulse will move away from
each other, and no shocks develop in this case.
(In hydrodynamics, solutions such as this one are called of rarefaction type.)
When the solution develops a shock, the hyperbolic system~\eqref{e:dam}
ceases to be valid, and the solution of the NLS equation in the 
weak dispersion limit must be obtained by properly regularizing the 
hyperbolic system, as we briefly discuss next.

\section{Regularization and the NLS-Whitham equations}
\label{s:nlswhitham}

The semiclassical limit of the NLS equation has been extensively studied
in the last fifteen years; e.g., see
Refs.~\fullcite{OTCMCC1986,CPAM52p613,PhysRep286p199,Kamchatnov,SJAM59p2162,FAA22p37,TMP71p584,CPAM52p655}
for different approaches, the connection with the integrable character 
and the multi-phase solutions of the NLS equation and with 
Whitham's averaging method.
A self-contained description of the regularization process can be found 
in Ref.~\fullcite{SJAM59p2162},
together with a detailed treatment of the situation in which the initial
condition contains only one frequency jump. 
Here we will limit ourselves to presenting a brief general overview 
and recalling some results which are relevant for the remainder of 
this work.
Some additional details (which are necessary to perform the calculations
described in sections~\ref{s:singlejump} and~\ref{s:multiphase})
are contained in Appendices~\ref{a:elliptic} and~\ref{a:NLS-Whitham}.

As we will see throughout this work,
for certain kinds of initial conditions
the solution of the NLS equation~\eqref{e:NLS} with small dispersion
develops high-frequency oscillations.
When this happens, the approximate hyperbolic system~\eqref{e:dam}
is inadequate to describe the dynamics of the solution,
because of strong dispersive effects appearing due to the
presence of high-frequency oscillations. 
An effective way to describe the behavior of the solution of the NLS equation 
in these situations is obtained by taking an average over these 
high-frequency oscillation via the Whitham technique,
which consists in locally approximating the the solution by
finite genus solutions of the NLS equation, then describing the
global behavior as a slow modulation of this local periodic or
quasi-periodic structure.  
The evolution of these modulations is then governed by the
Whitham equations, which are obtained by averaging the 
conservation laws of the NLS equation over one period of 
the fast oscillations, and which express the local average of 
the quasi-periodic solutions with respect to these fast oscillations
(cf.~Refs.~\fullcite{CPAM1980p739,JFM22p273,Whitham} for the Korteweg-de~Vries equation).

Recall that the hyperbolic system~\eqref{e:damRiemann} is a dispersionless limit
of the NLS equation. The presence of high-frequency oscillations in the solution of NLS
corresponds to the formation of a 
shock singularity in Eq.~\eqref{e:damRiemann} generating strong dispersion, and 
the Whitham averaging technique then provides an appropriate dispersive 
regularization of this singularity.
This regularization turns out to consist in enlarging
the system~\eqref{e:damRiemann} 
to include more than two Riemann invariants 
(as given by Eqs.~\eqref{e:nlsWhitham} below)
in such a way that the initial data for the enlarged system becomes
of rarefaction type, which in turn implies that the system possesses
a global solution for all values of time.
The solution of the original problem is then described in terms of 
the solution of an NLS-Whitham equation of finite genus. 
The dispersionless equation~\eqref{e:damRiemann} 
corresponds to the genus-zero NLS-Whitham equation.
The value of the genus is determined by the specifics of the initial 
condition considered, since these determine the number of 
Riemann invariants which are needed to regularize the 
hyperbolic system.

Let us briefly review some features of the 
regularization and introduce the NLS-Whitham equations
(see Appendices~\ref{a:elliptic} and~\ref{a:NLS-Whitham}
and Refs.~\fullcite{OTCMCC1986,CPAM52p613,SJAM59p2162,FAA22p37} 
for more details).
It is well-known that the NLS equation is integrable via the inverse
scattering transform.
The scattering problem associated with the defocusing 
NLS equation~\eqref{e:NLS} is given by the eigenvalue problem\cite{AblowitzSegur,JETP34p62}
${\cal L}\v= z\v$, 
where $\v$ is a two-component vector,
$z\in\Complex$ is the spectral parameter of the scattering problem,
and where the Lax operator ${\cal L}$ is defined by 
\begin{equation}
\label{e:scattering}
{\cal L}\v =\begin{pmatrix} -i\epsilon \partial_x & iq\\
-iq^* & i\epsilon\partial_x \end{pmatrix}\v\,. 
\end{equation}
A genus-$g$ solution of the NLS equation is associated to a genus-$g$
hyperelliptic Riemann surface $R:w^2=\mu_g(z)$, with 
\begin{equation}
\mu_g(z)= \prod_{k=1}^{2g+2}(z-\r_k)\,.
\label{e:mudef}
\end{equation}
The solution of the NLS equation~\eqref{e:NLS} corresponding to 
Eq.~\eqref{e:mudef}
is described in terms of a Baker-Akhiezer function constructed from 
Riemann theta~functions with $g$~phases
(e.g., see Ref.~\fullcite{BBEIM1994,GesztesyHolden} for details).
The branch points $\r_1,\dots,\r_{2g+2}$ of $R$ determine the spectrum 
of~${\cal L}$.
Since the Lax operator ${\cal L}$ is self-adjoint, these branch points are all real, and
we label them so that 
$\r_1<r_2<\cdots<\r_{2g+2}$.  
The spectrum of ${\cal L}$ is then given by 
$(-\infty,\r_1]\cup[\r_2,\r_3]\cup\cdots\cup
[\r_{2g},\r_{2g+1}]\cup[\r_{2g+2},\infty)$.

For an exact genus-$g$ solution of the NLS equation, 
the branch points $\r_1,\dots,\r_{2g+2}$ are obviously constant, 
independent of space and time, owing to the isospectrality
of the inverse scattering transform of the NLS equation.
The solution then gives a quasi-periodic solution with $g$ phases of the NLS
equation, and with a small dispersion of order $\epsilon^2$ implies that 
each period of the phase
is of order $\epsilon$ (i.e., high-frequency oscillations of order $1/\epsilon$).

Suppose now that the $g$-phase solution is slowly modulated. 
Then the spectral parameters $r_k$ are expected to shift slightly 
(and even additional small gaps are expected to open in general).
The shifted parameters are constant as the eigenvalues of the 
Lax operator $\cal L$.
We can treat the modulation problem using a singular perturbation method 
as follows:
if we consider modulations on the scale of $(x,t)$ in the NLS equation~\eqref{e:NLS},
the modulations are of order one, and the oscillations are of order $1/\epsilon$.
This implies that one can treat the modulation problem as a small perturbation 
of the $g$-phase solution, which is the key of the Whitham averaging method.
This method is based on an adiabatic assumption that
the leading-order solution of the NLS equation (i.e., the $g$-phase solution)
is preserved under the modulation, but with slowly changing parameters.
Employing a standard averaging perturbation method,
one introduces the fast and slow time and space scales respectively
as $(x'= x/\epsilon,\,t'=t/\epsilon)$ and $(x,t)$.
Then the first-order correction to the $g$-phase solution in the 
perturbation method compensates the motion of the parameters so that 
the parameters appear to be constant over regions of order one in $(x,t)$, 
but at the same time acquire small constant shifts due to the modulations 
in general.
The corresponding equations for the spectral parameters with respect to the 
slow scales $(x,t)$ are called the \textit{NLS-Whitham equations}, 
or simply the Whitham equations, and 
are obtained by averaging the conservation laws of NLS 
with respect to the fast oscillations at order $\epsilon$ 
(see Appendix \ref{a:NLS-Whitham} for more details).

It was shown in Refs.~\fullcite{OTCMCC1986,FAA22p37} that 
the genus-$g$ Whitham equations for the NLS equation can be 
written in the Riemann invariants form, in which the spectral parameters $r_k$ give
the Riemann invariants,
\begin{equation}
\partialderiv[]{\r_k}{t} = 
  s_k(\r_1,\dots,\r_{2g+2})\, \partialderiv[]{\r_k}x\,,
\label{e:nlsWhitham}
\end{equation}
for $k=1,\dots,2g+2$.
It then follows that for each $\epsilon\ne0$ the solution of the 
NLS-Whitham equations describes the slow modulation of
finite-genus solutions of the NLS equation.
It was also shown (cf.\ Lemma 4.1 in Ref.~\fullcite{SJAM59p2162}) 
the following, which is the most important property of
the NLS-Whitham equations:
\begin{proposition}
The characteristic velocities $s_k$ possess a double sorting property, 
i.e., $\forall k,l=1,\dots,2g+2$
they satisfy the two conditions
\begin{subequations}
\label{e:sorting}
\begin{align}
&\partialderiv{s_k}{\r_k}>0\,,\\
&\r_k<\r_l~\implies~s_k<s_l\,,
\end{align}
\end{subequations}
\label{p:doublesorting}
\end{proposition}

\par\vglue-\bigskipamount\noindent
Then, as a consequence of~\ref{p:doublesorting},
we have (Corollary 4.2 in Ref.~\fullcite{SJAM59p2162}):
\begin{corollary}
If the initial values of the $\r_k$ are each nonincreasing 
and if they satisfy the separability condition
\begin{equation}
\mathop\max\limits_{x\in\Real}\,\r_k^0(x)< \mathop\min\limits_{x\in\Real}\,\r_{k+1}^0(x)
\label{e:NLSnobreaking}
\end{equation}
$\forall k=1,\dots,2g+1$ at $t=0$,
the initial data is of rarefaction type, and therefore
the hyperbolic system of equations~\eqref{e:nlsWhitham} has a 
global solution for $t>0$, i.e., it is regular.
\label{c:uniquesolution}
\end{corollary}

In general, the initial conditions for the two Riemann invariants obtained 
from the dispersionless system~\eqref{e:NLSmodphase}
do not satisfy monotonicity and the separability condition~\eqref{e:NLSnobreaking}.
In other words, Eqs.~\eqref{e:g0invariants} 
(that is, the system~\eqref{e:nlsWhitham} with $g=0$)
in general do not have a global solution.
The regularization process then consists in enlarging 
the set of Riemann invariants so that the resulting 
NLS-Whitham equations have a global solution. 
This is done by representing the initial data for the NLS Eq.~\eqref{e:NLS}
in such a way that all the Riemann invariants $r_1,\dots,r_{2g+2}$ 
are monotonically decreasing functions of~$x$ at $t=0$.
Note that it is always possible to do so for piecewise-constant 
initial data, since in this case there is some ambiguity in how 
the spectrum of the Lax operator 
is represented in terms of the Riemann invariants.
This ambiguity can then be exploited to redefine the initial datum 
for the Whitham Eqs.~\eqref{e:nlsWhitham} by adding degenerate gaps 
(e.g., see Figs.~\ref{f:riemann1jump} and~\ref{f:riemann2jumps}).
All of the solutions discussed in sections~\ref{s:singlejump},
\ref{s:multiphase} and~\ref{s:arbitraryjumps} fall within the 
framework of piecewise-constant initial conditions.
In this case the data at $t=0$ are always genus-0, 
but highly degenerate, in the sense that they can be described 
in terms of a higher genus with degenerate gaps.

It should be noted that the adiabatic approximation 
(the use of the Whitham equations for the slow evolution) 
implies that the local genus of the solution is preserved.
A separate issue is how the genus of the solution changes 
from one region to the next.
This is a bifurcation problem through a critical point,
and can be approached by regularization, i.e., by trying to 
patch two Whitham systems with different genus at this point.
Our regularization acts as a ``globalization'',
in the sense that the Whitham equations with a proper (in general larger)
genus now describes a global behavior beyond the perturbation range. 
For fixed $x$, a regular scheme must solve a connection problem in $t$ 
because of the change of genus, and for fixed $t$ one also needs 
to solve a connection problem to match regions with different genus.

\begin{figure}[b!]
\vglue-2\bigskipamount
\centerline{\epsfxsize0.475\textwidth\epsfbox{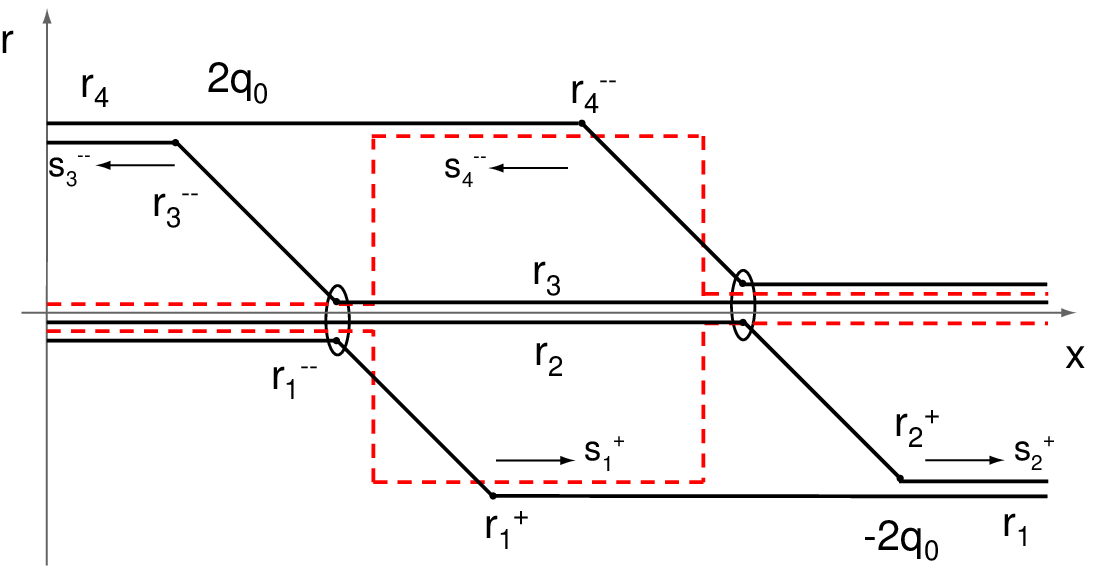}\quad\hss
\epsfxsize0.475\textwidth\epsfbox{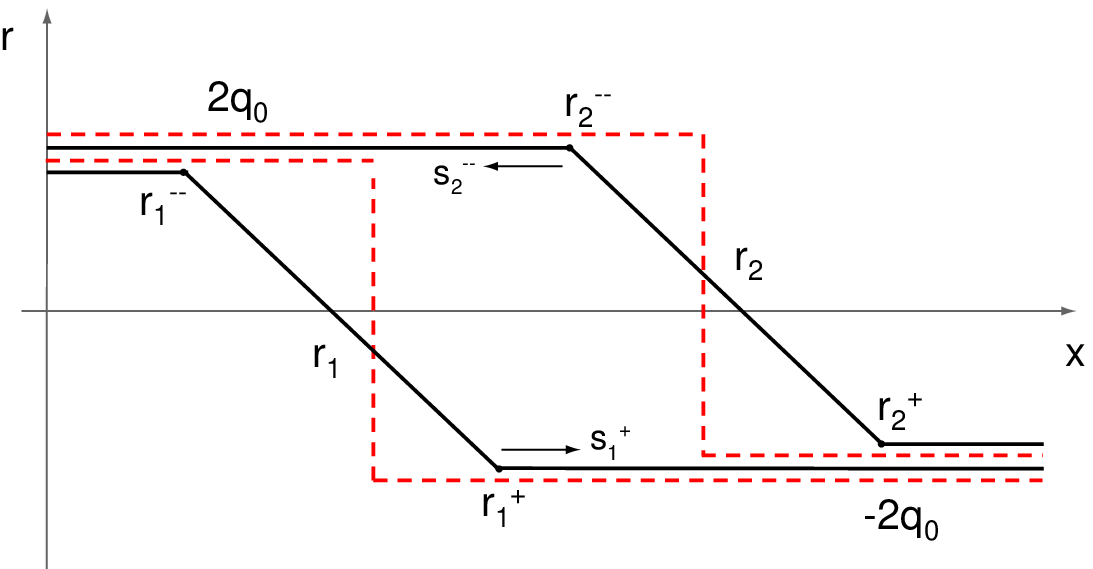}}
\medskip
\caption{Evolution of the 
Riemann invariants in two equivalent cases. 
Dashed lines: the Riemann invariants at $t=0$; 
solid lines: the invariants at $t\ne0$.
Figure~\ref{f:riemann0}a (left) 
corresponds to the square-wave initial datum 
in Eqs.~\eqref{e:dambreaking} and Fig.~\ref{f:NRZnojump},
regularized by genus-1 data;
Figure~\ref{f:riemann0}b (right) shows an equivalent diagram
of the left one, and it is given by genus-0 data.
Hereafter, the subscripts ``$-$'' and ``$+$'' refer to the value of
the invariants respectively to the left and to the right of 
their initial discontinuity.
The ellipses in Fig.~\ref{f:riemann0}a indicate ``locking'' points,
i.e., points $x_*$ for which $r_1(x_*,t)=r_2(x_*,t)=r_3(x_*,t)$
(on the left) and $r_2(x_*,t)=r_3(x_*,t)=r_4(x_*,t)$ (on the right) for all~$t$.
This implies that the regularization at those points is trivial.
Finally, note that the initial conditions $r_{1,2}^0(x)$ in the 
equivalent figure (Fig.~\ref{f:riemann0}b)
have been redefined whenever $\rho(x,0)=0$ 
(i.e., $u(x)$ itself is not defined in the NLS-Whitham equation when
$\rho(x)=0$, see Eqs.~\eqref{e:nlswhithamconsform}).}
\label{f:riemann0}
\end{figure}

It is also important to realize that for piecewise-constant initial data
there is more than one way to regularize the initial datum
(e.g., see Figs.~\ref{f:riemann0}a,b).
The minimum number of Riemann invariants that are necessary so that the system 
becomes regular is related to the genus of the solution of the 
NLS-Whitham equations.  That is, $2g+2$ invariants correspond to
a genus-$g$ solution of Eqs.~\eqref{e:nlsWhitham}.
The local genus of the solution of the NLS equation is roughly speaking 
the number of distinct frequencies which are locally present in the solution
over regions of order one in $(x,t)$.
Since the Riemann invariants are the branch points of the spectrum
of the finite-genus solution of NLS which locally approximates
the full solution,
it is then clear that the local genus is equal to the 
number of gaps determined by the local value of the
Riemann invariants (e.g., see Figs.~\ref{f:asymptoticinvariants}
and~\ref{f:asymptoticinvariants2}).
The opening or closing of one of the gaps for some values of $(x,t)$
corresponds to a local change of genus in the solution of the NLS equation.
Note however that all finite-genus solutions of the NLS equation 
with non-zero genus become singular in the limit~$\epsilon\to0^+$.
In this sense, the solution of the Whitham equations represents a
weak limit, since when $g\ne0$ the solutions of the NLS equation
only converge in an average sense (i.e., weak convergence).

Finally, with regards to Fig.~\ref{f:riemann0}
we should note that genus-1 data is necessary in order to preserve 
the value of $u(x,0)$.
With step initial data, however, no gap opens during propagation, 
which means that the data is degenerate, producing a genus-0 solution.
The situation would be different in the case of non-step initial data 
(e.g., if the transition from $\rho(x,0)=0$ to $\rho(x,0)=q_0^2$ were 
continuous); in that case oscillations would appear, as described
in Ref.~\fullcite{JNLSCI7p43}.

\section{Single-jump initial conditions}
\label{s:singlejump}

We now briefly summarize some results from Ref.~\fullcite{SJAM59p2162}
relative to a single-jump initial datum, since they provide the basis
for the framework that will be used to analyze the more complicated
scenarios discussed in the remainder of this work.
We will consider the constant-amplitude wave given by 
the (single-jump) initial condition in Eqs.~\eqref{e:rho0}
and~\eqref{e:g0u}, where we take $L\to\infty$.
The value of the original Riemann invariants in the genus-0 system
Eq.~\eqref{e:damRiemann} at $t=0$ is again given by
Eqs.~\eqref{e:riemann0onejump}, which are now valid $\forall x\in\Real$.
Four different situations arise depending on the size of the 
frequency jump~$2u_0$, as shown in Figs.~\ref{f:damshock}a--d:

\begingroup
\renewcommand\labelenumi{(\roman{enumi})}
\begin{enumerate}
\itemsep 0pt\parsep 0pt
\item
$u_0>2q_0>0$~
(Lemma~4.3 and Theorem~4.4 in Ref.~\fullcite{SJAM59p2162}).\\[1ex] 
Since $u_0>0$, the original Riemann invariants $r_{1,2}\o(x)$ 
are increasing functions of $x$, and therefore the genus-0 
system~\eqref{e:damRiemann} does not have a global solution.
In this case the problem is regularized by considering the 
genus-1 NLS-Whitham equations.
(That is, four invariants are necessary so that the resulting system has
a global solution.)  
The Riemann invariants are related to the solution at $t=0$ as follows:
\begin{align*}
&\r_1= -u_0-2q_0\,\quad\forall x\,,&
&\r_3= u_0\pm 2q_0\,\quad \mathrm{for}~x\lg 0\,,\qquad\qquad\\
&\r_2= -u_0\pm 2q_0\,\quad \mathrm{for}~x\lg 0\,,&
&\r_4= u_0+2q_0\,\quad\forall x\,,
\end{align*}
with the upper/lower signs corresponding to the upper/lower 
inequality for $x$, respectively.
The qualitative evolution of these Riemann invariants is shown in 
Fig.~\ref{f:riemann1jump}a.
The solution develops a region of genus-1 high-frequency oscillations
in the central portion of the pulse, surrounded by a genus-0 region, 
as illustrated in Fig.~\ref{f:damshock}a and Fig.~\ref{f:genus1jump}a.
As shown in Fig.~\ref{f:damshock}a, the genus-1 portion of the solution 
describes slow modulations of high-frequency oscillations, as would be
the case in a wave packet.
The genus-1 portion of the solution is located in the region $|x|<s_3^-t$.
The characteristic velocities 
are $s_3^-=-s_2^+$, $s_3^+=-s_2^-$, with $s_3^->s_3^+>0$ and
\begin{equation}
s_3^-= (u_0+q_0)
    \big[\, 1 + u_0q_0/(u_0+q_0)^2 \,\big]\,,
\qquad
s_3^+= u_0\,\big[\, 1-3a(1-a)K_1-2a^2K_2 \,\big]/\big(1-aK_1\big)
\label{e:s1jump_i}
\end{equation}
(cf.\ Eqs.~(4.16) and~(4.17) in Ref.~\fullcite{SJAM59p2162}),
where $a=2q_0/u_0<1$ and
\begin{equation*}
\def\txint{\mathop{\textstyle\int}\limits}
1 - K_n= \txint_0^{\pi/2}
  \frac{\sin^{2n+2}\theta}{ \sqrt{(1+a\,\sin^2\theta)(1+a\,\cos^2\theta)} }
  \,d\theta 
  \left/ \txint_0^{\pi/2}
  \frac{\sin^2\theta}{ \sqrt{(1+a^2\sin^2\theta)(1+a^2\cos^2\theta)} }
  \,d\theta
  \right.\,.
\end{equation*}
Hereafter, the superscripts ``$-$'' and ``+'' refer to the value 
of the Riemann invariants respectively to the left and to
the right of the discontinuities at $t=0$ (cf.~Fig.~\ref{f:riemann1jump}).
Note that, owing to Eq.~\eqref{e:nlsWhitham},
positive values of $s$ correspond to left-moving 
Riemann invariants, and that for regularized, non-increasing invariants
$r^-(x)\ge r^+(x)$ $\forall x\in\Real$.

\begin{figure}[b!]
\centerline{\epsfxsize0.495\textwidth\epsfbox{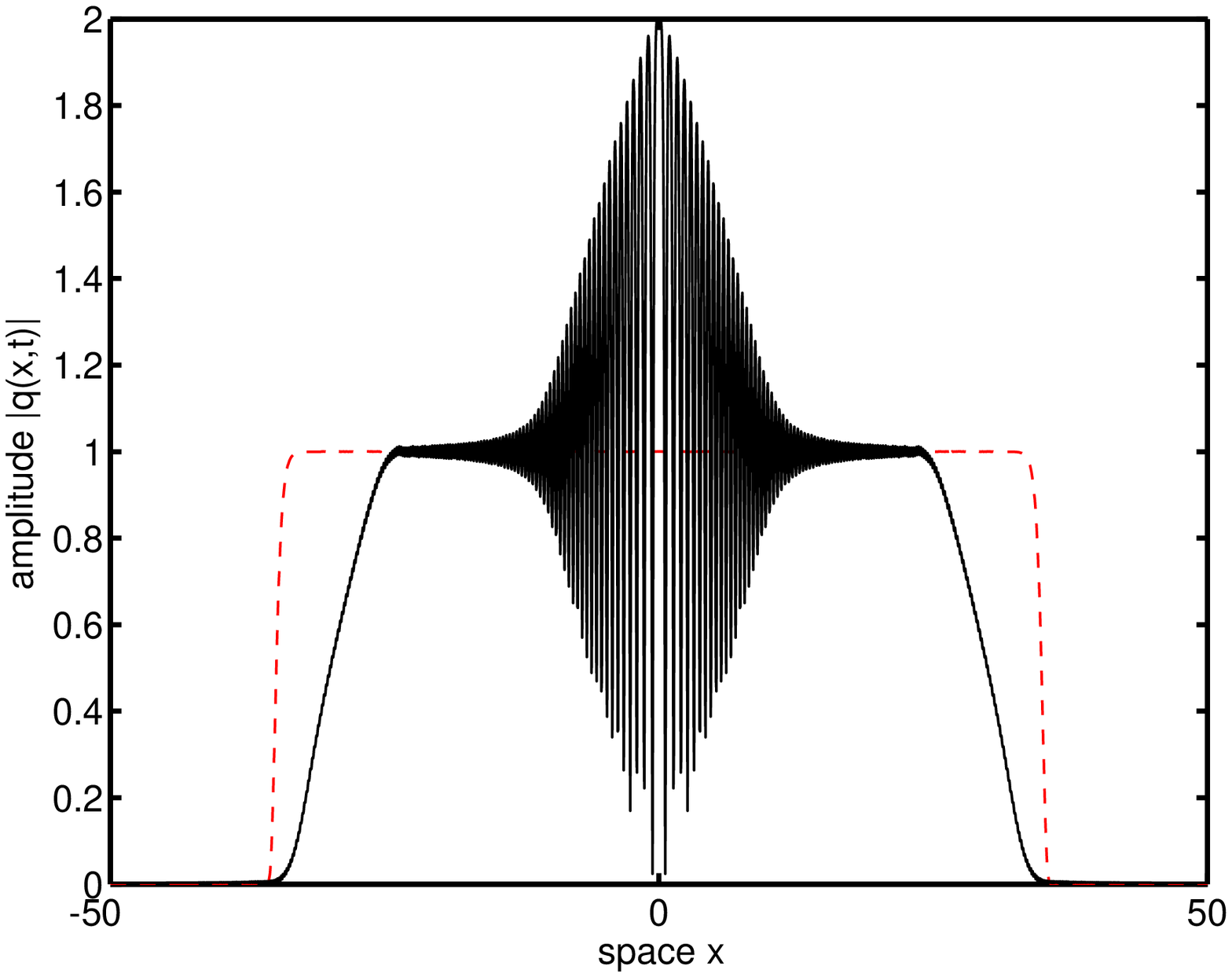}\hss
\epsfxsize0.495\textwidth\epsfbox{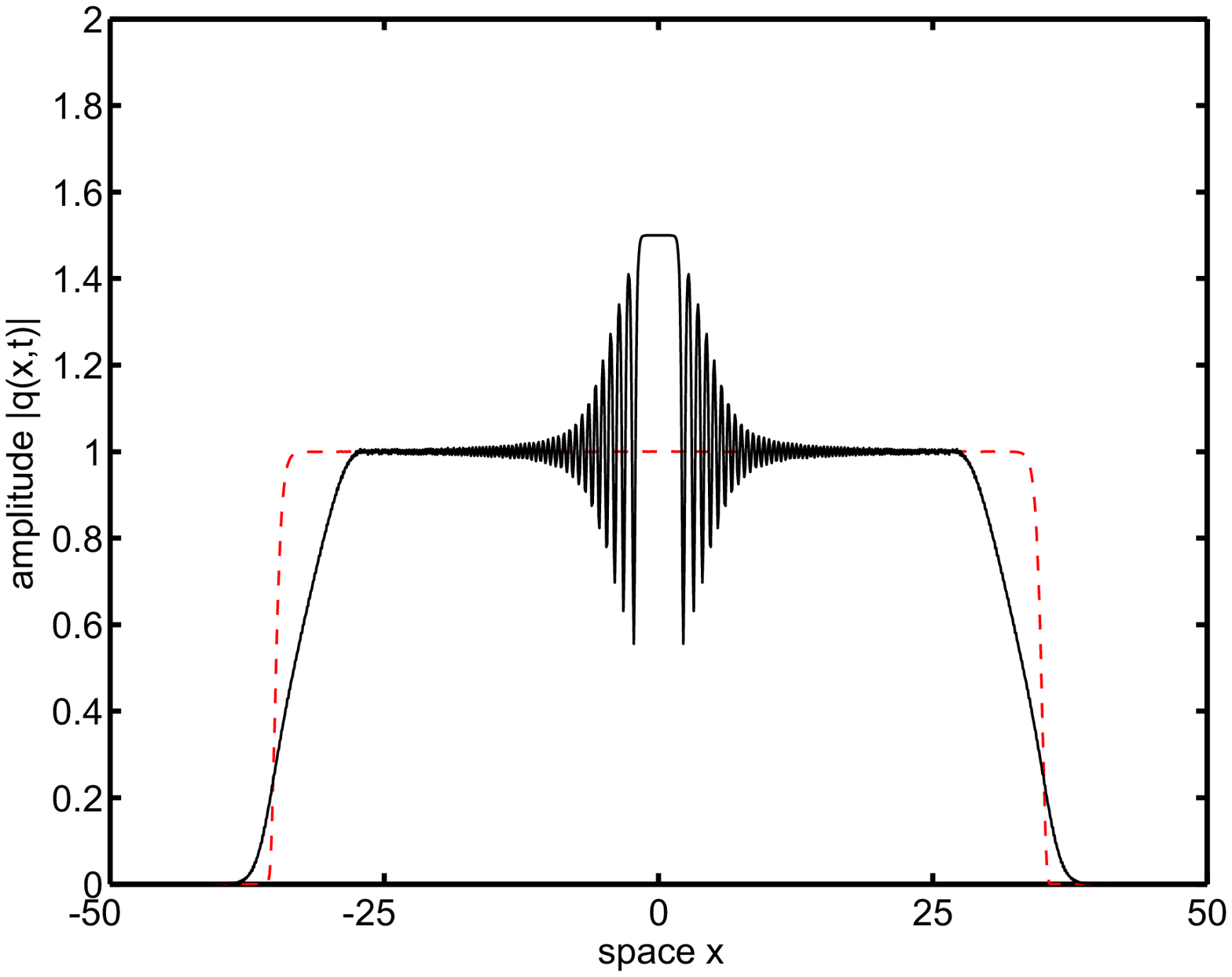}}
\centerline{\epsfxsize0.495\textwidth\epsfbox{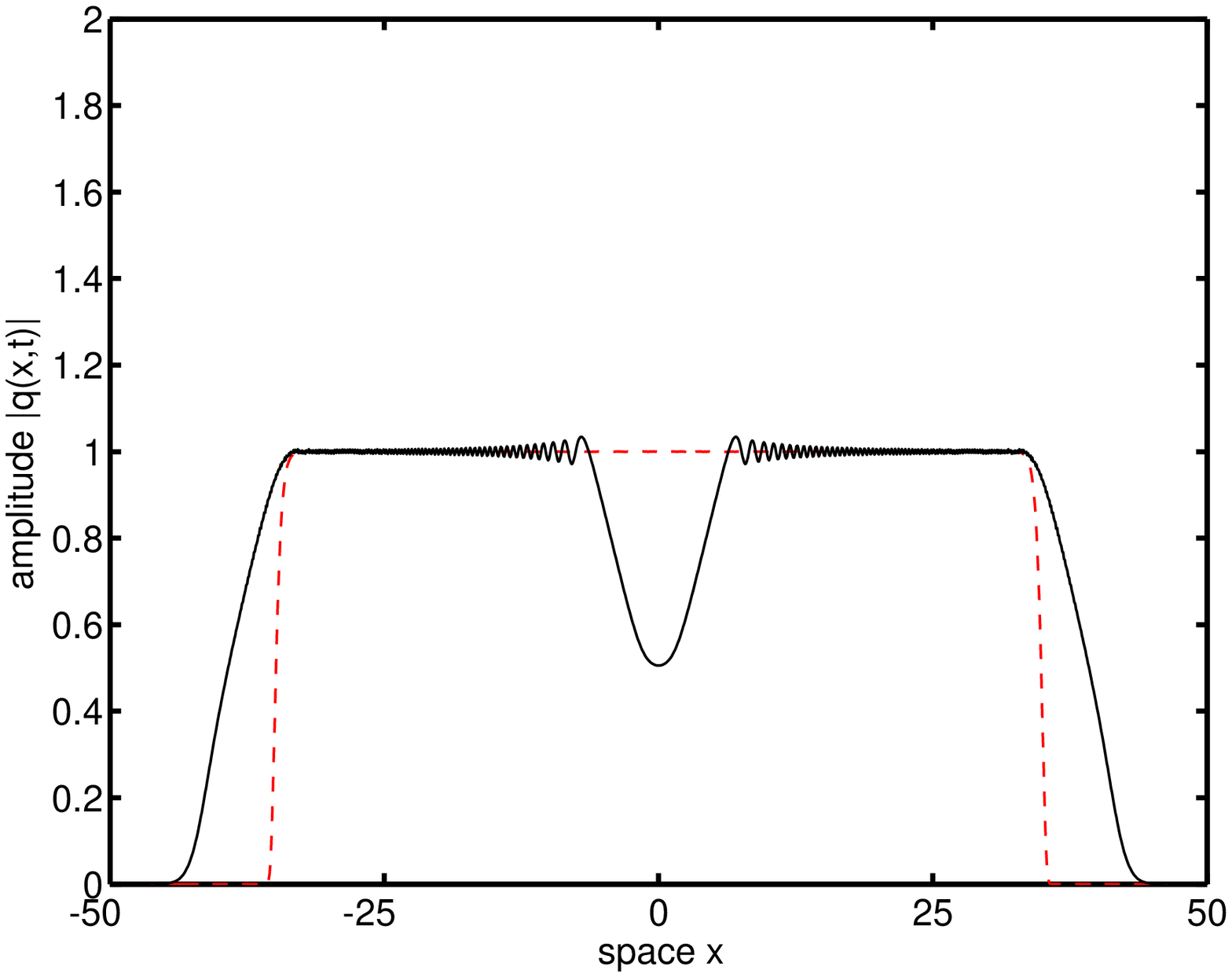}\hss
\epsfxsize0.495\textwidth\epsfbox{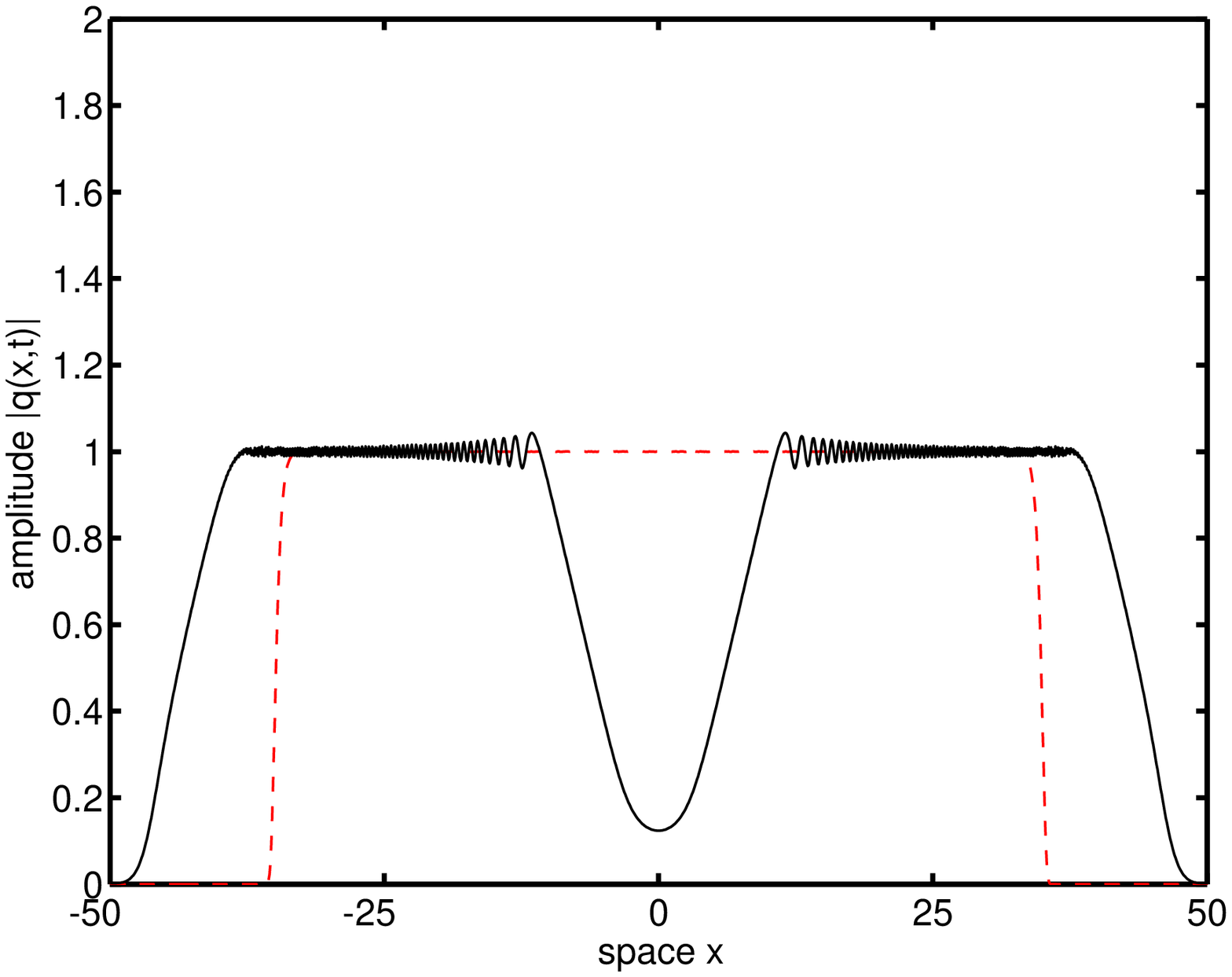}}
\caption{Deformation of an NRZ pulse 
in the presence of an initial frequency jump:
(a, top left)~$u_0=2.01$; (b, top right)~$u_0=1$; 
(c, bottom left)~$u_0=-1$; (d, bottom right)~$u_0=-2.5$.
The dotted line shows the initial condition; 
the solid line shows the result of numerical simulations of the NLS equation, 
performed as discussed in section~\ref{s:numerics}.
In all four cases it is $q_0=1$, $L=35$, and $\epsilon^2=0.1$, 
and the solution is shown at $t'=20$, where $t'=t/\epsilon$.
}
\vskip\medskipamount
\label{f:damshock}
\end{figure}

\begin{figure}[b!]
\centerline{\epsfxsize0.495\textwidth\lower1ex\hbox{\epsfbox{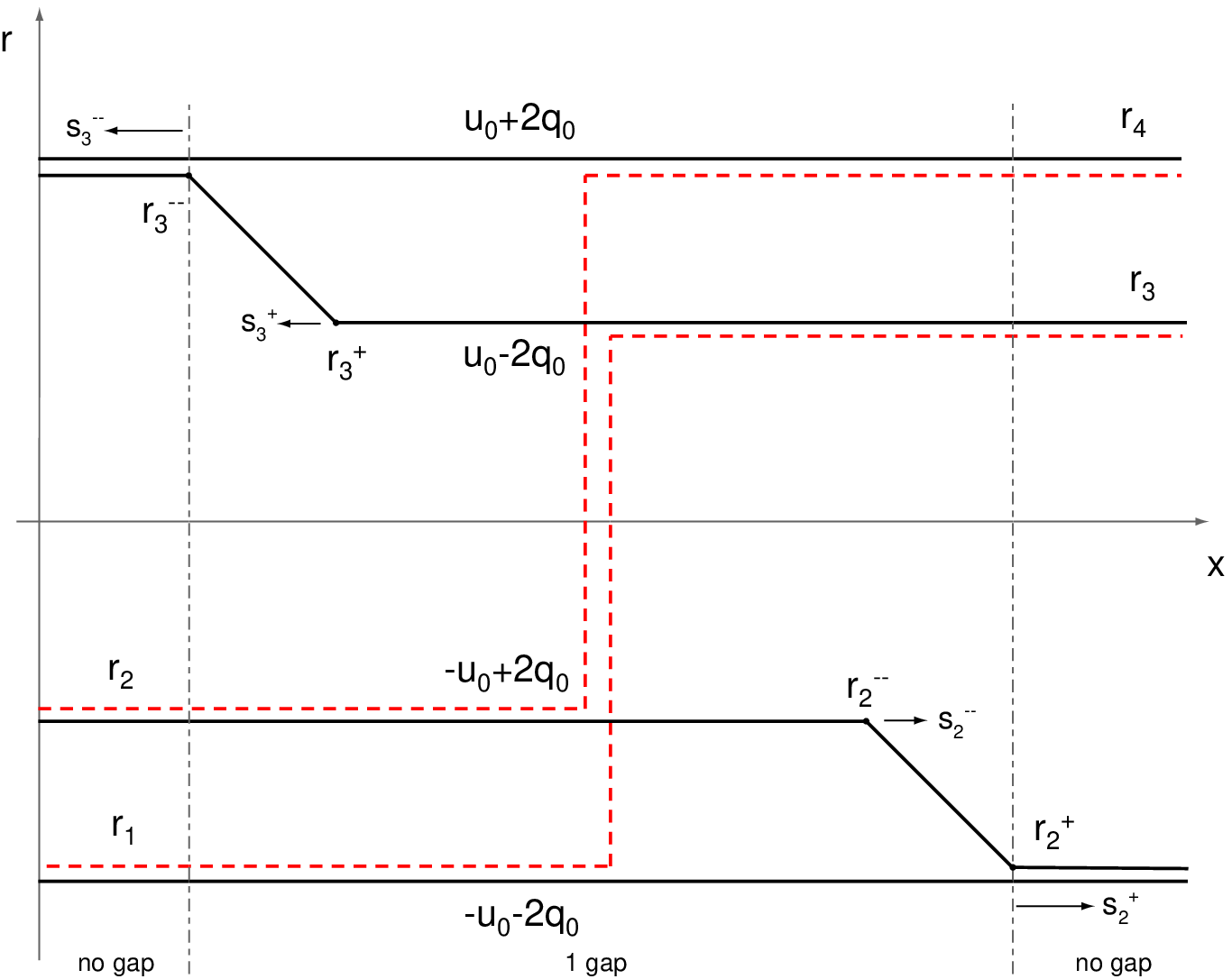}}\quad\hss
\epsfxsize0.495\textwidth\epsfbox{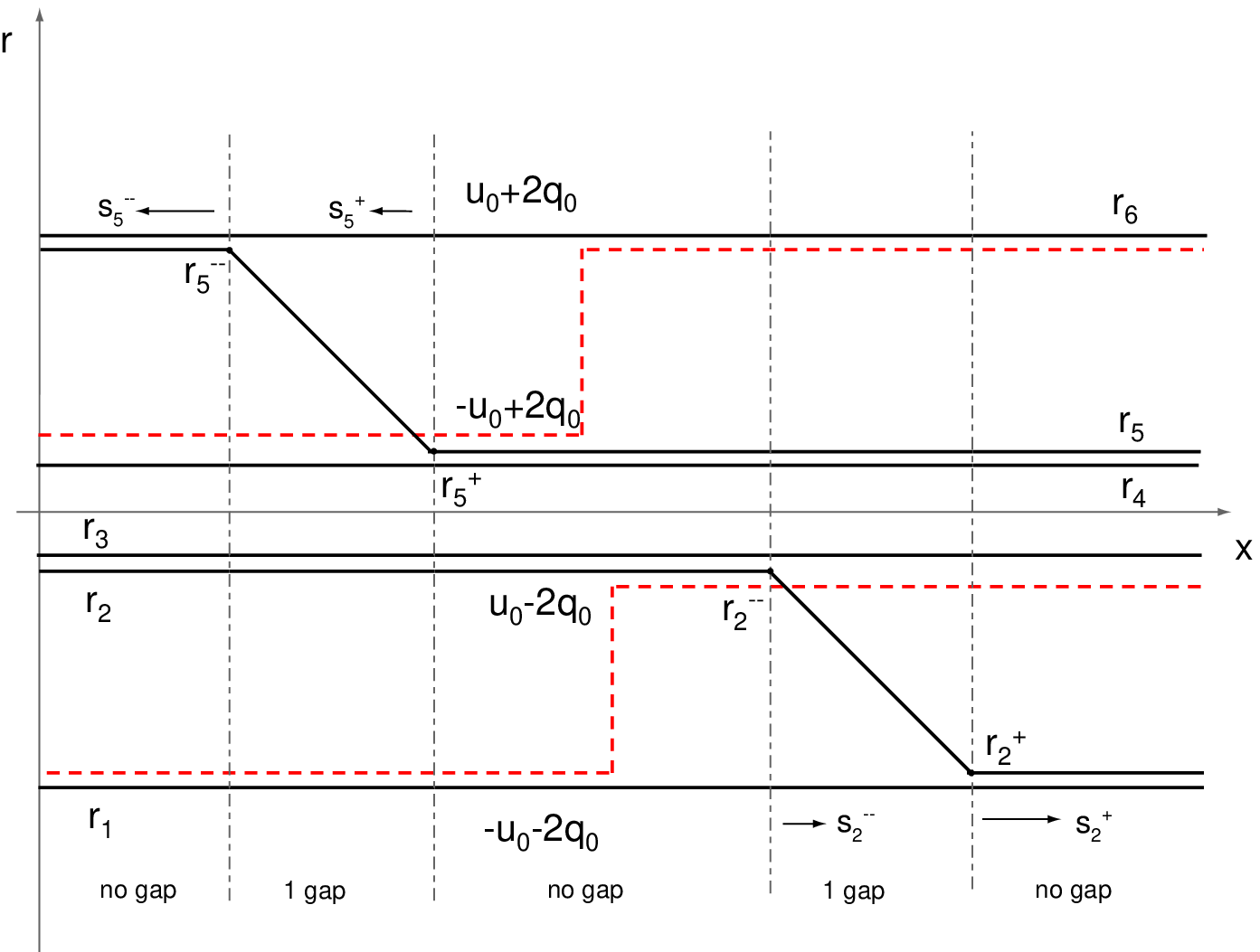}}
\medskip
\caption{Qualitative diagrams illustrating the evolution of the 
Riemann invariants: 
(a, left)~$u_0>2q_0$, with a single, expanding genus-1 region; 
(b, right)~$0<u_0<2q_0$, with an expanding genus-0 region surrounded by 
expanding genus-1 regions on either side.
Dashed lines: the original invariants $r_{1,2}^0(x)$ at $t=0$; solid lines: the regularized
invariants at $t\ne0$;
dot-dashed vertical lines: boundaries between regions of different genus.
Note that the connecting segments between 
$r_2^\pm$ in Fig.~\ref{f:riemann1jump}a,b, 
$r_3^\pm$ in Fig.~\ref{f:riemann1jump}a and 
$r_5^\pm$ in Fig.~\ref{f:riemann1jump}b are actually curved,
and are only represented here by straight lines for simplicity
(e.g., see Ref.~\protect\fullcite{SJAM52p909}).}
\label{f:riemann1jump}
\end{figure}

\item
$0<u_0<2q_0$~
(Lemma~4.5 and Theorem~4.6 in Ref.~\fullcite{SJAM59p2162}).\\[1ex]
Again, since $u_0>0$, the genus-0 Whitham equations do not have
a global solution.
This case is regularized by the genus-2 NLS-Whitham equations.
The six Riemann invariants are related to the solution at $t=0$ as follows:
\begin{align*}
&\r_1= -u_0-2q_0\,\quad\forall x\,,&
&\r_4= -u_0+2q_0\,\quad\forall x\,,\\
&\r_2= \pm u_0 - 2q_0\,\quad \mathrm{for}~x\lg 0\,,&
&\r_5= \pm u_0 + 2q_0\,\quad \mathrm{for}~x\lg 0\,,\qquad\qquad\\
&\r_3= u_0 - 2q_0\,\quad\forall x\,,&
&\r_6= u_0+2q_0\,\quad\forall x\,.
\end{align*}
The qualitative evolution of these Riemann invariants is shown in 
Fig.~\ref{f:riemann1jump}b.
Even though $g=2$ is necessary to regularize the initial data,
no genus-2 region appears.
The solution develops a genus-0 flat region (a degenerate genus-2 region)
in the central portion of the pulse, surrounded by genus-1 
high-frequency oscillations, as illustrated in Fig.~\ref{f:damshock}b.
The genus-0 and genus-1 portions of the solution are respectively located 
in the regions $|x|<s_5^+t$ and $s_5^+t<|x|<s_5^-t$,
with $s_5^+=-s_2^-$, $s_5^-=-s_2^+$, and where $s_5^->s_5^+>0$.
The characteristic velocities are (Eqs.~(4.28) in Ref.~\fullcite{SJAM59p2162})
\begin{equation}
s_5^+= q_0-\txtfrac12 u_0\,,\qquad
s_5^-= (u_0+q_0)
    \big[\, 1 + {u_0q_0}/{(u_0+q_0)^2} \,\big]\,.
\label{e:s1jump_ii}
\end{equation}
Furthermore, in the genus-0 region, the solution 
takes the value $\rho=\frac1{16}(r_2\o-r_1\o)^2$, i.e.,
\begin{equation}
\rho= q_0^2\, [\, 1 + {u_0}/{2q_0} \,]^2\,.
\end{equation}
Note that the amplitude of the genus-0 (non-oscillatory) region 
at the center increases with increasing modulation strength~$u_0$.
At the same time, however, the dependence of $s_5^+$ on $u_0$ implies 
that the width of the genus-0 region decreases with increasing
modulation strength, and the region ceases to exist when $u_0>2q_0$.
The maximum possible amplitude that can be obtained in this way
is thus $\rho_\max=4q_0^2$, obtained for $u_0=2q_0$.

This case, $0<u_0<2q_0$, is the most interesting case for applications 
because, unlike the high-frequency oscillations in the genus-1 portion, 
the high-amplitude genus-0 region can survive an appropriate 
filtering and produce short, high-intensity optical pulses.
As mentioned above, however, the maximum amplitude that can
be obtained with this arrangement is limited,
i.e., $q=(1+u_0/q_0)q_0<2q_0$.
In the next section we will see how this limit on the 
maximum pulse amplitude can be overcome
by employing more than one frequency jump.

\item
$-2q_0<u_0<0$~
(Theorem~4.7 in Ref.~\fullcite{SJAM59p2162}).\\[1ex]
Since $u_0<0$, the Riemann invariants $r_{1,2}\o(x)$ are decreasing functions
of $x$, and therefore the genus-0 NLS-Whitham equations~\eqref{e:dam}
with the Riemann invariants defined as in Eq.~\eqref{e:g0invariants}
has a global solution, and no regularization is necessary.
The solution develops a depression zone in the center, as illustrated in
Fig.~\ref{f:damshock}c.
More precisely (Eqs.~(4.31) and (4.33) in Ref.~\fullcite{SJAM59p2162}),
\[
\rho(x,t)= \left\{\!\!\begin{array}{ll}
  q_0^2\big[\, 1 +  u_0/(2q_0) \,\big]^2 	 &|x|<s_2^+t\,,\\[0.4ex] 
  \big[\, x+(u_0+2q_0)t \,\big]^2/9t^2\qquad	 &s_2^+t<|x|<s_2^-t\,,\\[0.4ex] 
  q_0^2 	                	 &|x|>s_2^-t\,,
\end{array}\right.
\]
where 
\begin{equation}
s_2^+= \txtfrac12 (u_0+2q_0)\,,\qquad
s_2^-= q_0-u_0\,.
\label{e:s1jump_iii}
\end{equation}

\begin{figure}[b!]
\begingroup
\hbox to \textwidth{\hss\epsfxsize\smallerfigwidth\epsfbox{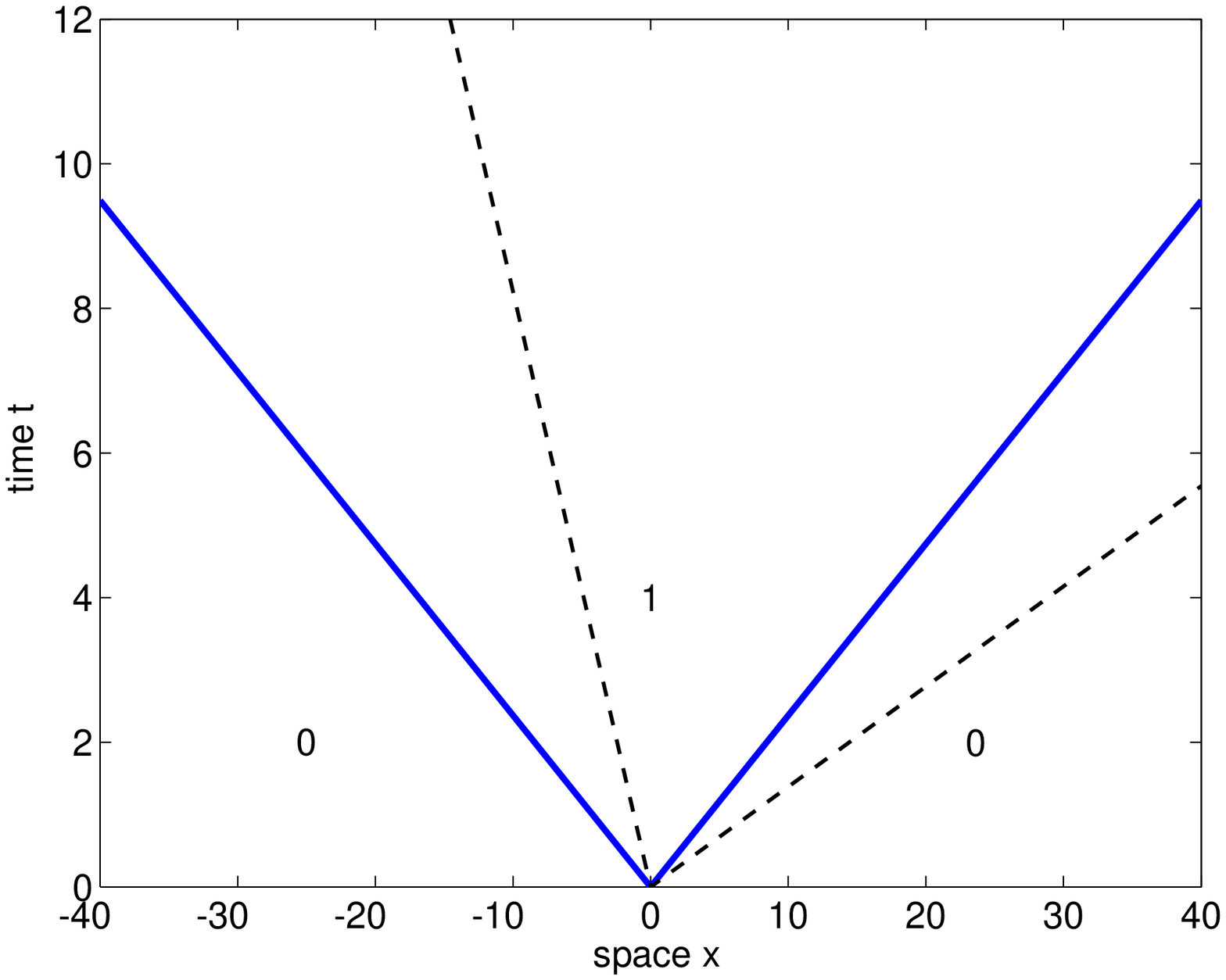}\qquad\quad  
\epsfxsize\smallerfigwidth\epsfbox{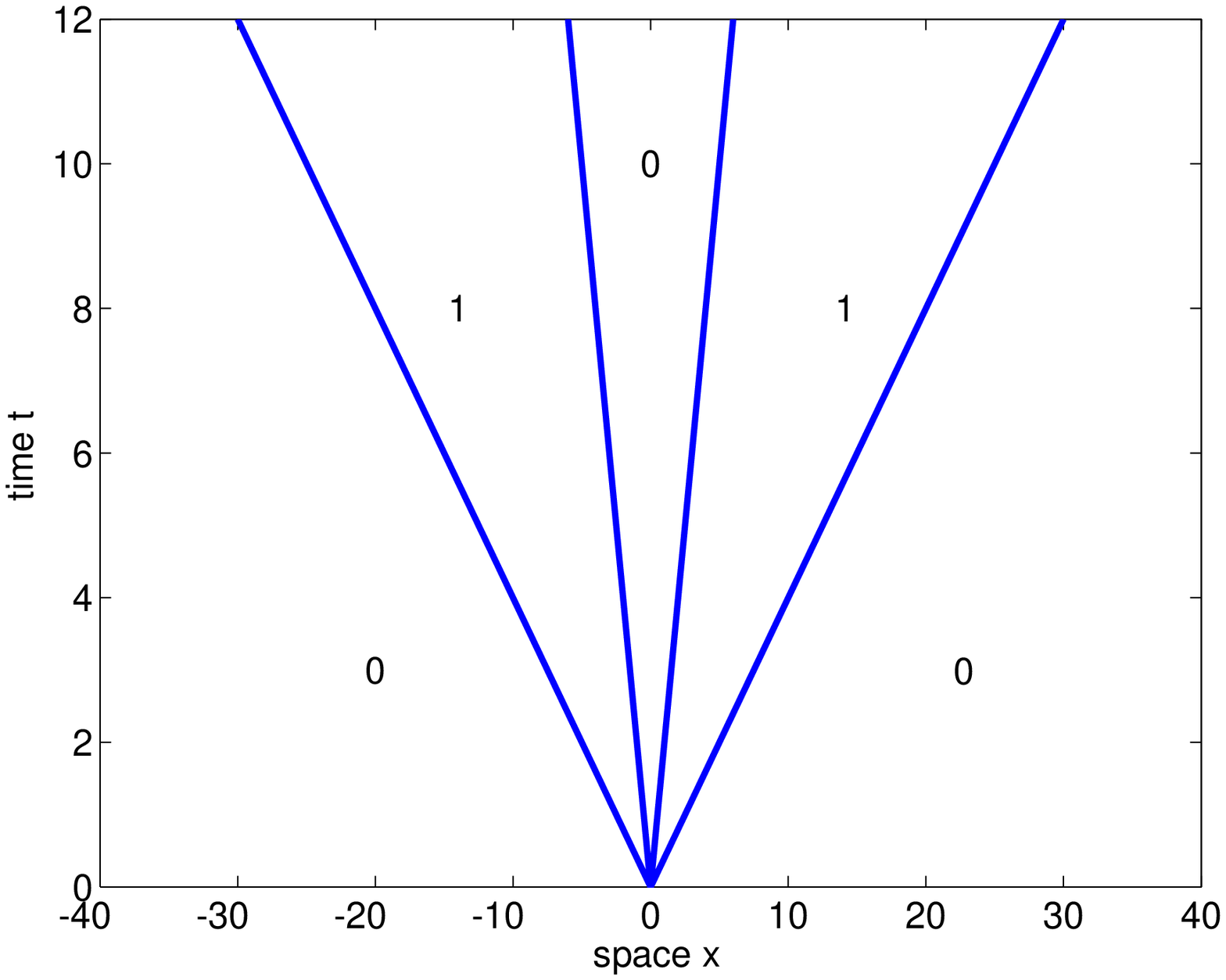}\hss}
\endgroup
\caption{The boundaries between regions of genus-0 and 
genus-1 in the $(x,t)$-plane corresponding the cases shown in 
Fig.~\ref{f:riemann1jump}:
(a, left)~$u_0=5/2$, corresponding to case~(i) and 
Figs.~\ref{f:damshock}a and~\ref{f:riemann1jump}a;
(b, right)~$u_0=1$, corresponding to case~(ii)
and Figs.~\ref{f:damshock}b and~\ref{f:riemann1jump}b.
Dashed lines: the same boundaries after adding a constant 
frequency offset $u_\mathrm{avg}=3$ to both sides of the jump, 
as discussed in section~\ref{s:multiphase}.}
\label{f:genus1jump}
\end{figure}

\item
$u_0<-2q_0$~
(Lemma~4.8 and Theorem~4.9 in Ref.~\fullcite{SJAM59p2162}).\\[1ex]
This case is regularized by the genus-1 NLS-Whitham equations~\eqref{e:dam}.
The Riemann invariants at $t=0$ are defined by:
\begin{align*}
&\r_1= u_0\pm 2q_0\,\quad \mathrm{for}~x\lg 0\,,\qquad\qquad&
&\r_3= -u_0-2q_0\,\quad\forall x\,\\
&\r_2= u_0+2q_0\,\quad\forall x\,,&
&\r_4= -u_0\pm 2q_0\,\quad \mathrm{for}~x\lg 0\,,
\end{align*}
The corresponding solution is shown in Fig.~\ref{f:damshock}d.
Even though $g=1$ is necessary to regularize the initial data,
the gap never opens, and this case produces a degenerate 
genus-0 solution.
The solution is similar to the one in the previous case, 
except that $q(x,t)$ tends to zero at $x=0$.  
More precisely, $\rho(x,t)=q_0^2$ for $|x|>s_4^-t$ 
and $\rho(x,t)=0$ in the limit $\epsilon\to0^+$
in the region $|x|<s_4^+t$, where (Eqs.~(4.36) in Ref.~\fullcite{SJAM59p2162})
\begin{equation}
s_4^+= -u_0-2q_0\,,\qquad
s_4^-= -u_0+2q_0\,.
\label{e:s1jump_iv}
\end{equation}
(The nonzero value of $q(0,t)$ in Fig.~\ref{f:damshock}d is due to the 
finiteness of $\epsilon$ in the numerical simulations.)
Furthermore, the solution does not develop genus-1 regions
(the high-frequency oscillations visible in Fig.~\ref{f:damshock}d 
disappear in the limit $\epsilon\to0^+$ in the case of step initial data).
\end{enumerate}
\endgroup

The calculation of the characteristic speeds for the Riemann invariants 
is based on the formulation of the NLS-Whitham equations. 
We refer the reader to 
appendices~\ref{a:NLS-Whitham} and~\ref{a:characteristicspeeds}
and to Refs.~\fullcite{BlochKodama,SJAM52p909,SJAM59p2162}
for further details.
Note that the outer boundaries of the genus-1 region are given
by $x= \pm s_\mathrm{outer}t$, where 
$s_\mathrm{outer}$ is the same in case~(i) and case~(ii):
$s_\mathrm{outer}= (u_0+q_0)\,\big[\, 1 + u_0q_0/(u_0+q_0)^2 \,\big]$
(cf.\ $s_3^-$ in Eq.~\eqref{e:s1jump_i} and $s_5^-$ 
in Eq.~\eqref{e:s1jump_ii}).
Note also that the location of the boundaries between regions
of genus-0 and genus-1 in the numerical simulations shown in 
Figs.~\ref{f:damshock}a--d agrees very well with the analytical results 
just presented, even though the value of $\epsilon$ used is not very small.

Case~(ii) is the most interesting for applications because of the
high-amplitude genus-0 region.
Also, in cases~(iii) and~(iv) (i.e., when $u_0<0$)
the solution is only of genus-0, but lower-amplitude.  
For this reason we will mainly focus our attention to the case of 
positive frequency jumps.
Hereafter, we refer to case~(ii) as a ``sub-critical'' frequency jump
and to case~(i) as a ``super-critical'' frequency jump.
More precisely, 
\begin{definition}
We say that a single frequency jump located at $x=x_0$ is super-critical 
if $r_1\o(x_0^+)>r_2\o(x_0^-)$, sub-critical if $r_1\o(x_0^+)<r_2\o(x_0^-)$, 
where as usual $r_k^0(x^\pm)=\lim_{\Delta x\to0^\pm}r_k^0(x+\Delta x)$.
\label{d:criticalsinglejump}
\end{definition}
In the following sections we generalize the above solutions and discuss
the behavior of solutions of the finite-genus NLS-Whitham equations 
in the presence of an arbitrary number of jumps in the initial data.

\section{Behavior of finite-genus solutions: Two frequency jumps}
\label{s:multiphase}

More complicated situations than those described in the previous section
arise when the initial conditions contain more than one frequency jump.
The main purpose of this and the following section is to study
the interaction among finite-genus solutions of the NLS-Whitham
equations generated by those frequency jumps.
First of all, however, let us briefly discuss the effect of adding 
a non-zero average frequency across the jump.
The Galilean invariance of the NLS equation implies that it is possible 
to redefine the local frequency $u(x,0)$ up to an arbitrary additive constant.
The effect of such a constant, representing a global frequency translation, 
is just a change in the overall group velocity of the solution.
More precisely, if $q_o(x,t)$ is a solution of Eq.~\eqref{e:NLS},
so is
\begin{equation}
q_c(x,t)= \exp[i(cx-c^2t/2)/\epsilon]\,\,q_o(x+c t, t)\,,
\label{e:Galilean}
\end{equation}
for any constant~$c$.  
Corresponding to Eq.~\eqref{e:Galilean} there exists a Galilean symmetry of
the NLS-Whitham equations~\eqref{e:nlsWhitham} for the Riemann invariants:
namely, 
if $r_j(x,t)$ is a solution of the Whitham equations~\eqref{e:nlsWhitham},
then so is the translated solution $r_j(x+ct,t)+c$. 
Note from Eq.~\eqref{e:Galilean} that a positive frequency~$c/\epsilon$ 
corresponds to a negative shift~$-c$ in the group velocity~$dx/dt$.
Also, as an effect of the Galilean transformation~\eqref{e:Galilean}
the rescaled frequency of the transformed solution is shifted by $c$, i.e., 
$u_c(x,t)= \epsilon\,\partial_x[\arg\,q_c(x,t)]=
  c+\epsilon\partial_x[\arg\,q_o(x+ct,t)]$.
In terms of the decomposition of the $(x,t)$-plane into regions of different
genus, we thus have the following:
\begin{lemma}
\label{l:Galilean}
The boundaries between genus-0 and genus-1 regions for a single 
frequency jump upon adding a nonzero average frequency $u_\mathrm{avg}$
to the initial condition are given by $x=s_\pm t$, where 
$s_\pm= \pm s_\mathrm{old} + u_\mathrm{avg}$\,, and where
$s_\mathrm{old}$ is still given by 
Eqs.~\eqref{e:s1jump_i}, \eqref{e:s1jump_ii}, \eqref{e:s1jump_iii}
and~\eqref{e:s1jump_iv} in cases (i--iv), respectively.
\end{lemma}
Thus, adding a nonzero frequency offset to both sides of a single
frequency jump
has the effect of tilting the corresponding boundaries between 
regions of different genus, as shown in Fig.~\ref{f:genus1jump}a.
If only one jump is present, it is obviously possible to choose 
the average frequency so that both of the boundaries of the
outermost genus-0 region move in the same direction.
In the presence of several frequency jumps, however,
is not always possible to do so, as will be discussed later.
Because of the Galilean invariance, we will sometimes 
describe the initial condition for $u(x,0)$ in terms of the 
frequency jumps, defined as $C_j= u(X_j^+,0)-u(X_j^-,0)$ 
for all $j=1,\dots,N$, where $N$ is the total number of jumps
and $X_1,\dots,X_N$ are the jump locations.

We now turn to initial conditions with two frequency jumps.
We will consider constant-amplitude initial 
conditions with $|q(x,0)|=q_0$ $\forall x\in\Real$.
For simplicity we will take symmetric jumps of size $C_1=C_2=:2u_0$ 
located at $X_2=-X_1=\Delta X/2>0$,
so that $2u_0$ is the size of each frequency jump,
as in the single-jump case.
That is, let
\begin{equation}
u(x,0)= \left\{\!\!\begin{array}{ll}
  -2u_0 	 &x<-\Delta X/2\,,\\
  0\qquad	 &|x|<\Delta X/2\,,\\
  2u_0         	 &x>\Delta X/2\,,
\end{array}\right.
\end{equation}
again with $|q(x,0)|=q_0>0$ $\forall x\in\Real$.
In the absence of either one of the two jumps, 
the solution would behave according to the theory described in 
Section~\ref{s:singlejump}, with the only exception that
the average frequency across the jump is now nonzero.
More precisely, since the average frequency at $x=\pm\Delta X/2$ is~$\pm u_0$,
the solutions described in the previous section would move with
velocity $\mp u_0$.
In other words, the central portion of the solutions described in 
Section~\ref{s:singlejump} will tend to move towards each other
if $u_0>0$ and away from each other if $u_0<0$.
As we will see, this description provides an accurate picture of the 
overall solution at sufficiently small propagation times.
After this initial stage, however, significant interaction effects appear.

In analogy with the calculations described in Section~\ref{s:singlejump}, 
let us now proceed to analyzing the solution by looking at the Riemann 
invariants.  The initial values $r_{1,2}\o(x)$ of the non-regularized
Riemann invariants 
for the genus-0 NLS-Whitham equations~\eqref{e:damRiemann} are
\begin{gather}
r_1\o(x)= 
  \left\{\!\!  \begin{array}{lll}
    -2q_0 -2u_0  &\quad x<-\Delta X/2\,,\\
    -2q_0        &\quad |x|<\Delta X/2\,,\\ 
    -2q_0 +2u_0 &\quad x>\Delta X/2\,,\end{array}
  \right.
\qquad
r_2\o(x)= 
  \left\{\!\!  \begin{array}{lll}
    2q_0 - 2u_0 &\quad x<-\Delta X/2\,,\\
    2q_0        &\quad |x|<\Delta X/2\,,\\ 
    2q_0 + 2u_0 &\quad x>\Delta X/2\,.\end{array}
  \right.
\end{gather}
These invariants are shown as dashed lines in Figs.~\ref{f:riemann2jumps}a--d.
Since expressing the values of all the regularized Riemann invariants 
at $t=0$ in terms of the initial datum would be rather tedious,
and since all the values can be easily inferred from Figs.~\ref{f:riemann2jumps}a--d,
we omit the formulae for brevity.

Based on Proposition~\ref{p:doublesorting} and Corollary~\ref{c:uniquesolution},
the following four scenarios arise depending on the size of the 
frequency jumps:

\begingroup
\renewcommand\labelenumi{(\roman{enumi})}
\begin{enumerate}
\itemsep 0pt\parsep 0pt

\item
$u_0>2q_0>0$:\\
This case is regularized by the genus-2 NLS-Whitham equations
(i.e., with six invariants),
as illustrated in Fig.~\ref{f:riemann2jumps}a,
where the (two) original and (six) regularized Riemann invariants 
are shown respectively by dashed and solid lines.

The regions of genus-0, 1 and 2 for the solution of the NLS equation
are shown in the bifurcation diagram in Fig.~\ref{f:genus2jumps}a.
Both jumps are super-critical individually, so 
two genus-1 regions open up, one at the location of each frequency jump.
These two regions interact and they create a genus-2 region
at the center of the solution.
The genus-2 region then expands forever
(cf.~Propositions~\ref{p:s0before} and~\ref{p:g0outer}).
In terms of the numerical simulations described in the next section,
this case corresponds to Fig.~\ref{f:twojumps}a.

\item
$q_0<u_0<2q_0$:\\
This case is regularized by the genus-4 NLS-Whitham equations
(i.e., with ten invariants),
as illustrated in Fig.~\ref{f:riemann2jumps}c,d,
where the original and regularized Riemann invariants are shown at two
different values of $t$. 

The regions of genus-0, 1 and 2 for the solution of the NLS equation
are shown in the bifurcation diagram in Fig.~\ref{f:genus2jumps}b.
Both jumps are sub-critical individually; so
two genus-0 regions open up initially at the location of each frequency jump
(each surrounded by genus-1 regions, as in case~(ii) of section~\ref{s:singlejump}).
As these two portions of the solution interact, a genus-0 region
forms temporarily in the central portion of the solution.
This region disappears after a while, however, and a genus-2 region 
forms which expands forever as in the previous case
(cf.~Propositions~\ref{p:s0before}, \ref{p:g0outer} and~\ref{p:g0inner}).
In terms of the numerical simulations,
this case corresponds to Fig.~\ref{f:twojumps}b and 
Figs.~\ref{f:jumpsize}c,d.

\begin{figure}[b!]
\bigskip
\hbox to \textwidth{\hss\epsfxsize\figwidth\epsfbox{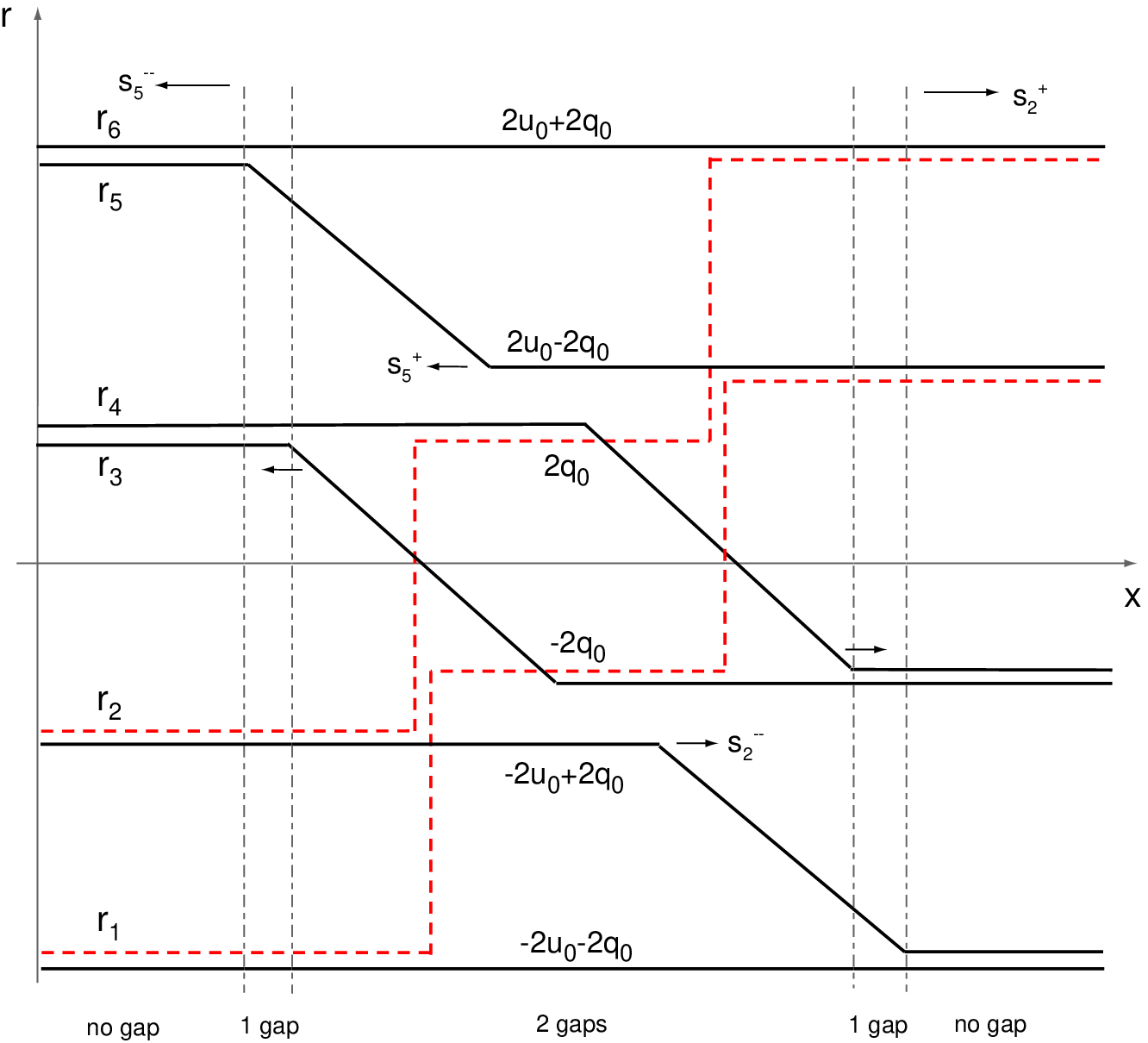}\quad\hss
\epsfxsize\figwidth\epsfbox{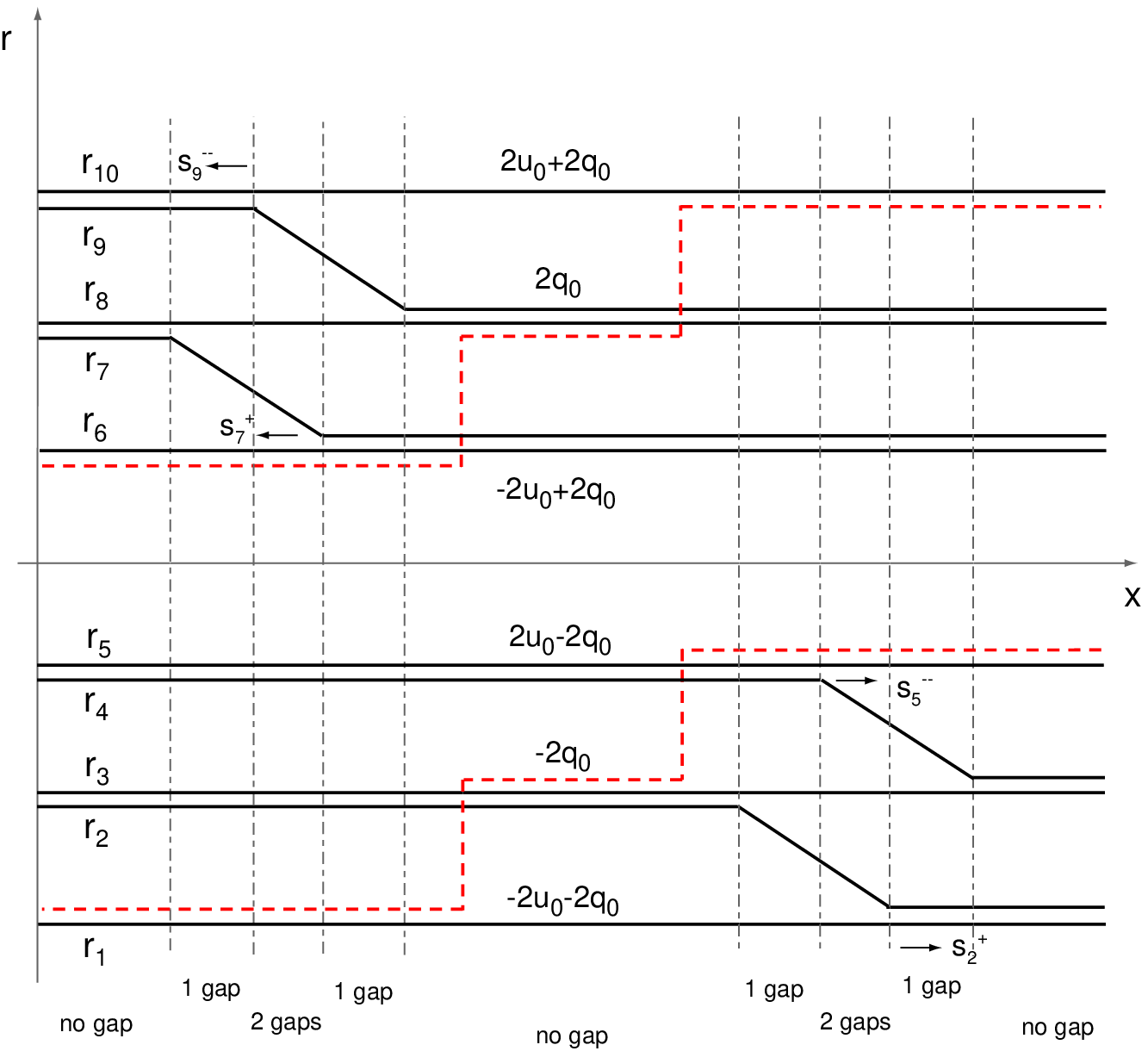}\hss}
\bigskip
\hbox to \textwidth{\hss\epsfxsize\figwidth\epsfbox{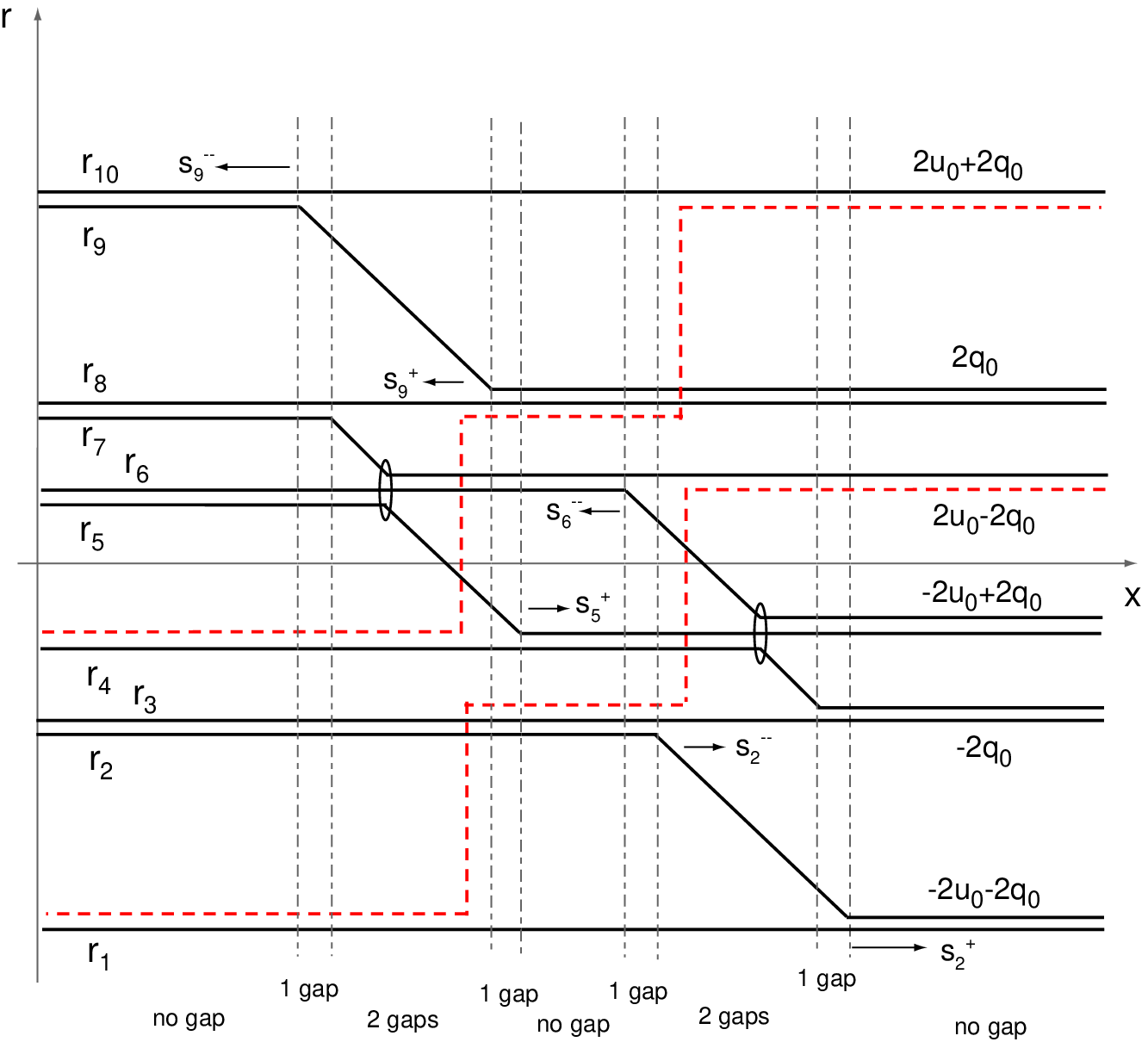}\quad\hss
\epsfxsize\figwidth\epsfbox{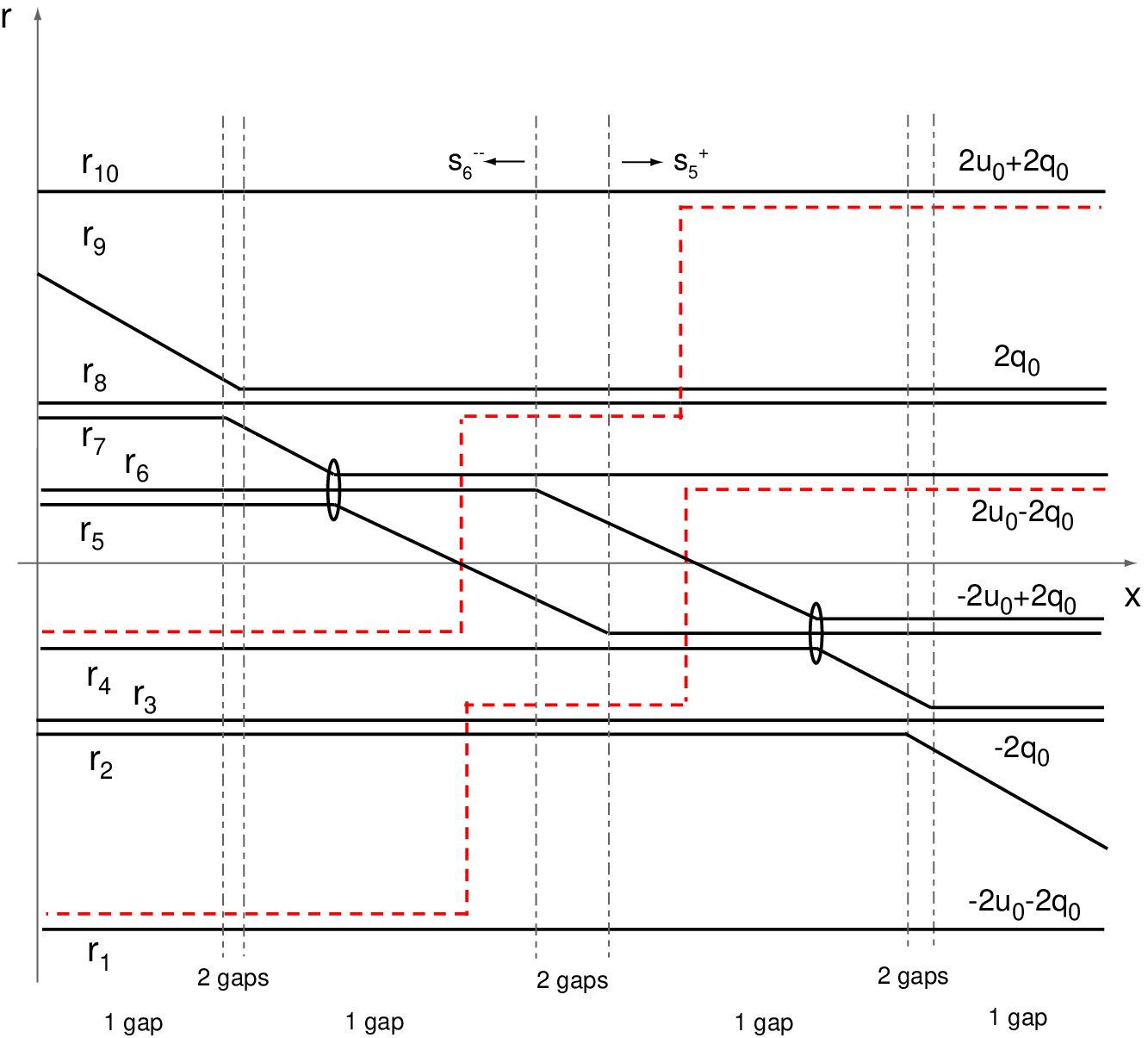}\hss}
\bigskip
\caption{Qualitative diagrams illustrating the evolution of the 
Riemann invariants:
(a, top left)~$u_0>2q_0$, corresponding to case~(i) and showing expanding
genus-1 and genus-2 regions; 
(b, top right)~$0<u_0<q_0$, corresponding to case~(iii) and showing an expanding
genus-0 central region;
(c, bottom left)~$q_0<u_0<2q_0$ corresponding to the intermediate case~(ii), 
and showing the invariants at small time~$t$ and a temporary
genus-0 region in the center;
(d, bottom right)~the same intermediate case (i.e., $q_0<u_0<2q_0$ as in (c)),
but now showing the solution at large time~$t$ and an expanding
genus-2 region in the center.
Dashed lines: the original invariants $r_{1,2}^0(x)$ at $t=0$; 
solid lines: the regularized invariants at $t\ne0$;
dot-dashed vertical lines: boundaries between regions of different genus.
As in section~\ref{s:nlswhitham},
the ellipses in Figs.~\ref{f:riemann2jumps}c,d indicate ``locking points'',
i.e., locations corresponding to the same value of~$x$ $\forall t$
(cf.~Remark~\ref{r:locking}.5).
Finally, note that, as in Fig.~\ref{f:riemann1jump}, those portions of the 
invariants that coincide in the limit $\epsilon\to0^+$ can be omitted
when calculating the local genus.}
\label{f:riemann2jumps}
\end{figure}

\item
$0<u_0<q_0$:\\
This case is also regularized by the genus-4 NLS-Whitham equations.
The original and regularized Riemann invariants are shown in 
Fig.~\ref{f:riemann2jumps}b,
and the regions of genus-0, 1 and~2 for the solution of the NLS equation
are shown in the bifurcation diagram in Fig.~\ref{f:genus2jumps}d.

As in the previous case, the individual jumps are sub-critical, so
two genus-0 regions open up initially at the location of each frequency jump.
Also similarly to the previous case,
a genus-0 region forms in the central portion of the solution 
as the two genus-0 portions of the solution interact.
Contrary to the previous case, however, the genus-0 portion of the
solution now persists and expands forever
(cf.~Propositions~\ref{p:s0before}, \ref{p:g0outer} and~\ref{p:g0inner}).
Numerical simulations corresponding to this case are shown in
Fig.~\ref{f:twojumps}c and Figs.~\ref{f:jumpsize}a,b.

\item
$u_0<0$:\\
As in the single-jump case, two expanding depression regions form at the 
location of each frequency jump.  Eventually, these regions merge to form 
a unique depression zone.
As in the single-jump case, the details of the regularization process vary 
depending on the value of $u_0$.
Because the solution is always genus-0, however, this case is not as
interesting as the previous ones in terms of applications,
and therefore it will not be investigated further.
\end{enumerate}
\endgroup

Let us now discuss more quantitatively the behavior of the solutions in
each of the above scenarios.
When $u_0>0$ as in cases (i)--(iii), the initial values of
the non-regularized Riemann invariants $r_{1,2}\o(x)$ are 
increasing functions of~$x$, and one has:
\begin{subequations}
\label{e:r12infty}
\begin{gather}
\mathop\max\limits_{x\in\Real}r_1\o(x)= 
  r_1\o(\infty)= \lim_{x\to\infty}r_1\o(x)= 2(u_0-q_0)\,,
\\
\mathop\min\limits_{x\in\Real}r_2\o(x)= 
  r_2\o(-\infty)= \lim_{x\to-\infty}r_2\o(x)= -2(u_0-q_0)\,.
\end{gather}
\end{subequations}
As we will see in section~\ref{s:numerics}, cases~(ii) and~(iii) offer 
the most useful behavior for practical applications.
In particular, the case $u_0=q_0$, which is the separatrix
between cases~(ii) and~(iii), is especially interesting
since it produces a high-amplitude genus-0 region of constant width
at the center of the pulse, as shown in Fig.~\ref{f:genus2jumps}c.
This case is regularized by the genus-3 NLS-Whitham equations,
corresponding to Fig.~\ref{f:riemann2jumps}b with $r_5$ and $r_6$ deleted
and where now $r_1\o(\infty)=r_2(-\infty)$, since $u_0=q_0$.
Hereafter, 
we will call the case $u_0=q_0$ the ``critical'' two-jump case, 
and we will refer to cases~(ii) and~(iii) as the super-critical
and sub-critical two-jump cases, respectively.
More precisely, 
\begin{definition}
We say that an arbitrary collection of positive frequency jumps 
$C_1,\dots,C_N$ located at positions
$X_1,\dots,X_N$ is collectively super-critical 
if $r_1\o(\infty)>r_2\o(-\infty)$, 
collectively sub-critical if $r_1\o(\infty)<r_2\o(-\infty)$, 
and collectively critical if $r_1\o(\infty)=r_2\o(-\infty)$,
where as usual $r_k\o(\pm\infty)=\lim_{x\to\pm\infty}r_k\o(x)$.
\label{d:criticaljumpcollection}
\end{definition}
Note the difference between the above and Definition~\ref{d:criticalsinglejump}, 
which distinguishes whether a single jump is \textit{individually} 
sub-critical or super-critical.

\begin{figure}[b!]
\bigskip
\begingroup
\kern\smallskipamount
\hbox to \textwidth{\hss\epsfxsize\smallerfigwidth\epsfbox{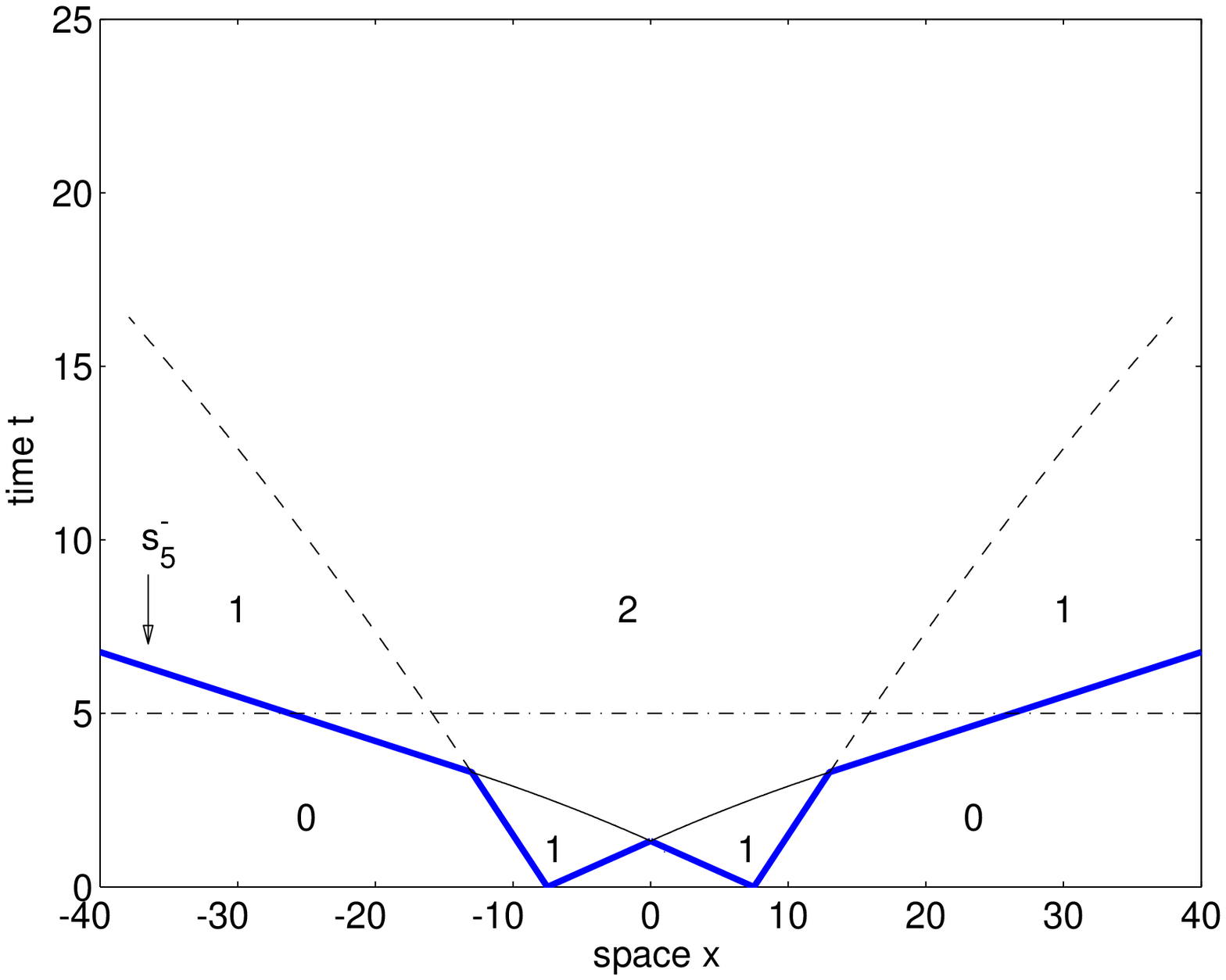}\qquad\quad  
\epsfxsize\smallerfigwidth\epsfbox{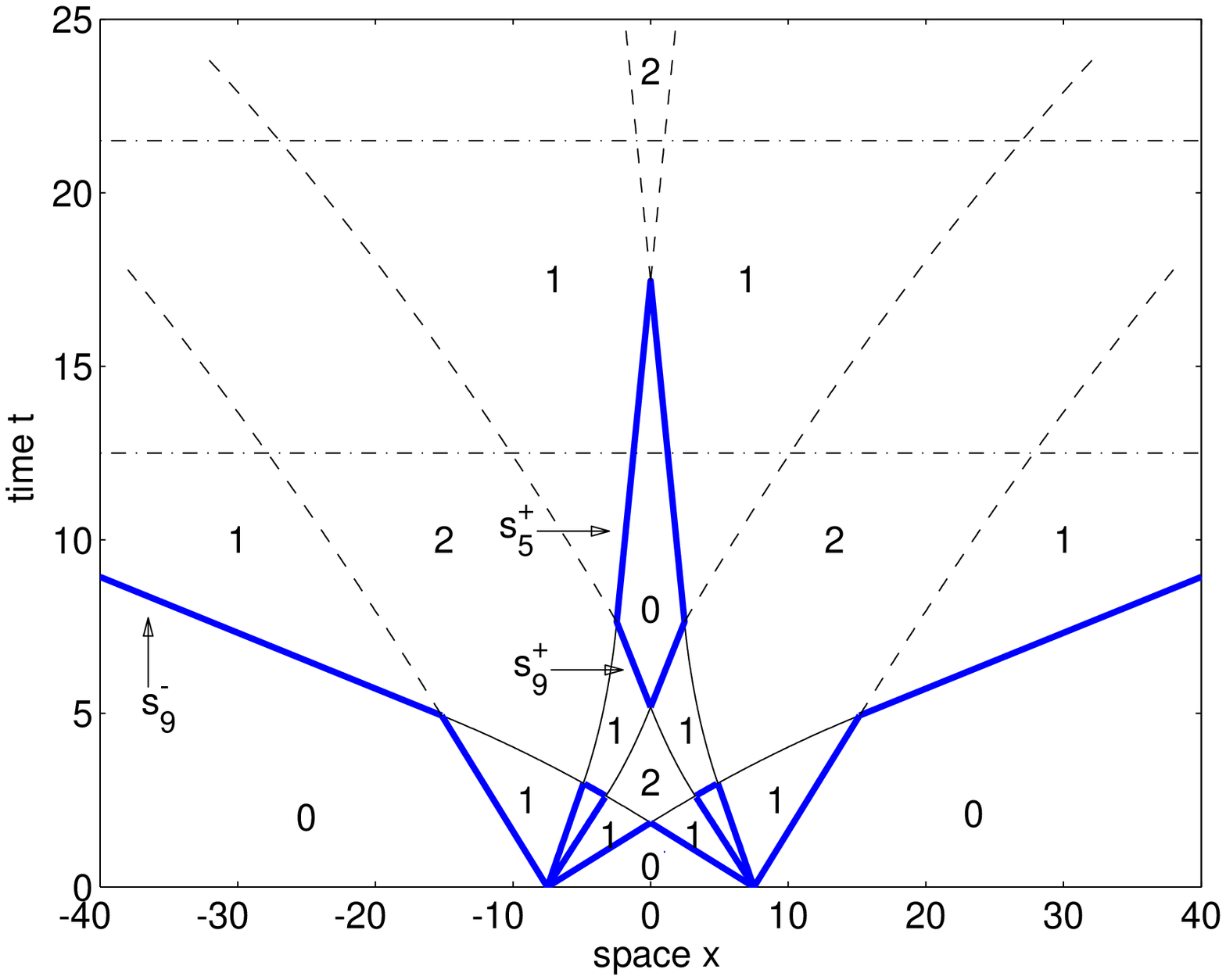}\hss}
\kern\bigskipamount
\hbox to \textwidth{\hss\epsfxsize\smallerfigwidth\epsfbox{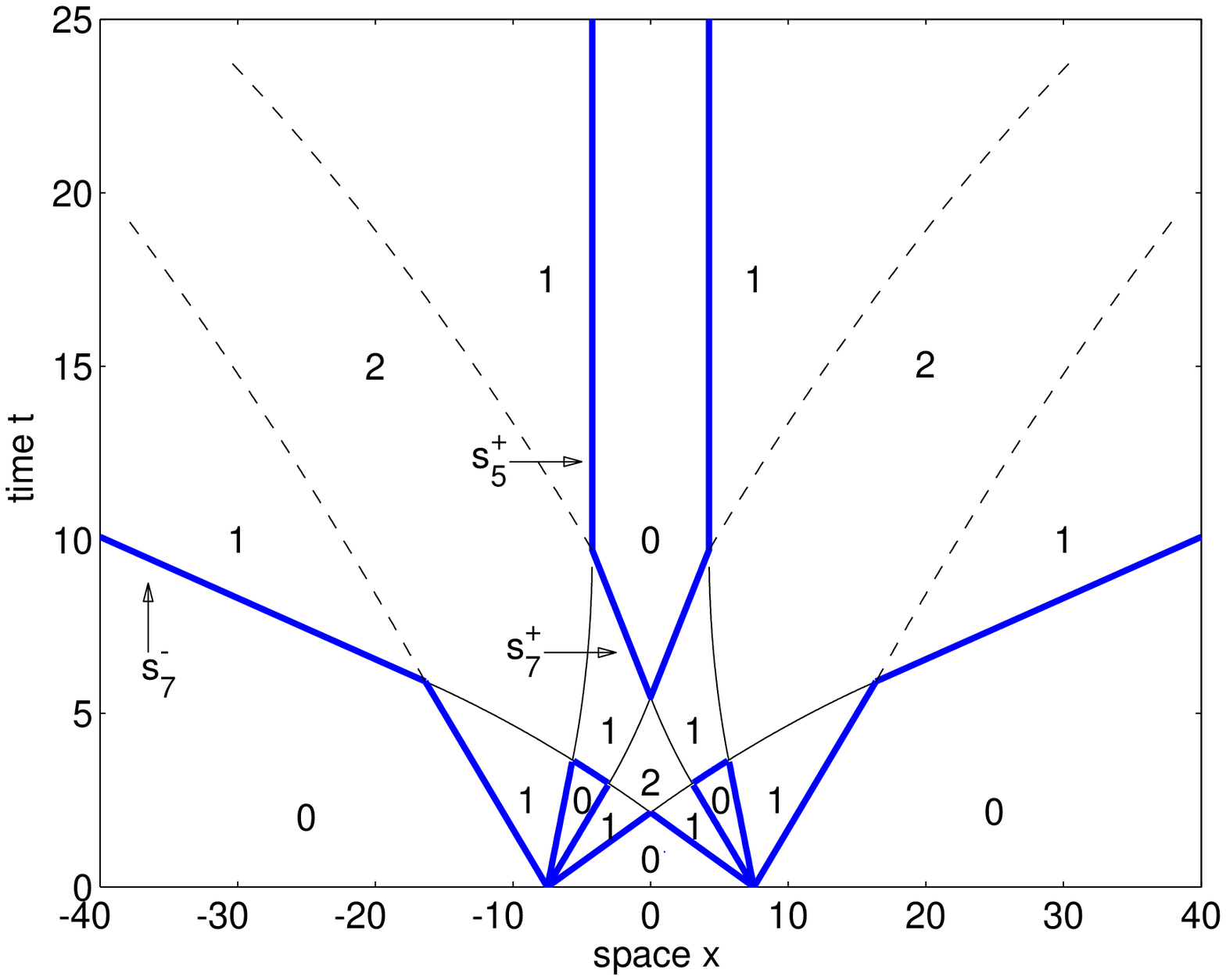}\qquad\quad
\epsfxsize\smallerfigwidth\epsfbox{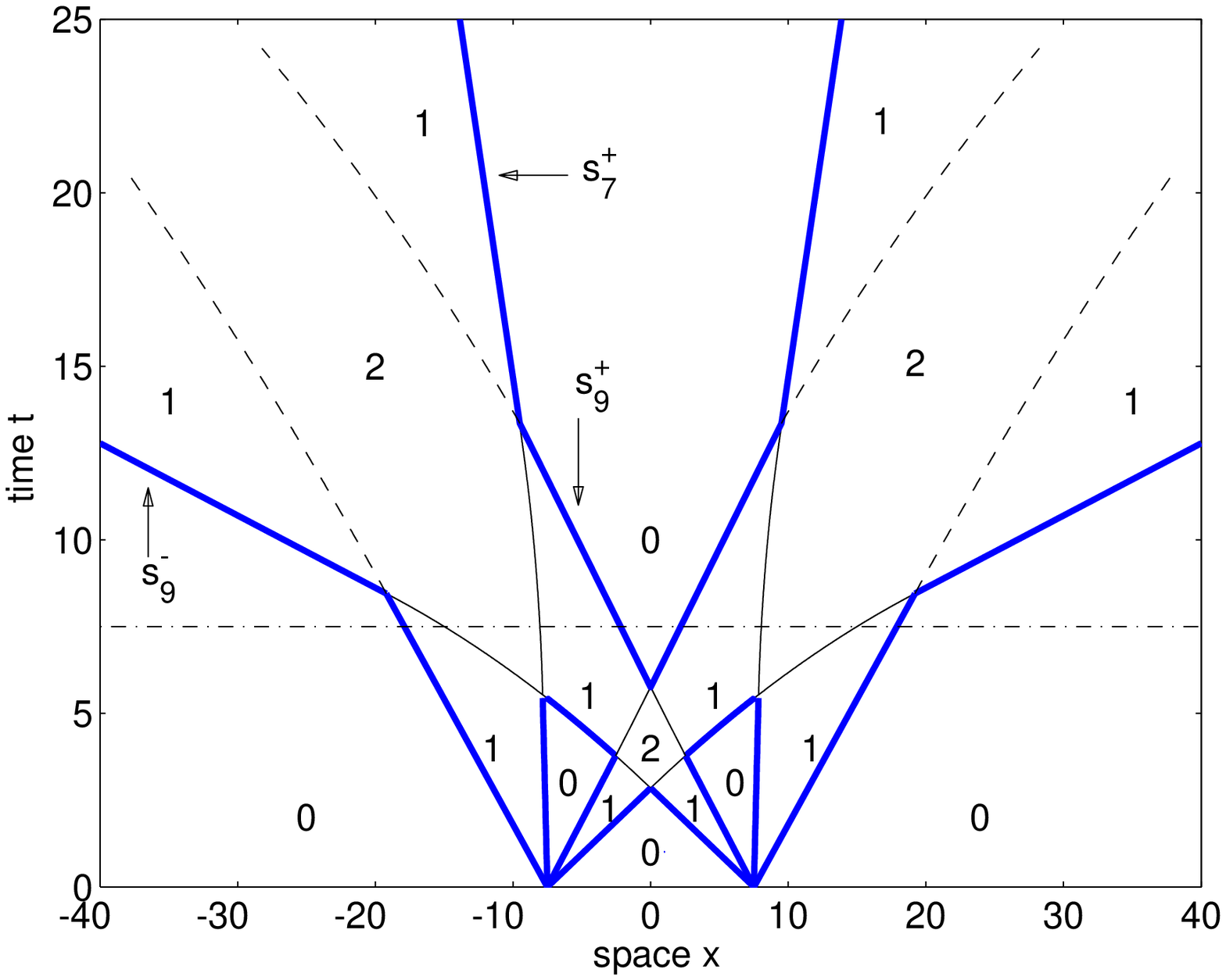}\hss}
\kern\smallskipamount
\endgroup
\caption{Bifurcation diagrams illustrating the regions of 
genus-0, genus-1 and genus-2 
in the $(x,t)$-plane for the cases shown in Fig.~\ref{f:riemann2jumps},
namely: 
(a, top left)~$u_0=2001/1000 ~(u_0>2q_0)$, corresponding to case~(i) and Fig.~\ref{f:riemann2jumps}a;
(b, top right)~$u_0=5/4 ~(2q_0>u_0>q_0)$, corresponding to case~(ii)
and Figs.~\ref{f:riemann2jumps}c,d;
(c, bottom left)~the critical case $u_0=1$, which is the separatrix between 
cases~(ii) and~(iii); and 
(d, bottom right)~$u_0=5/8~(q_0>u_0>0)$, corresponding to case~(iii) and Fig.~\ref{f:riemann2jumps}b.
The horizontal axis is position,~$x$, and the vertical axis is time,~$t$.
In all cases, $q_0=1$.
The thick (blue) lines indicate the boundaries between regions of
genus-0 and genus-1
(cf.~Propositions~\ref{p:s0before}, \ref{p:g0outer} and~\ref{p:g0inner}).
The thin (black) lines indicate boundaries between genus-1 and genus-2 regions,
and were computed by approximating them with circular arcs
of given starting point and given initial and final slopes.
(These conditions determine the circular arc uniquely.)
The thin dashed lines indicate boundaries between genus-1 and genus-2 regions
in those cases when the final slope is unknown.
The dot-dashed horizontal lines indicate the values of $t$ 
corresponding to the diagrams of the Riemann invariants in 
Figs.~\ref{f:riemann2jumps}a--d.}
\label{f:genus2jumps}
\end{figure}

It should be noted that, even though the NLS-Whitham equations with
$g=3,4$ are necessary to regularize the data in some cases,
only the genus $g\le2$ appears in the solution with two frequency jumps
(similarly to the scenario with a single frequency jump).
Note also that, unlike the case when a single phase is present, 
calculating analytically the precise location of the boundaries 
between different multi-phase regions after the individual portions 
of the solutions have come into contact is a highly nontrivial task,
and we have not attempted to do so.
However, it is possible to calculate the velocity of all 
boundaries between genus-0 and genus-1 regions.
The main reason for the difference between the two types of boundaries
is that, upon removal of degenerate gaps, 
the calculation of characteristic speeds in regions of genus-1
is expressed in terms of elliptic integrals and the boundary with genus-0
turns out to be a straight line (see Lemma \ref{01boundary} below), 
whereas the characteristic speeds in regions of genus-2 require
the evaluation of hyperelliptic integrals 
(e.g., see Ref.~\fullcite{CPAM55p1569}). 

The explicit calculation of a few characteristic speeds is reported
in appendix~\ref{a:characteristicspeeds}.
Since such calculations are rather lengthy, however, and since 
the methods used are described in Refs.~\fullcite{BlochKodama,SJAM52p909,SJAM59p2162},
here we omit the details relative to all the different cases, 
and we limit ourselves to summarizing the main results.

\begin{lemma}\label{01boundary}
All boundaries between genus-0 and genus-1 regions are given by straight lines.
\end{lemma}
\begin{proof}
The result follows from Eqs.~\eqref{e:nlsWhitham} and by 
noting that the characteristic speeds are constant since 
all the invariants are constant in the genus-0 portion.
\end{proof}
This result is obviously independent of the number of frequency
jumps present, and is just a special case of a generic feature of 
hyperbolic systems:
if one side of the initial condition is constant, there is a solution 
in terms of simple waves, for which the characteristics are straight lines
\unskip\cite{Whitham}.

\begin{proposition}
\label{p:s0before}
The boundaries between genus-0 and genus-1 regions before the interaction 
are given by straight lines with the same velocities as in 
Eq.~\eqref{e:s1jump_i} and Eqs.~\eqref{e:s1jump_ii}
upon application of the appropriate Galilean shifts.
That is, $s_{i,\pm}= \pm s_i+u_0$ for the regions relative to the
frequency jump to the left and $s_{i,\pm}= \pm s_i-u_0$ for those 
relative to the frequency jump to the right.
\end{proposition}
\begin{proof}
The result follows by direct computation of the characteristic speeds
in the NLS-Whitham equations~\eqref{e:nlsWhitham},
but is consistent with Lemma~\ref{l:Galilean} regarding 
the application of a constant offset to a single frequency jump,
showing that before interaction the behavior of the solution in
each region is unaffected by the presence of the other jump.
\end{proof}

\begin{remark}
\label{r:locking}
In case~(ii) 
(cf.~Figs.~\ref{f:riemann2jumps}c,d and Fig.~\ref{f:genus2jumps}b), 
it is $r_5(x_*,t)=r_6(x_*,t)=r_7(x_*,t)$ for all~$t$,
where $x_*(t)=\max\{x\in\Real:r_5(x,t)=2(u_0-q_0)\}$.
That is, the three invariants are ``locked'' together at that point.
Similarly, 
if one defines $x_{**}(t)=\min\{x\in\Real:r_6(x,t)=-2(u_0-q_0)\}$, 
it is $r_4(x_{**},t)=r_5(x_{**},t)=r_6(x_{**},t)$ for all~$t$. 
This locking phenomenon is identified by the two ellipses in 
Figs.~\ref{f:riemann2jumps}c,d (see also Fig.~\ref{f:riemann0}a).
Moreover, 
since two of the three invariants always coincide in a neighborhood 
of~$x_*$ or $x_{**}$, in each case 
we locally obtain a trivial regularization.
That is, near $x=x_*$, the three Riemann invariants $r_5$, $r_6$ and $r_7$
are reduced to a single invariant $\hat r(x,t)$ given by 
$\hat r(x,t)= r_7(x,t)$ for $x<x_*(t)$ and 
$\hat r(x,t)= r_5(x,t)$ for $x>x_*(t)$,
and a similar situation arises for $r_4$, $r_5$ and $r_6$ near $x=x_{**}$.
Those trivial regularizations are however necessary in order for 
all the Riemann invariants to be nonincreasing functions in $x$ 
(unlike the case in Fig.~\ref{f:riemann0}a, where trivial ones can be 
removed).
\end{remark}
\begin{conjecture}
The width of the outermost genus-2 regions present in cases~(ii) 
and~(iii) after interaction tends asymptotically to zero.
\label{c:shrinking}
\end{conjecture}
In both case~(ii) and case~(iii), the speeds of the left- and 
right-boundaries of the genus-2 region in question are respectively 
given by $s_9^+$ and~$s_7^-$
(cf. Figs~\ref{f:riemann1jump} and~\ref{f:genus2jumps}), where
as before the superscripts ``$-$'' and ``+'' refer to the values 
of the Riemann invariants respectively to the left and to the right of 
their discontinuities at $t=0$.
The fact that the width decreases monotonically and 
tends asymptotically to a constant follows from 
the double sorting property of the characteristic velocities. 
It is not possible, however, to prove whether the asymptotic
width is zero without explicitly calculating the speeds of the 
boundaries between genus-1 and genus-2 regions, which is beyond the 
scope of this work.
(The appearance of this phenomenon was first noted by F.-R.~Tian for the
case of the KdV-Whitham equations with two jumps\cite{FRTian}.)

\begin{proposition}
\label{p:g0outer}
After interaction, the boundary between the outermost genus-1 regions 
and the surrounding genus-0 regions is given by 
$x=\pm(x_0+s_\mathrm{outer}t)$, where 
$s_\mathrm{outer}=s_5^-$
in case (i) and $s_\mathrm{outer}=s_9^-$ in cases (ii) and (iii).
In all these cases, the resulting velocity is
\end{proposition}
\begin{equation}
s_\mathrm{outer}= (2u_0+q_0)
    \big[\, 1 + {2u_0q_0}/{(2u_0+q_0)^2} \,\big]\,.
\label{e:g0outerspeed}
\end{equation}
The characteristic speed can be obtained from degenerate $g=2$ calculations
(cf.\ Appendix~\ref{a:characteristicspeeds}).
Note however that the appropriate values of $s_5^-$ and $s_9^-$ however
can also be obtained from $s_3^+$ in Eq.~\eqref{e:s1jump_i} 
and $s_5^+$ in Eq.~\eqref{e:s1jump_ii} upon $u_0\to2u_0$.
(Cf.~Figs.~\ref{f:riemann1jump} and~\ref{f:riemann2jumps}
and discard the Riemann invariants corresponding to degenerate gaps).

\begin{proposition}
\label{p:g0inner}
After interaction, the boundary between the innermost genus-0 region
and the surrouding genus-1 regions in cases~(ii) and~(iii) 
can be written as
\begin{equation}
\smash{x=\pm(x_1+s_\mathrm{inner}\1 t)}\,,\qquad
\smash{x=\pm(x_2+s_\mathrm{inner}\2 t)}\,,
\end{equation}
respectively for the initial portion and
for the final portion,
where the characteristic speeds 
are:
$\smash{s_\mathrm{inner}\1=s_9^+}$ both in case~(ii) and case~(iii);
$\smash{s_\mathrm{inner}\2=s_5^+}$ in case~(ii),
and 
$\smash{s_\mathrm{inner}\2=s_7^+}$ in case~(iii).
Furthermore, in both case~(ii) and case~(iii) 
the values of these speeds are:
\end{proposition}
\begin{equation}
\smash{s_\mathrm{inner}\1= q_0\,,\qquad
s_\mathrm{inner}\2= q_0-u_0\,.}
\label{e:g0innerspeed}
\end{equation}
As before, the characteristic speeds can be obtained via
degenerate $g=2$ calculations
(see Appendix~\ref{a:characteristicspeeds} for more details).
As before, however, the appropriate values of $s_5^+$, $s_7^+$ and $s_9^+$
can also be obtained from $s_5^+$ in case~(ii) of section~\ref{s:singlejump} 
upon rescaling
$q_0\to q_0+\frac12 u_0$ and $u_0\to2u_0$.
More precisely, neglecting degenerate gaps, 
the Riemann invariants that determine the value of 
$\smash{s_\mathrm{inner}\1=s_5^+}$ in case~(ii) and 
$\smash{s_\mathrm{inner}\1=s_7^+}$ in case~(iii) 
are obtained from those that determine~$s_5^+$ in Fig.~\ref{f:riemann1jump}b
upon rescaling $u_0\to 2u_0$.
(Cf.~Figs.~\ref{f:riemann1jump}b and~\ref{f:riemann2jumps}b,c.)
Similarly, the Riemann invariants
that determine the value of $\smash{s_\mathrm{inner}\1=s_9^+}$ 
both in case~(ii) and in case~(iii)
are obtained from those that determine~$s_5^+$ in Fig.~\ref{f:riemann1jump}b
(case~(ii) in section~\ref{s:singlejump})
upon rescaling $q_0\to q_0+\frac12 u_0$.
(Again, cf.~Figs.~\ref{f:riemann1jump}b and~\ref{f:riemann2jumps}b,c.)

\begin{remark}
Equations~\eqref{e:g0outerspeed} and~\eqref{e:g0innerspeed} 
also describe the motion of the boundaries of the genus-0 regions 
in the critical case $u_0=q_0$, upon proper relabeling of the 
Riemann invariants.  In particular, they predict that in the 
critical case the width of the inner genus-0 region is constant
(i.e., $s_\mathrm{inner}\2=0$).
\end{remark}

The above results, and in particular Eqs.~\eqref{e:g0outerspeed} 
and~\eqref{e:g0innerspeed}
show that the speed of the boundaries between regions of
genus-0 and genus-1 after the individual perturbations have come 
into contact is significantly altered as a result of the interaction.
In other words, the interaction between different genus-1 regions
results in significant bending.
This can perhaps be best appreciated in the critical case 
(i.e., $u_0=q_0$, cf.~Fig.~\ref{f:genus2jumps}c).

Finally, we note that, as in the case of one frequency jump, 
all of the above values for the speeds of the boundaries between 
genus-0 and genus-1 regions agree very well with the results
of numerical simulations of the NLS equation, to be discussed in
section~\ref{s:numerics}.

\section{Behavior of finite-genus solutions: Arbitrary number of jumps}
\label{s:arbitraryjumps}

Similar calculations to the ones described in section~\ref{s:multiphase}
can be repeated for jumps of arbitrary size and/or when more than 
two frequency jumps are present.  
In particular, from the generalization of the above calculations to 
an arbitrary number of jumps, it is possible to extract some general 
features.  
Thus, we consider an initial condition with constant-amplitude and
with initial frequency expressed in terms of $N$ frequency jumps each 
of size $C_j$ and located at $x=X_j$, $j=1\,\dots,N$: 
\begin{gather}
u(x,0)= - C_\mathrm{tot}/2 + \sum_{j=1}^kC_j
\qquad \mathrm{for}~~X_k<x<X_{k+1}\,,
\label{e:ugeneral}
\end{gather}
where $X_0=-\infty$ and $X_{N+1}=\infty$, and where 
$C_\mathrm{tot}=\sum_{j=1}^N C_j =u(\infty,0)-u(-\infty,0)$ 
is the total jump size.
The constant frequency offset $C_\mathrm{tot}/2$ in $u(x,0)$
does not affect the qualitative behavior of solution, and is chosen so that
the mean frequency of the solution is zero, i.e., so that the mean position
of the solution is constant.

For simplicity, hereafter we will limit ourselves to
consider a symmetric collection of positive frequency jumps.
That is, we will assume that $C_j>0$ $\forall j=1,\dots,N$
and that if a given jump of amplitude $C=C_*$ exists at~$x=X_*$, 
another jump with the same amplitude exists at~$x=-X_*$.
(Or, in other words, the initial condition $u(x,0)$ for the frequency possesses 
reflection symmetry with respect to the origin.)\,\ 
Since $C_j>0$ $\forall j=1,\dots,N$, the analog of Eqs.~\eqref{e:r12infty}
holds:
\begin{subequations}
\label{e:rpminftyNjumps}
\begin{gather}
\mathop\max\limits_{x\in\Real}r_1\o(x)= 
  r_1\o(\infty)= \lim_{x\to\infty}r_1\o(x)= -2q_0 - C_\tot/2\,,
\\
\mathop\min\limits_{x\in\Real}r_2\o(x)= 
  r_2\o(-\infty)= \lim_{x\to-\infty}r_2\o(x)= 2q_0 + C_\tot/2\,,
\end{gather}
\end{subequations}
Note that again $r_1\o(x)$ and $r_2\o(x)$ are the Riemann invariants 
before regularization.
Owing to Eqs.~\eqref{e:rpminftyNjumps} and Definition~\ref{d:criticaljumpcollection}, 
a collection of jumps will be super-critical if $C_\tot>2q_0$, 
sub-critical if $C_\tot<2q_0$ and
critical if $C_\tot=2q_0$.

\begin{proposition}
\label{p:genus0region?}
In the case of $N$ equal frequency jumps each of size $C=2u_0$, 
no individual genus-0 regions develop if the jumps
are individually super-critical, i.e., $u_0>2q_0$.
\end{proposition}
Recall that the size of each of the jumps is $2u_0$ and that 
the jumps are individually super-critical according to Definition~\ref{d:criticalsinglejump}
when the original Riemann invariants overlap at a single location in space,
that is, if $r_2\o(X_j^-)<r_1\o(X_j^+)$ for $j=1,\dots,N$.
The result applies independently of the number of jumps.
More in general, however for an arbitrary collection of frequency jumps
we have:

\begin{figure}[b!]
\medskip
\begingroup
\smallerfigwidth= 0.905\smallerfigwidth
\hbox to \textwidth{\hss\epsfxsize\smallerfigwidth\epsfbox{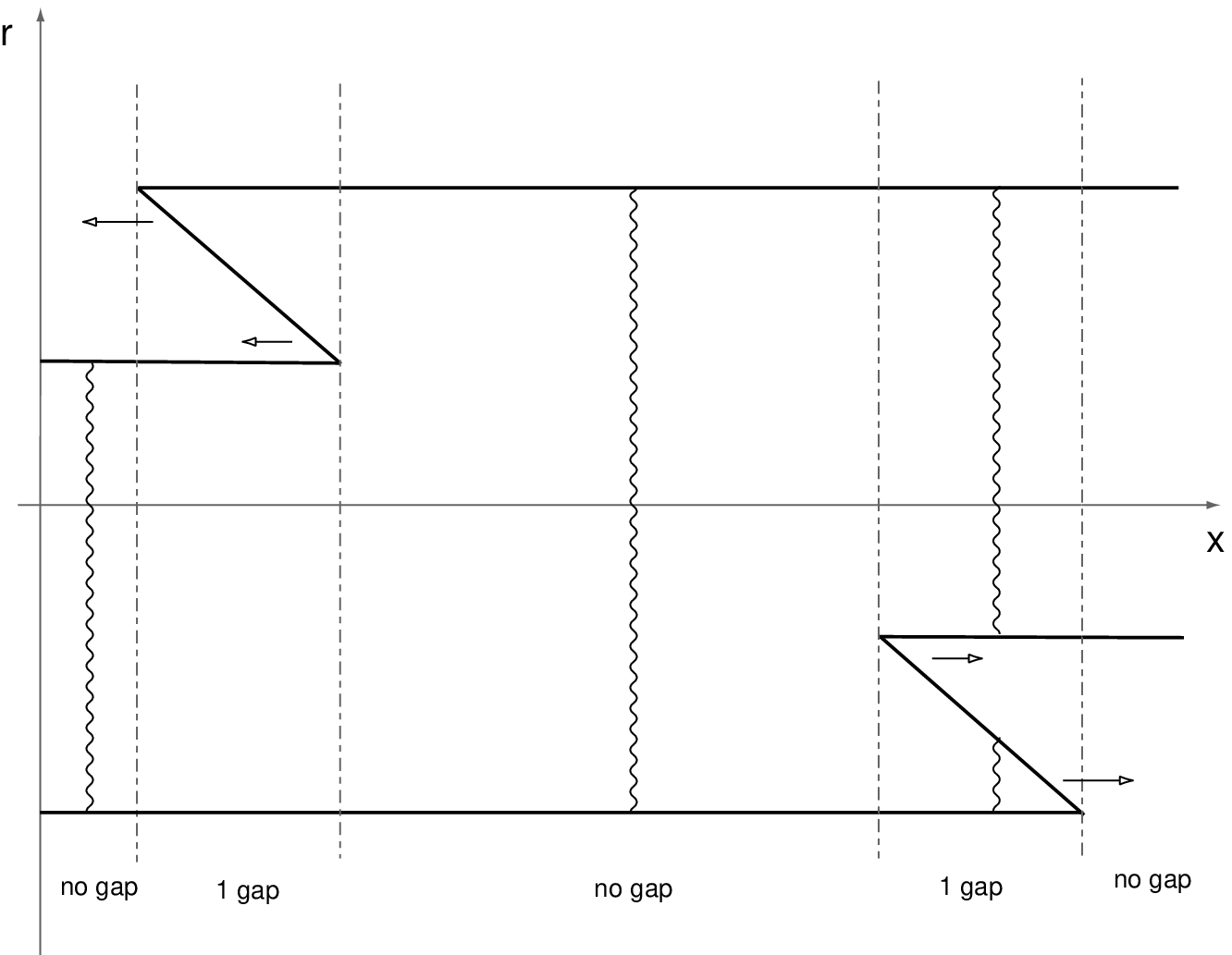}\qquad
\epsfxsize\smallerfigwidth\epsfbox{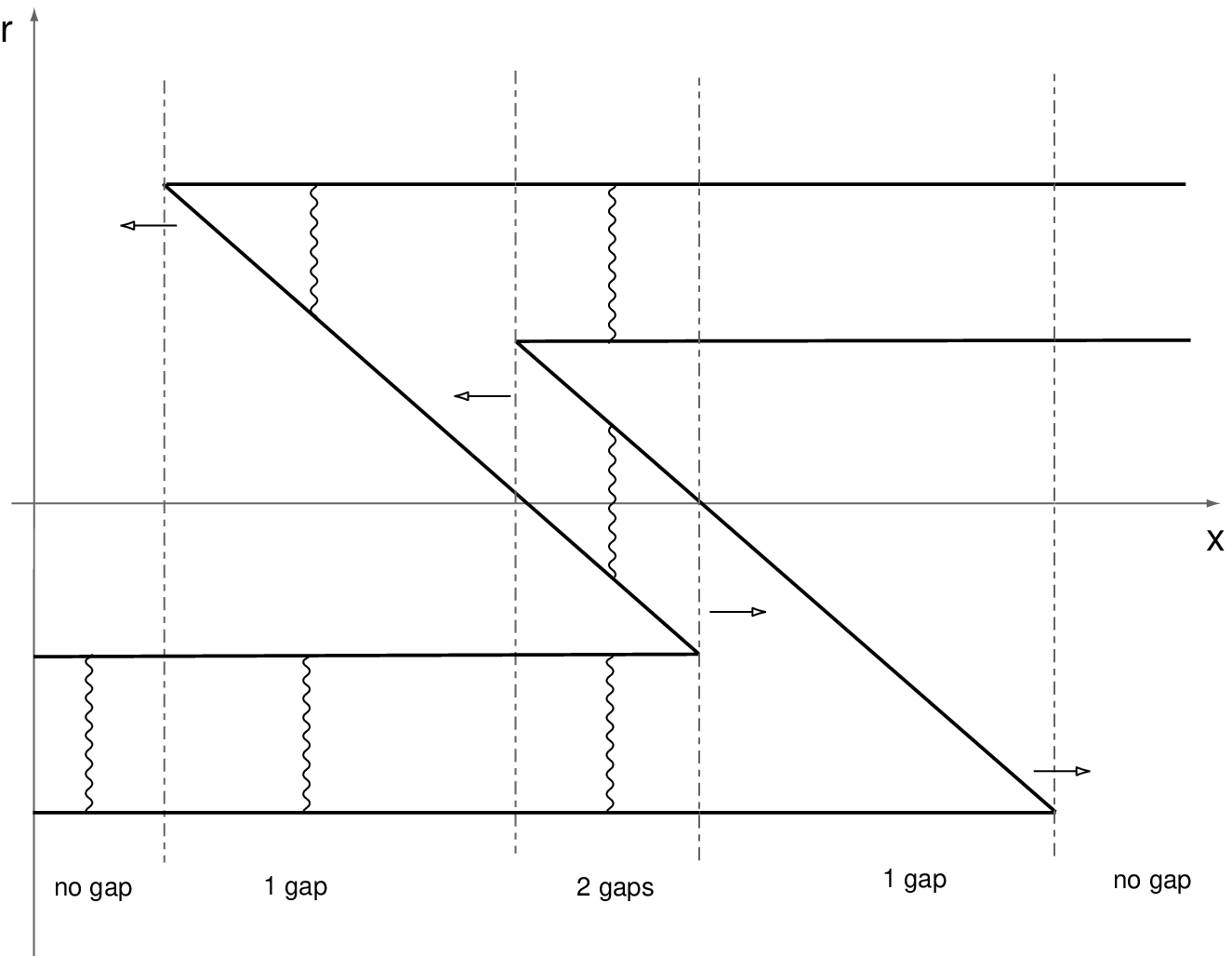}\hss}
\kern\bigskipamount
\hbox to \textwidth{\hss\epsfxsize\smallerfigwidth\epsfbox{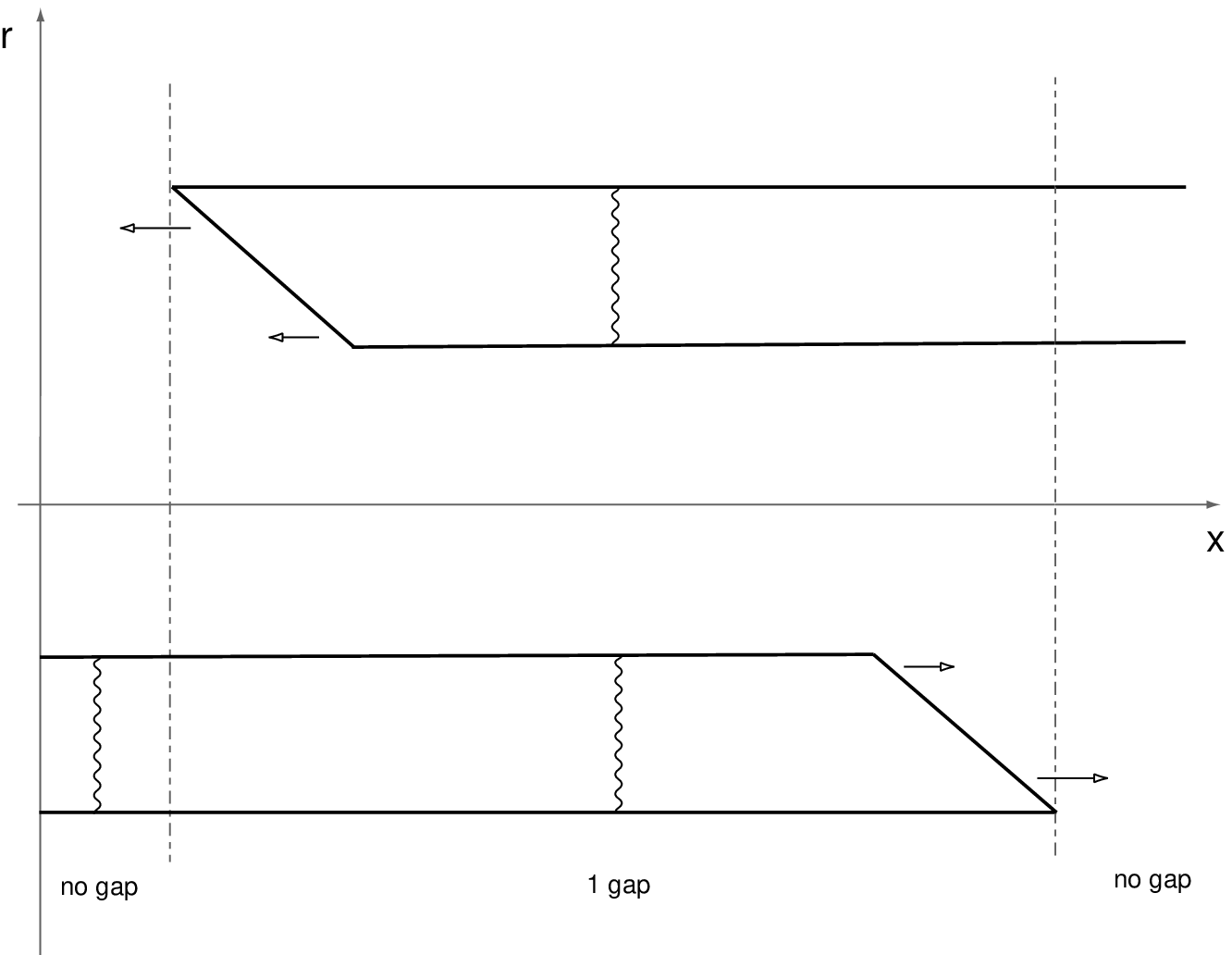}\qquad
\epsfxsize\smallerfigwidth\epsfbox{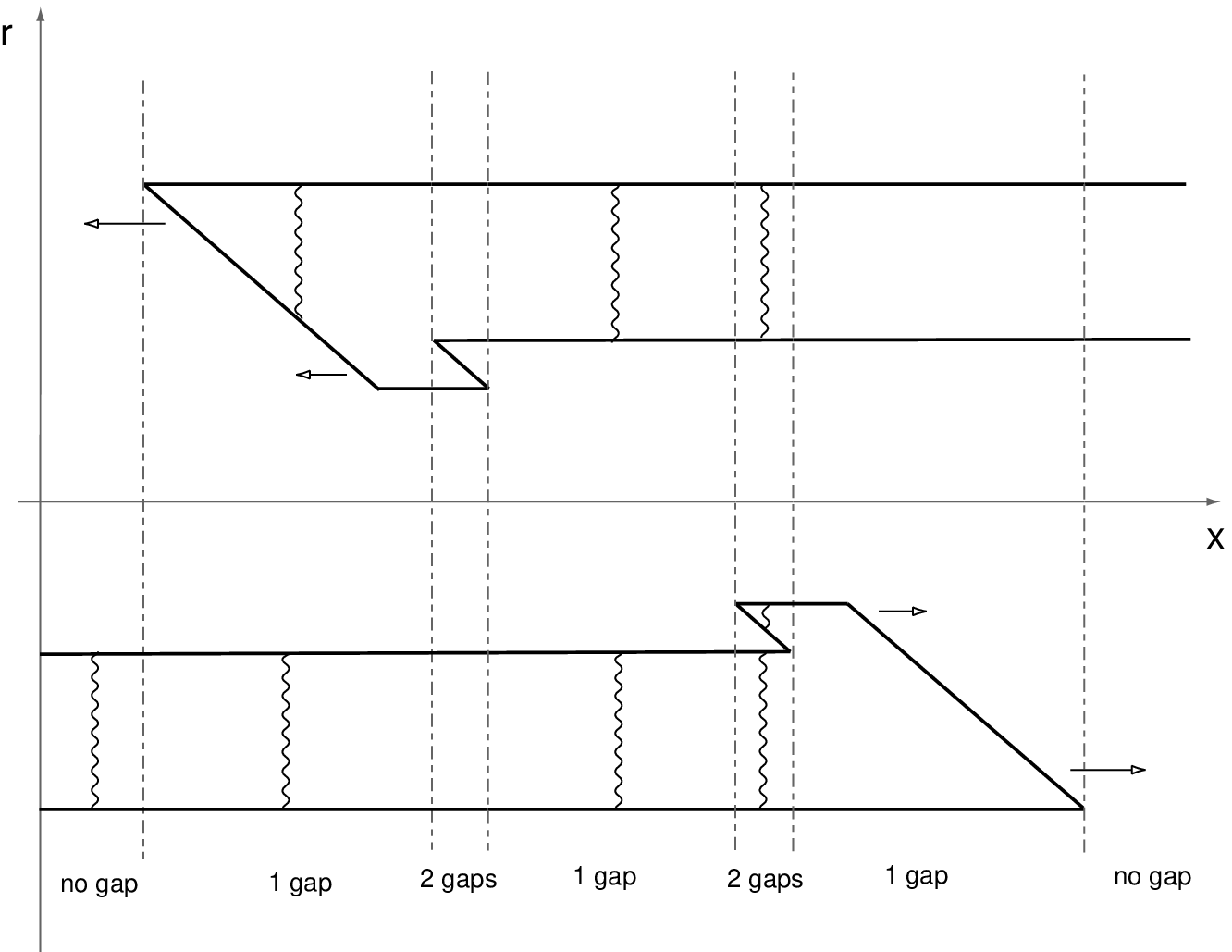}\hss}
\kern\medskipamount
\endgroup
\caption{Qualitative diagrams illustrating topologically distinct cases
for the asymptotic values of the Riemann invariants
after removal of the degenerate portions,
together with the corresponding separation between regions of genus-0, 
genus-1 and genus-2: 
(a, top left)~genus-0 region at the center, corresponding to 
Figs.~\ref{f:riemann1jump}b and~\ref{f:riemann2jumps}b;
(b, top right)~genus-2 region at the center, corresponding to
Fig.~\ref{f:riemann2jumps}d; 
(c, bottom left)~genus-1 region at the center, corresponding to 
Fig.~\ref{f:riemann1jump}a; and 
(d, bottom right)~originated by a super-critical jump surrounded
by sub-critical ones.
In each case, the wavy lines show the branch cuts in the Riemann surface 
$y^2=\prod_{j=1}^{2g+2}(z-r_j(x,t))$~ $\forall x\in\Real$,
corresponding to the local spectrum of the Lax operator 
of the NLS equation.}
\label{f:asymptoticinvariants}
\kern-\medskipamount
\end{figure}

\begin{figure}[b!]
\medskip
\begingroup
\smallerfigwidth= 0.905\smallerfigwidth
\hbox to \textwidth{\hss\epsfxsize\smallerfigwidth\epsfbox{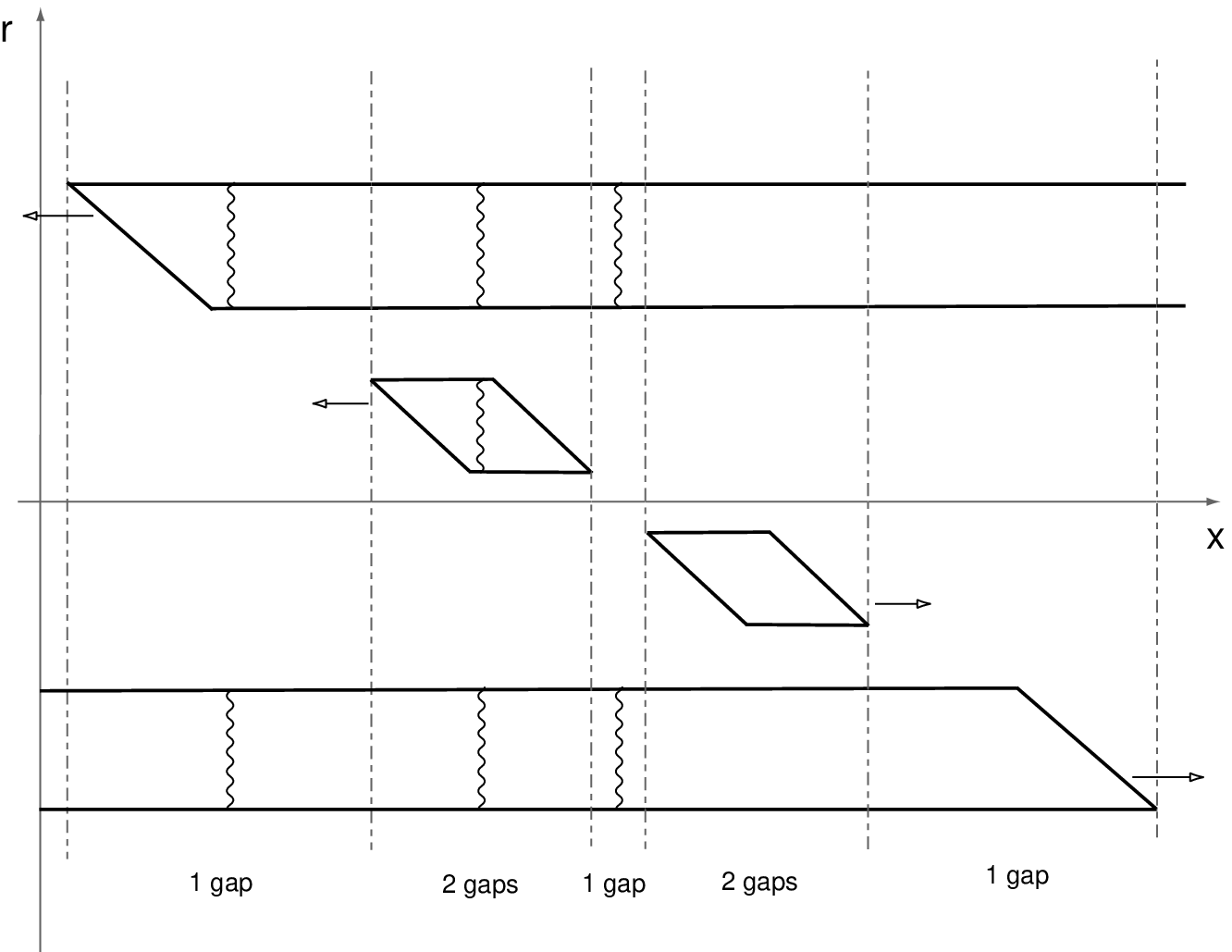}\qquad
\epsfxsize\smallerfigwidth\epsfbox{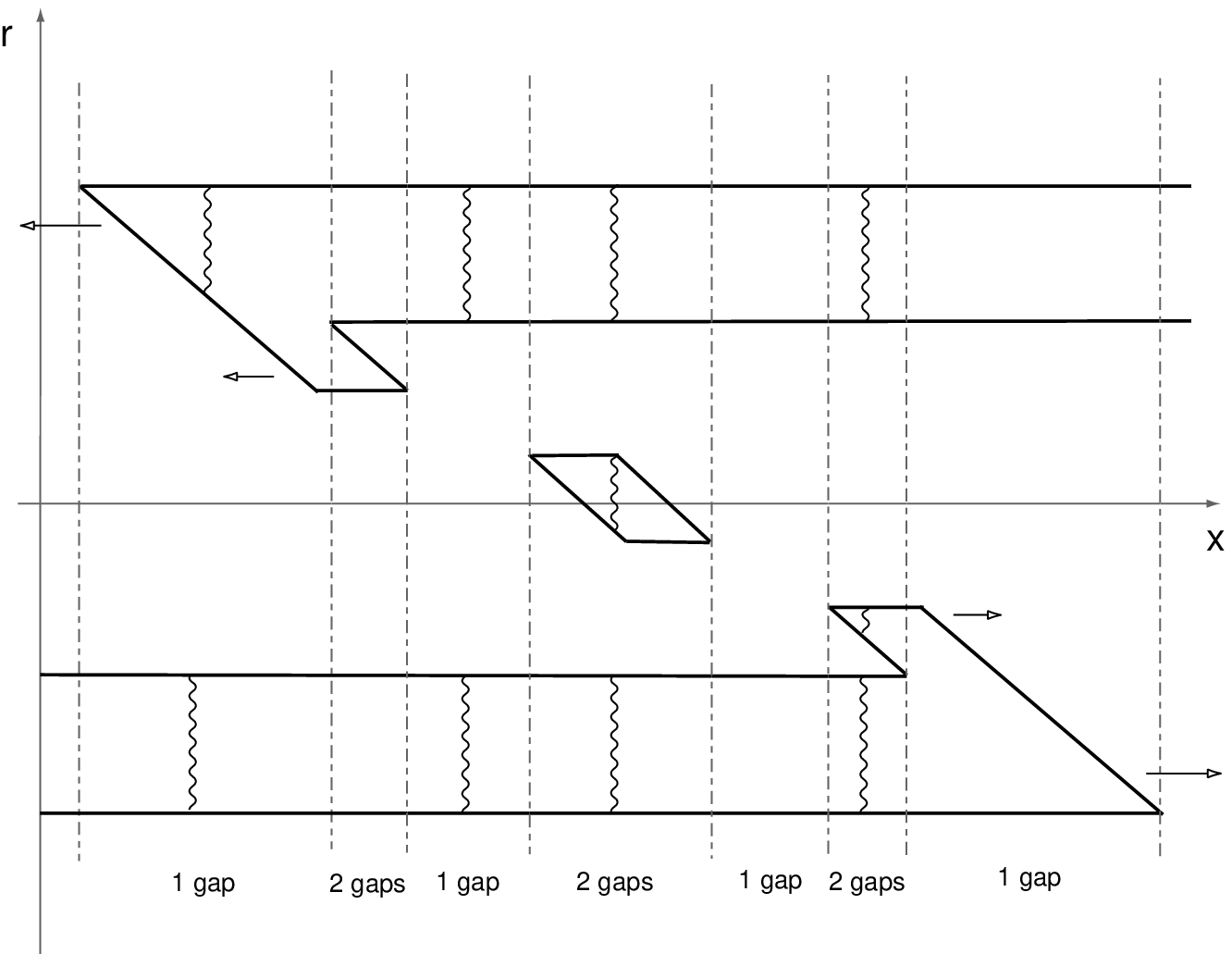}\hss}
\kern\medskipamount
\endgroup
\caption{Additional diagrams of asymptotic configurations of Riemann
invariants, illustrating the possible presence of 
separate ``islands'', each of which produces one additional region of genus-2.
Figure~\ref{f:asymptoticinvariants2}a (left): a variant of 
 Fig.~\ref{f:asymptoticinvariants}c
 corresponding four supercritical jumps.
Fig.~\ref{f:asymptoticinvariants2}b (right): a variant of 
 Fig.~\ref{f:asymptoticinvariants}d
 corresponding to two super-critical jumps surrounded by two sub-critical ones.}
\label{f:asymptoticinvariants2}
\kern-\medskipamount
\end{figure}

\begin{theorem}
\label{t:asymptoticgenus}
It is possible to obtain arbitrarily large genera for finite times
by considering appropriate collections of frequency jumps.
However, asymptotically as $t\to\infty$, the only expanding regions
in the solution are of genus-0, genus-1 and genus-2.
\end{theorem}
\begin{proof}
The result follows directly from Proposition \ref{p:doublesorting} (double sorting property of
the characteristic speeds). Let us first note that the non-regularized 
system has two Riemann invariants, and therefore, 
independently of the behavior at intermediate values of~$x$,
the regularized system will have only two non-degenerate branches 
as $x\to\pm\infty$, i.e., $r_1^0(x)<r_2^0(x)$.
The only topologically distinct ways to connect these two branches 
are shown in Figs.~\ref{f:asymptoticinvariants}a-d
(some of these possibilities were also studied in Ref.~\fullcite{PHYSD1995v87p186}, but not for genus-2 cases),
where we have neglected the possible presence of one or more shrinking regions
such as the outermost genus-2 regions in Fig.~\ref{f:riemann2jumps}d
(see Remark~\ref{r:g3} later).
The distinction between the different cases in
Figs.~\ref{f:asymptoticinvariants}a--d
is based on the size of the frequency jumps across the pulse.
More precisely:
\begingroup
\begin{enumerate}
\itemsep0pt\parsep0pt
\item[(i)] 
 Fig.~\ref{f:asymptoticinvariants}a is obtained when
 $\r_1\o(\infty)<\r_2\o(-\infty)$, i.e., when the jumps are
 collectively sub-critical
 (as in Figs.~\ref{f:riemann1jump}b and~\ref{f:riemann2jumps}b).
\item[(ii)] 
 Fig.~\ref{f:asymptoticinvariants}b is obtained when 
 $\r_1\o(\infty)>\r_2\o(-\infty)$
 but $\r_1\o(x^+)<\r_2\o(x^-)~\forall x\in\Real$;
 that is, when the jumps are collectively super-critical but
 every jump is individually sub-critical
 (as in Fig.~\ref{f:riemann1jump}d).
\item[(iii)] 
 Fig.~\ref{f:asymptoticinvariants}c is obtained when there is only one
 jump, but that jump is super-critical; that is, a single jump at~$x=t_*$,
 such that $\r_1\o(x^+_*)>\r_2\o(x^-_*)$
 (as in Fig.~\ref{f:riemann1jump}a). 
\item[(iv)] 
 Fig.~\ref{f:asymptoticinvariants}d is obtained 
 when there is one super-critical jump at the center, 
 surrounded by others jumps which are all sub-critical.
 That is, when there is one jump at
 $x=0$  such that $\r_1\o(0^+)>\r_2\o(0^-)$,
 surrounded by other jumps at $x=t_j$ such that 
 $\r_1\o(x_j^+)<\r_2\o(x_j^-)~\forall x_j\ne 0$.
\end{enumerate}
\endgroup\noindent
Of course, many variations of these basic configurations are possible
depending on the details of the arrangement of the frequency jumps.
In addition, cases (iii) and~(iv)
(corresponding to Figs.~\ref{f:asymptoticinvariants}c,d),
admit a variant in which one or more localized ``islands'' 
exist between the two outer portions, as shown 
in Fig.~\ref{f:asymptoticinvariants2}a,b (cf.\ Fig.~\ref{f:riemann2jumps}a).
One of these islands is obtained whenever two additional frequency jumps 
are inserted which are both individually super-critical,
that is, when 
$\r_1\o(x^+_*)>\r_2\o(x^-_*)$ for more than one value of $x_*\in\Real$.
For example, 
Fig.~\ref{f:asymptoticinvariants2}a is produced when there are four
super-critical jumps (cf.~Proposition~\ref{p:genus0region?});
Fig.~\ref{f:asymptoticinvariants2}b is produced when there are two
super-critical jumps surrounded by two sub-critical ones.
From these examples it should be clear that the above situations can be 
easily generalized to cases where an arbitrary number of islands is present
by adding a proper number of super-critical frequency jumps.
Note that the range of values for the Riemann invariants
covered by each of these islands is disjoint from that of all of the other
islands.
(That is, the islands are not lying side by side.
Rather, they are arranged by increasing values of the invariants.)
Thanks to the double sorting property possessed by the Riemann invariants 
(namely, Eqs.~\eqref{e:sorting} in Proposition \ref{p:doublesorting}), islands characterized by larger values 
of the invariants 
propagate faster to the left, and therefore, if several islands exist, 
different islands will temporarily end up stacked on top of 
each other upon propagation.
It is then clear that by properly choosing the number and
size of frequency jumps, it is possible to produce regions of 
arbitrarily high genus for finite values of $t$.
However, the sorting property of the Riemann invariants 
also implies that, eventually, 
each of the islands will separate from the others.
Thus, asymptotically in $t$, at most one island will be 
present inbetween the top and bottom branches for each value of~$x$,
implying that corresponding to each value of~$x$ there will be at most 
two gaps.
\end{proof}

\begin{figure}[b!]
\medskip
\begingroup
\smallerfigwidth= 0.905\smallerfigwidth
\kern-\medskipamount
\hbox to \textwidth{\hss\raise2ex\hbox{\epsfxsize\smallerfigwidth\epsfbox{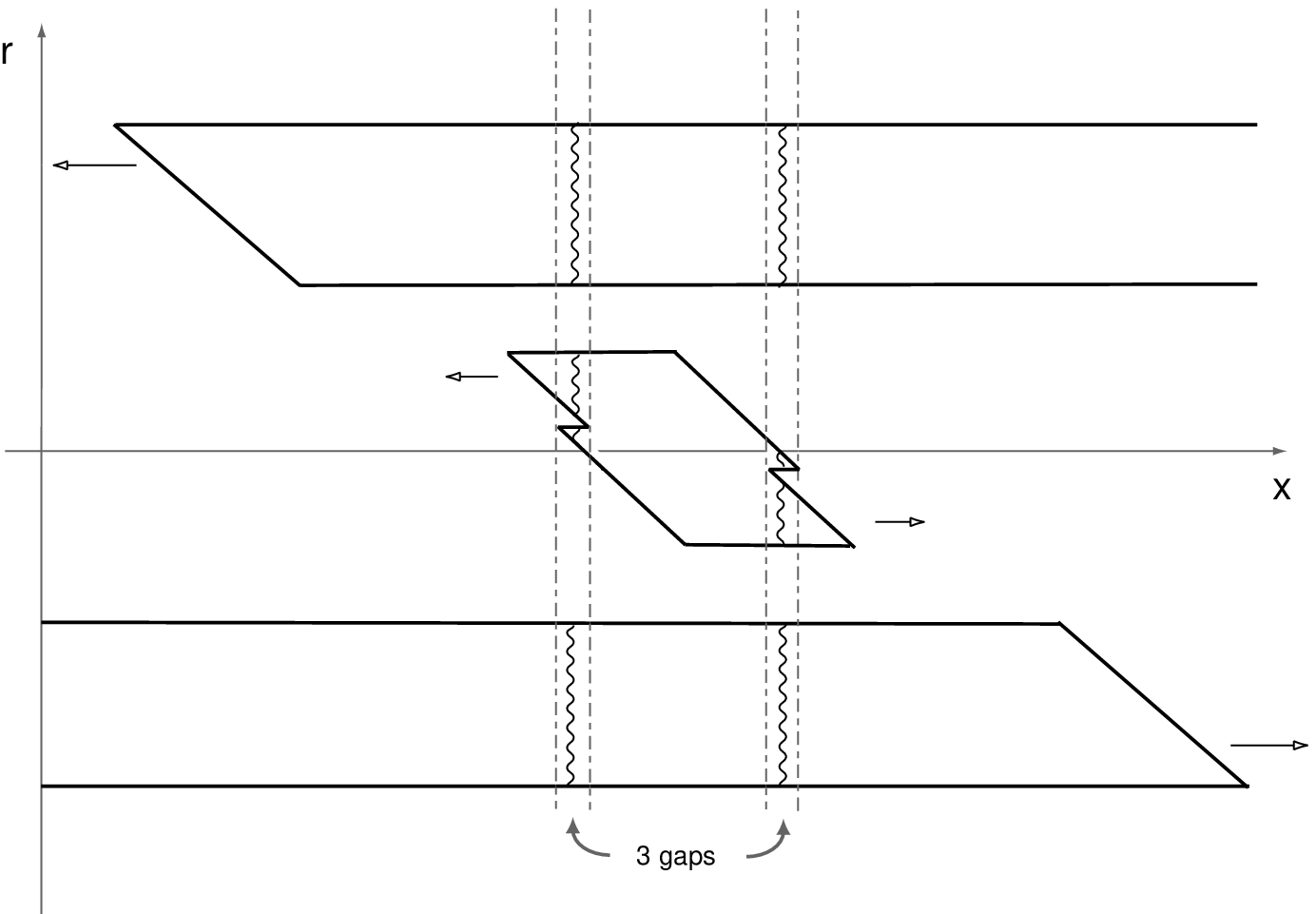}}\qquad
\epsfxsize\smallerfigwidth\epsfbox{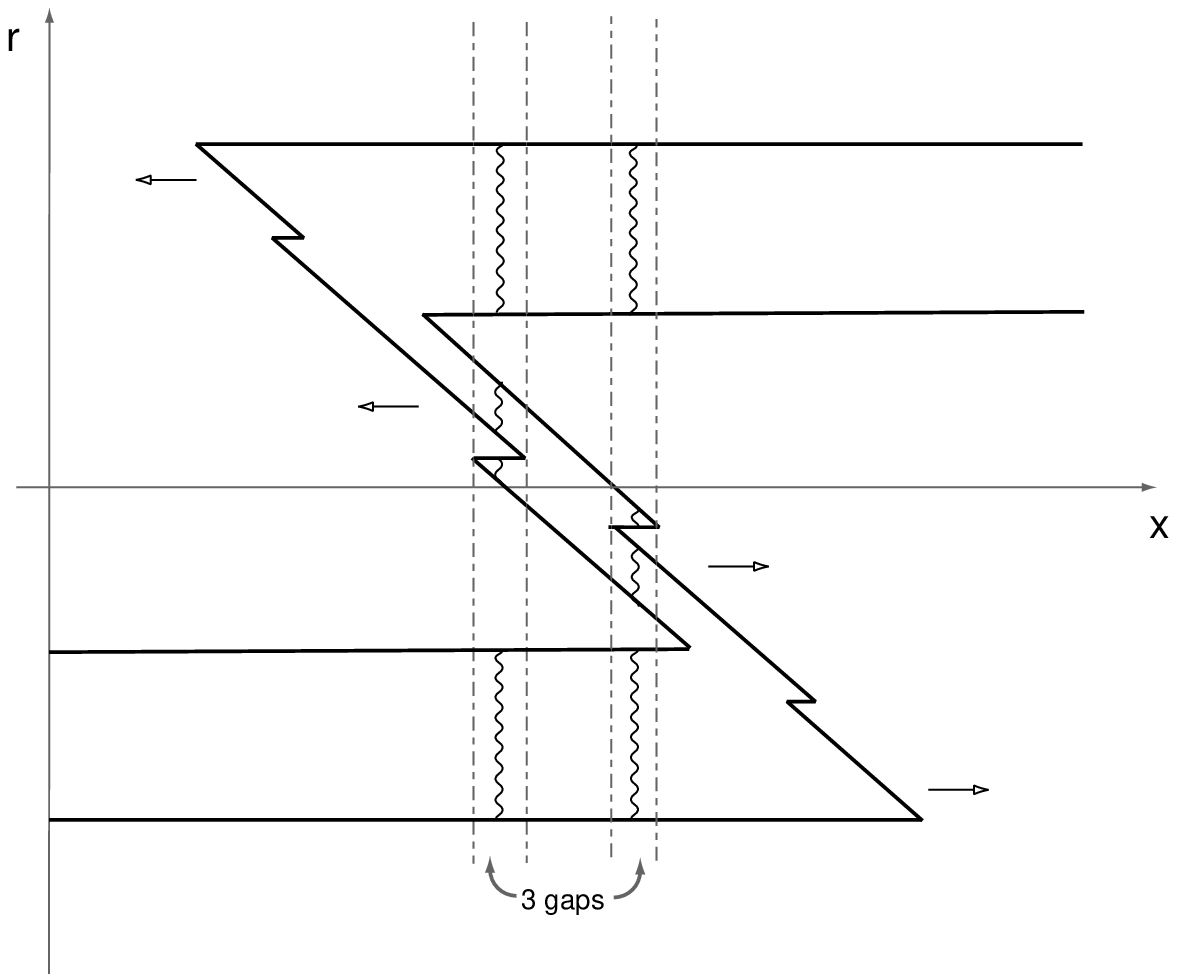}\hss}
\endgroup
\caption{Additional diagrams of asymptotic configurations of Riemann invariantes
showing explicitly the presence of shrinking genus-3 regions;
Fig.~\ref{f:genus3}a is the analog of Fig.~\ref{f:asymptoticinvariants2}b;
Fig.~\ref{f:genus3}b the analog of Fig.~\ref{f:asymptoticinvariants}b (see text for details).}
\label{f:genus3}
\kern-\medskipamount
\end{figure}

\begin{remark}
The above decomposition of the solution of the NLS equation~\eqref{e:NLS}
into regions of genus-0, 1 and 2 asymptotically with time
should be compared to the corresponding result for the 
Korteweg-de~Vries (KdV) equation $u_t+6uu_x+u_{xxx}=0$, 
for which it was shown that the solution decomposes into expanding regions
of genus-0 and genus-1 asymptotically in time.\cite{CPAM55p1569}
\end{remark}
\begin{remark}
It is important to note that regions of genus-3 may also exist 
in the solution of NLS for all finite values of time.
For example, 
Fig.~\ref{f:genus3}a shows the invariants produced 
by a configuration with two collectively sub-critical jumps 
surrounded by two individually super-critical ones,
while Fig.~\ref{f:genus3}b shows the invariants produced by a collection
of three jumps which are individually sub-critical but collectively
supercritical.
Due to the double sorting property of the Riemann invariants, 
however, the width of these regions decreases monotonically,
and tends to a constant asymptotically with time. 
Therefore, the presence of these regions
does not invalidate the statement of Theorem~\ref{t:asymptoticgenus}.
Moreover, we expect the width of these genus-3 regions to tend
asymptotically to zero with time (cf.~Conjecture~\ref{c:shrinking}).
The situation is similar to the case of the KdV equation,
where, in addition to expanding regions of genus-0 and genus-1,
shrinking regions of genus-2 may also exist asymptotically.
\label{r:g3}
\end{remark}
\noindent
From the proof of Theorem~\ref{t:asymptoticgenus} we also have:

\begin{corollary}
\label{c:genus0amplitude}
Whenever an overall genus-0 region develops at the center of the pulse
as a result of the interaction among all frequency jumps, 
its amplitude 
is always given by the analogue of Eq.~\eqref{e:g0amplitude}, namely
$\rho_\max=\frac1{16}\big(r_2\o(\infty)-r_1\o(-\infty)\big)^2$, 
in a similar way as for the single-jump case.
\end{corollary}
In other words, the above value is the largest amplitude of a genus-0
region that can be achieved with any number~$N$ of frequency jumps
(cf.~Figs.~\ref{f:riemann1jump}b and~\ref{f:riemann2jumps}b,c).
In the case of equal frequency jumps of size $C=2u_0$, it is
$\rho_\max=q_0^2(1+Nu_0/2q_0)^2$.

\begin{corollary}
\label{c:genus0asymptotic?}
Asymptotically as $t\to\infty$, the solution in the central portion of 
the pulse will be a region of genus-0 iff the jumps are collectively
sub-critical, i.e., iff
\setbox0=\hbox{$r_1\o(\infty)< r_2\o(-\infty)$.}
\dp0=0pt
\box0
\end{corollary}
In the case of equal-size frequency jumps, this condition is
satisfied iff $u_0\le 2q_0/N$.
Thus, the maximum amplitude of a stable genus-0 region is always
$\rho_\max=4q_0^2$, again as in the single-jump case, 
and independently of the number of frequency jumps.

All of the results in this work are relative to initial conditions with
constant amplitude and piecewise-constant frequency. 
If the initial amplitude and frequency vary continuously,
some differences can obviously be expected in the solution.
Because the quasilinear system of PDEs for the regularized Riemann invariants
is always strictly hyperbolic, 
it is stable with respect to small changes in the initial conditions.
Thus, even though the previous results are relative to the case of 
discontinuous amplitude and frequency jumps, they describe 
generic features of the solution.
Consider for example the dam breaking problem.
If the amplitude varies continuously from 0 to its maximum value,
oscillations appear near the edges of the pulse.
(Indeed, these oscillations are visible in the numerical simulations.)
In the absence of frequency jumps,
this scenario was studied by Forest and McLaughlin using
the Lax-Levermore theory\cite{JNLSCI7p43}.
As long as the transition region is narrow, however, 
the amplitude of these oscillations is small, and therefore 
the solution will bear a close resemblance to the solutions described here.
Similar considerations apply if the frequency transitions are not discontinuous,
as will be discussed in section~\ref{s:numerics}.

The above results are derived assuming a symmetric collection of
positive frequency jumps.
The presence of negative frequency jumps at the edges or inbetween 
other jumps would obviously affect the quantitative details of the 
picture.
We expect that the qualitative features, however, and in particular
the asymptotic decomposition into regions of genus-0, 1 and~2,
to remain valid.

\section{Behavior of finite-genus solutions: Numerical simulations}
\label{s:numerics}

Even though the analytical calculations described in the previous sections
provide a general picture of the behavior of finite genus solutions of 
the NLS-Whitham equations,
there are obviously limits to what can be done analytically.
In this section and the following one we therefore complement 
those analytical results by presenting numerical simulations of 
the defocusing NLS equation with small dispersion.

\begin{figure}[b!]
\bigskip
\hbox to \textwidth{\hss\epsfxsize\figwidth\epsfbox{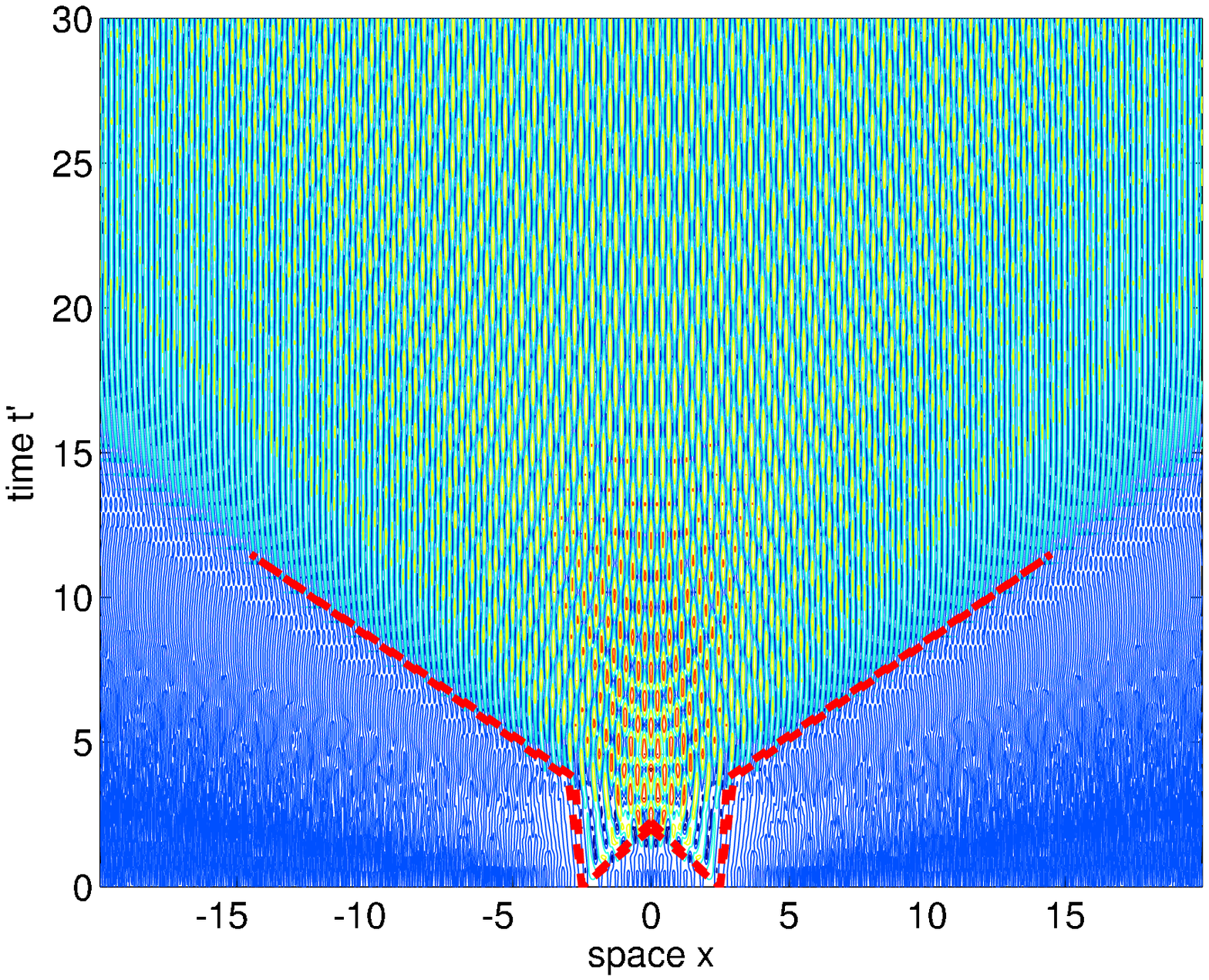}\hss
\epsfxsize\figwidth\epsfbox{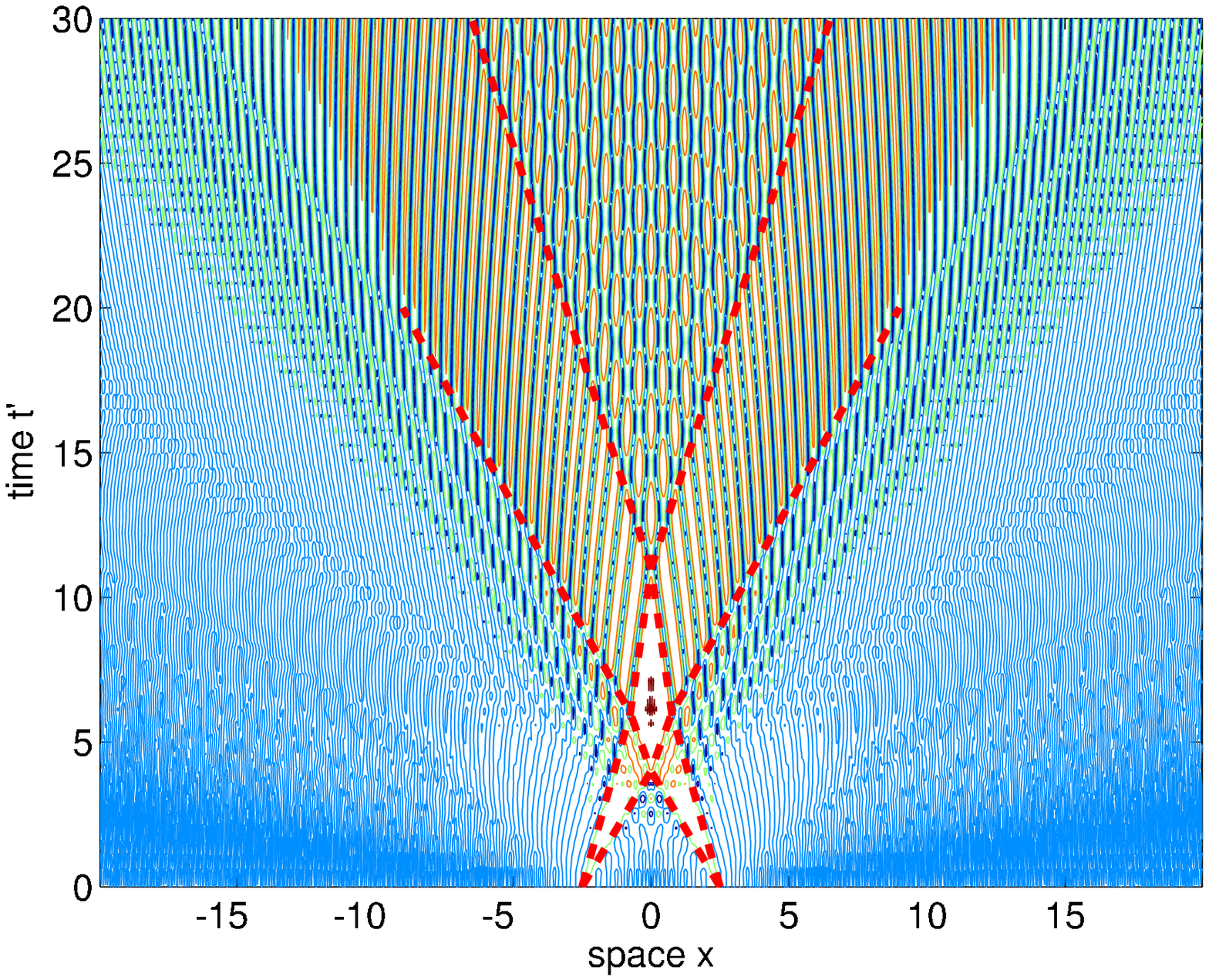}\hss}
\hbox to \textwidth{\hss\epsfxsize\figwidth\epsfbox{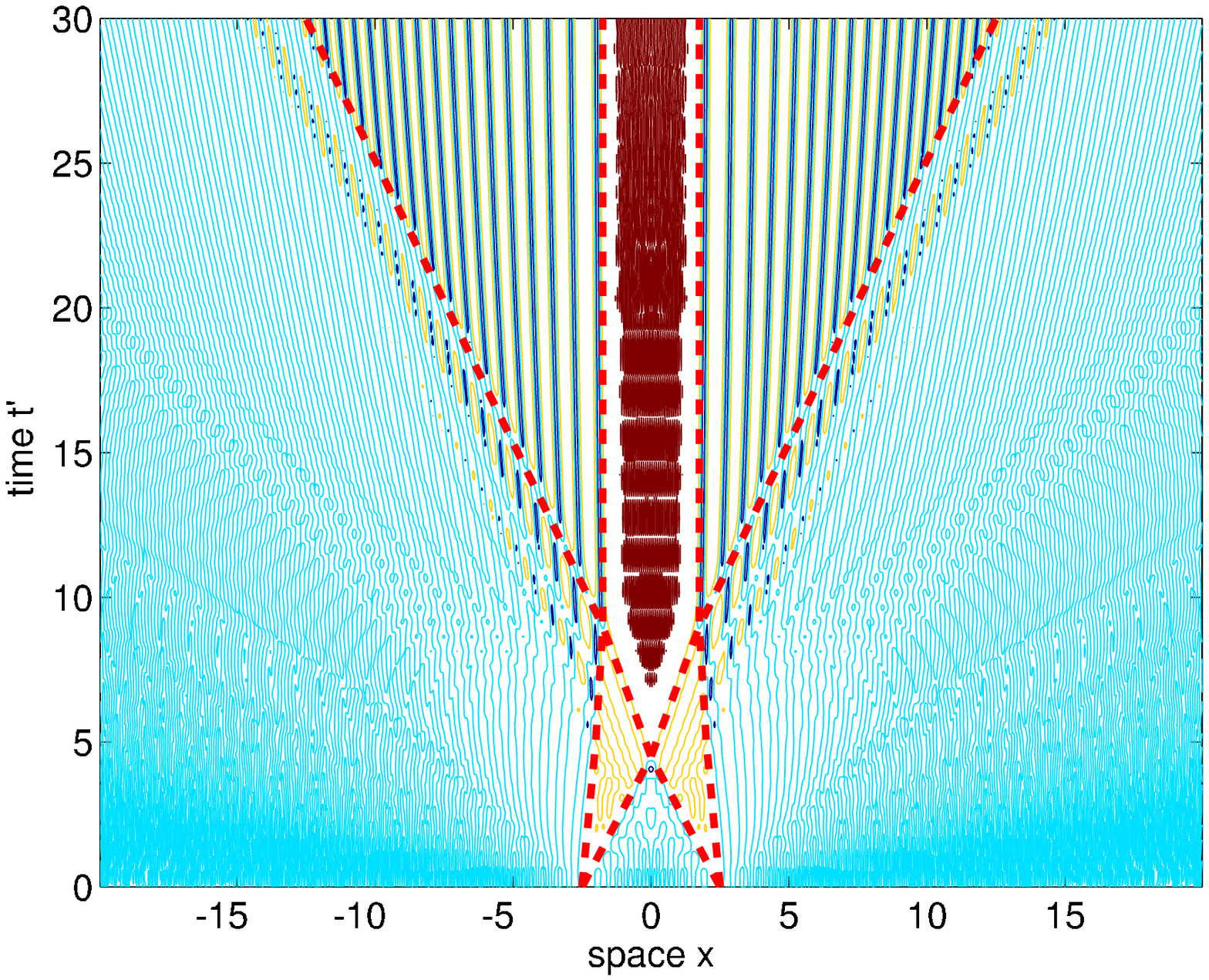}\hss
\epsfxsize\figwidth\epsfbox{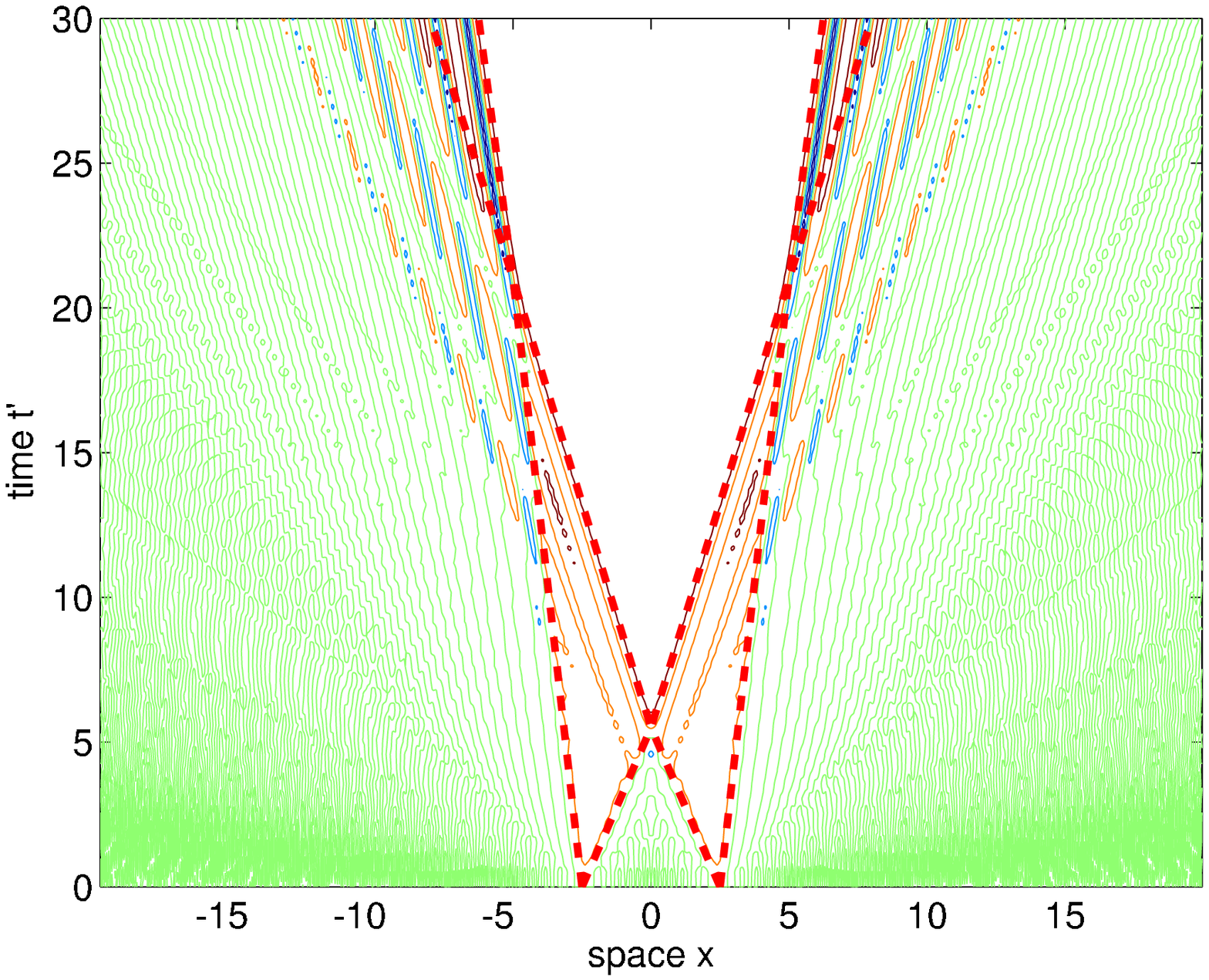}\hss}
\vskip0.25\bigskipamount
\caption{Contour plots of numerical simulations of the NLS equation 
\unskip~\eqref{e:NLS} for a two-jump initial condition
and for different values of the jump size parameter~$C$: 
(a, top left)~$C=5$, corresponding to Fig.~\ref{f:riemann2jumps}a and 
Fig.~\ref{f:genus2jumps}a, showing two jumps that are both individually 
super-critical; 
(b, top right)~$C=3$ corresponding to Figs.~\ref{f:riemann2jumps}c,d and 
Fig.~\ref{f:genus2jumps}b, showing two jumps which are individually 
sub-critical but super-critical in combination; 
(c, bottom left)~the case 
$C=2$, corresponding to Fig.~\ref{f:genus2jumps}c
in which the two jumps are collectively critical; 
(d, bottom right)~$C=1$, corresponding to Fig.~\ref{f:riemann2jumps}b and 
Fig.~\ref{f:genus2jumps}d and showing a case
in which the jumps are collectively sub-critical.
In each case, the dashed lines demarcate the recognizable regions 
of different genus.
In particular, the speeds of the boundaries between genus-0 and genus-1 
regions are in very good agreement with the analytical results described 
in the previous sections.}
\label{f:twojumps}
\vglue-\medskipamount
\end{figure}

\begin{figure}[b!]
\hbox to \textwidth{\hss\epsfxsize\figwidth\epsfbox{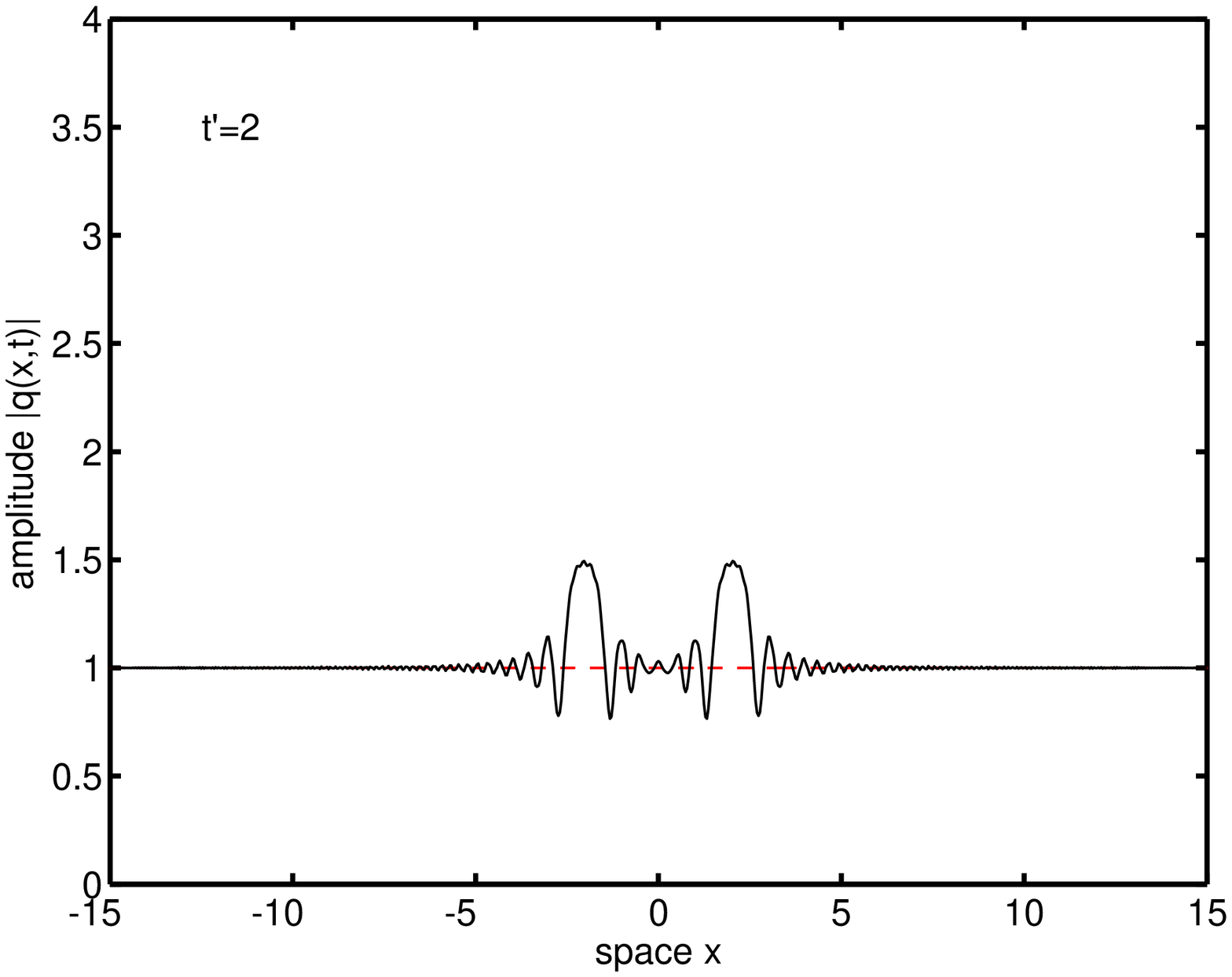}\hss
\epsfxsize\figwidth\epsfbox{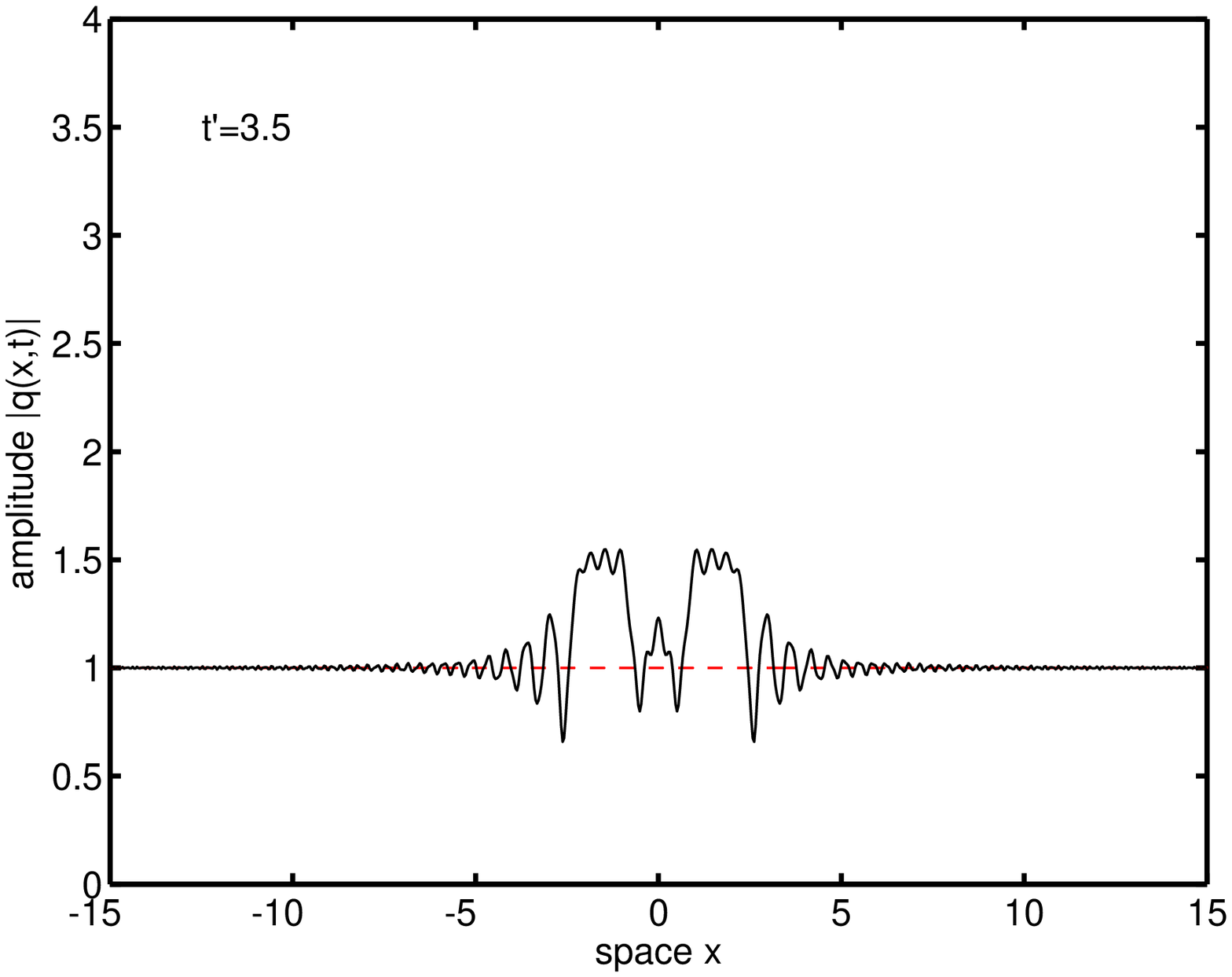}\hss}
\hbox to \textwidth{\hss\epsfxsize\figwidth\epsfbox{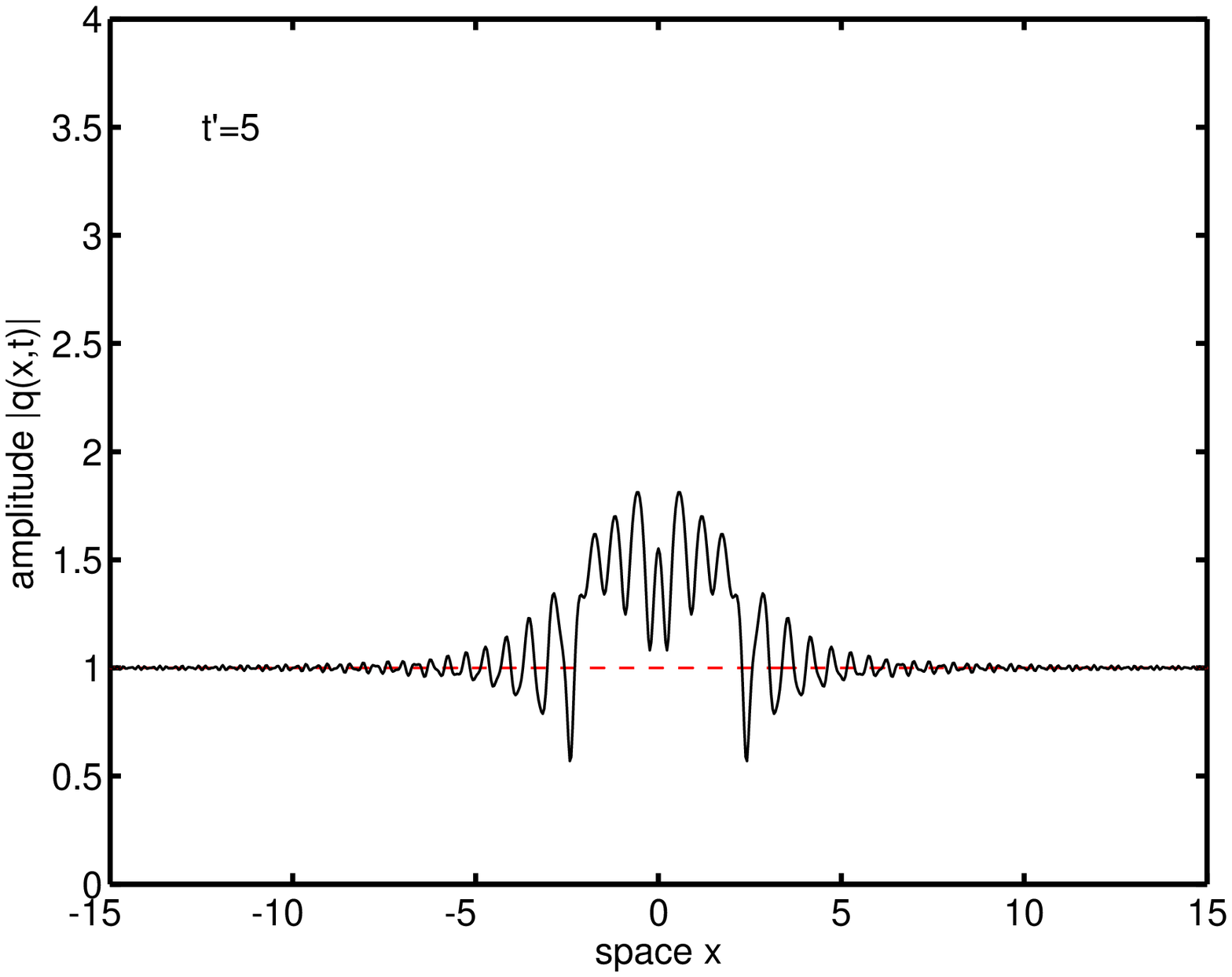}\hss
\epsfxsize\figwidth\epsfbox{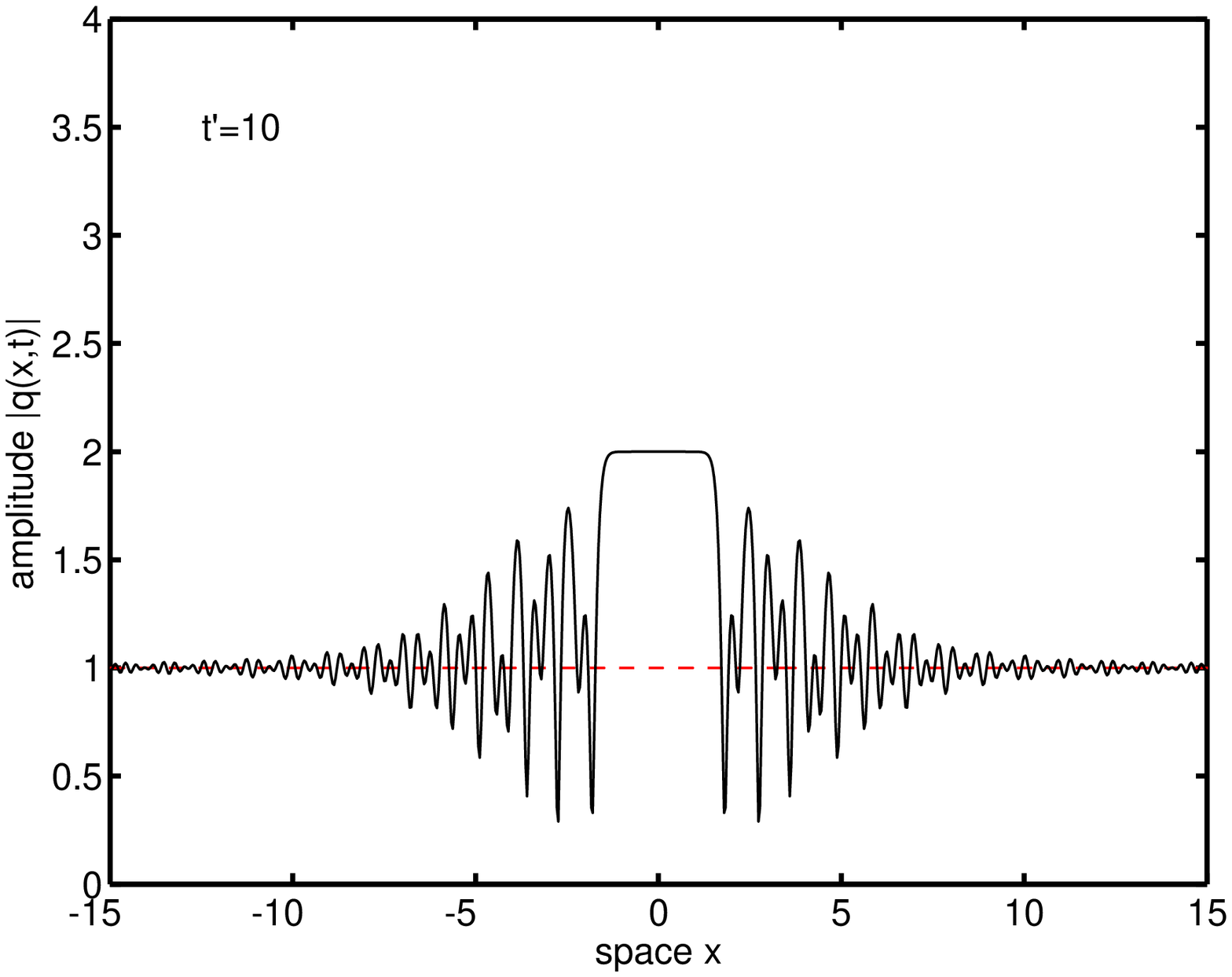}\hss}
\vskip0.25\bigskipamount
\caption{Numerical simulations of the NLS equation illustrating the
evolution with time of a critical two-jump initial condition.  
The frequency jumps, located at $X_2=-X_1=5/2$
have amplitude $C_1=C_2=2$, so that their interaction
produces a genus-0 region of constant width in the center
(as shown in Fig.~\ref{f:twojumps}c). 
The solution is shown at: 
(a, top left)~$t'=2$, (b, top right)~$t'=3.5$, 
(c, bottom left)~$t'=5$, (d, bottom right)~$t'=10$. 
This case corresponds to the contour plot in Fig.~\ref{f:twojumps}c and 
the bifurcation diagram in Fig.~\ref{f:genus2jumps}c.}
\label{f:evolution2jumps}
\kern-0.5\medskipamount
\end{figure}

\begin{figure}[b!]
\hbox to \textwidth{\hss\epsfxsize\figwidth\epsfbox{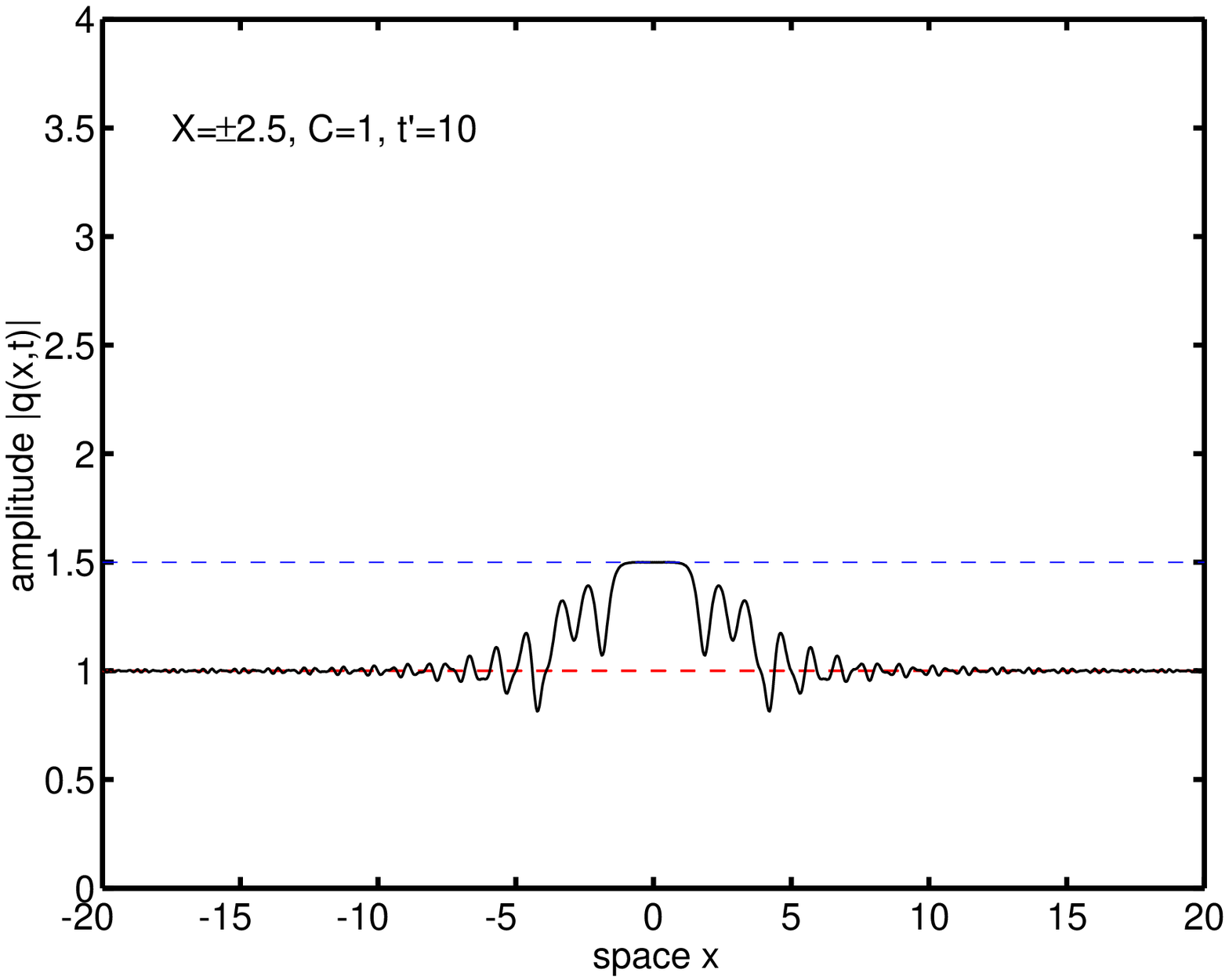}\hss
\epsfxsize\figwidth\epsfbox{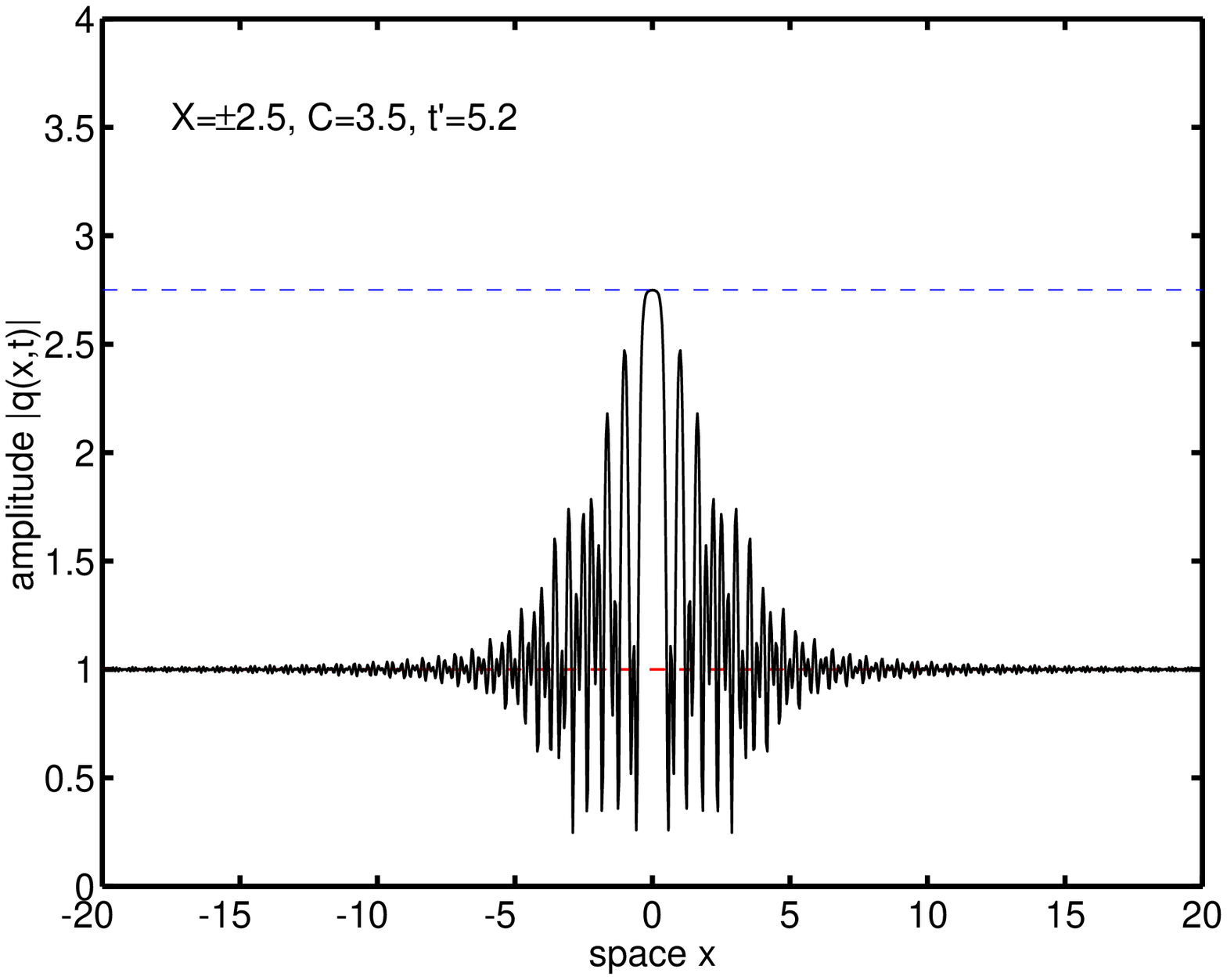}\hss}
\vskip0.2\bigskipamount
\caption{Numerical simulations of a two-jump initial condition
with $\Delta X=5$, 
for different values of the jump size $C_1=C_2=C$:
(a, left)~$C=1$; 
(b, right)~$C=7/2$.
The critical case $C=2$ is also shown in Fig.~\ref{f:evolution2jumps}d.
In Fig.~\ref{f:jumpsize}a the solution is subcritical 
(the genus-0 region expands forever, see Fig.~\ref{f:genus2jumps}d), 
and the solution is shown at $t'=10$, 
whereas in Fig.~\ref{f:jumpsize}b the solution is supercritical 
(see Fig.~\ref{f:genus2jumps}b),
and it is shown at the time when the genus-0 portion
has maximum width, $t'=5.2$.
The horizontal dashed line at $|q|=1$ represents the amplitude of
the initial condition, and the other horizontal dashed line identifies
the amplitude of the genus-0 region.
Note that:
(i)~the amplitude of the genus-0 portion of the solution increases linearly
with increasing value of $C$.
(ii)~for increasingly large values of $C$, the genus-0 solution tends to close
more rapidly, and, above a certain threshold, does not open at all.}
\label{f:jumpsize}
\vglue3\medskipamount
\hbox to \textwidth{\hss\epsfxsize\figwidth\epsfbox{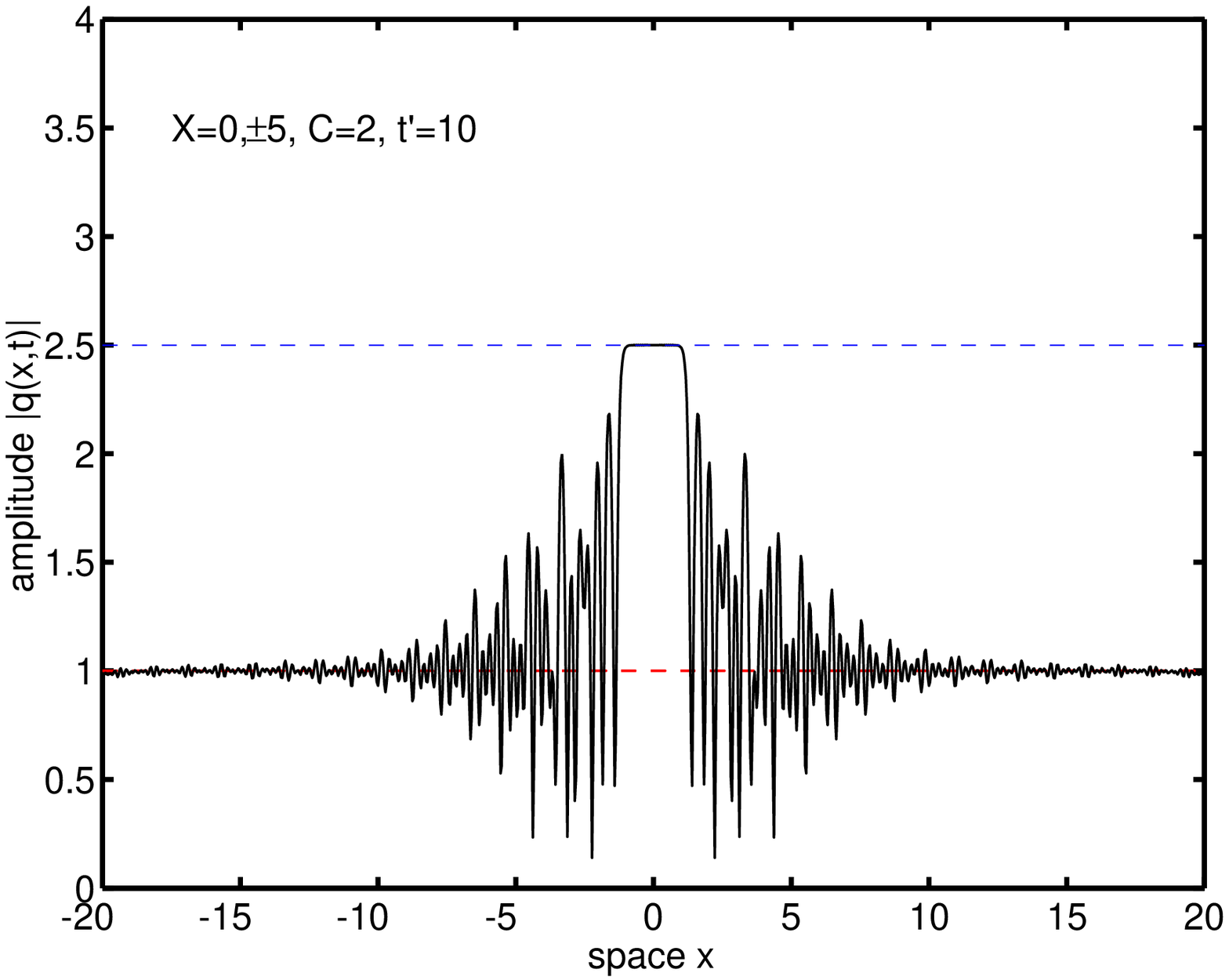}\hss
\epsfxsize\figwidth\epsfbox{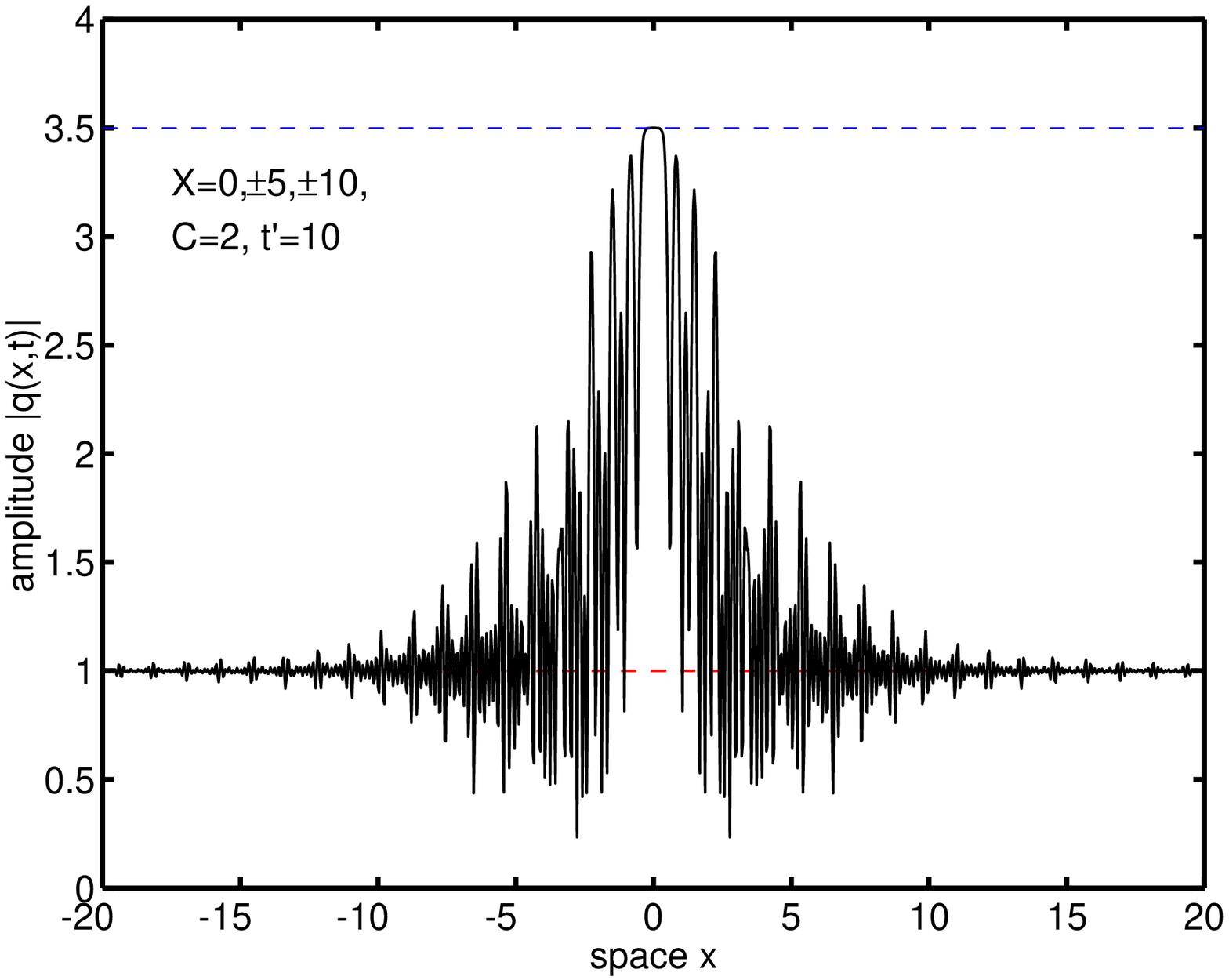}\hss}
\vskip0.2\bigskipamount
\caption{Numerical simulations for different numbers $N$ of frequency jumps, 
with the size of each frequency jump fixed at $C_i=2$, $i=1,\dots,N$:
(a, left)~$N=3$; 
(b, right)~$N=5$.
In both cases the solution is shown at $t'=10$.
See also Fig.~\ref{f:damshock}b, illustrating the case $N=1$,
and Fig.~\ref{f:evolution2jumps}d, illustrating the (critical) case $N=2$.
Note that for $N=1$ the central genus-0 region expands
and for $N=2$ it has a steady width,
whereas for $N>2$ it exists only for limited times. 
In all cases the amplitude of this central genus-0 portion increases 
linearly with the number of jumps.
However, for larger number of jumps, this genus-0 portion tends to close
more rapidly, and above a certain number $N$ it does not open at all,
at least for a fixed value of $\Delta X$.}
\vglue-0.5\medskipamount
\label{f:numberofjumps}
\end{figure}

\begin{figure}[b!]
\hbox to \textwidth{\hss\epsfxsize\figwidth\epsfbox{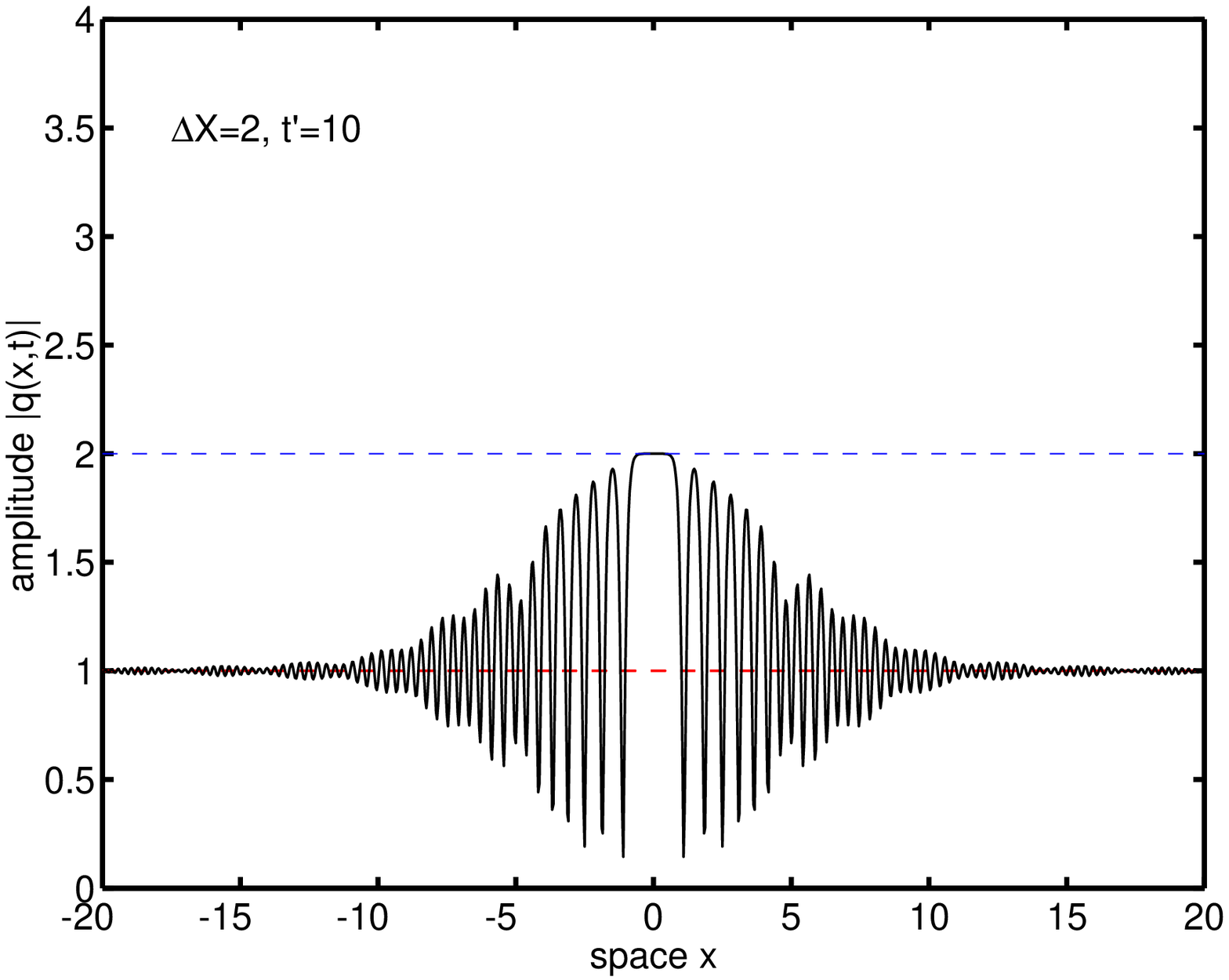}\hss
\epsfxsize\figwidth\epsfbox{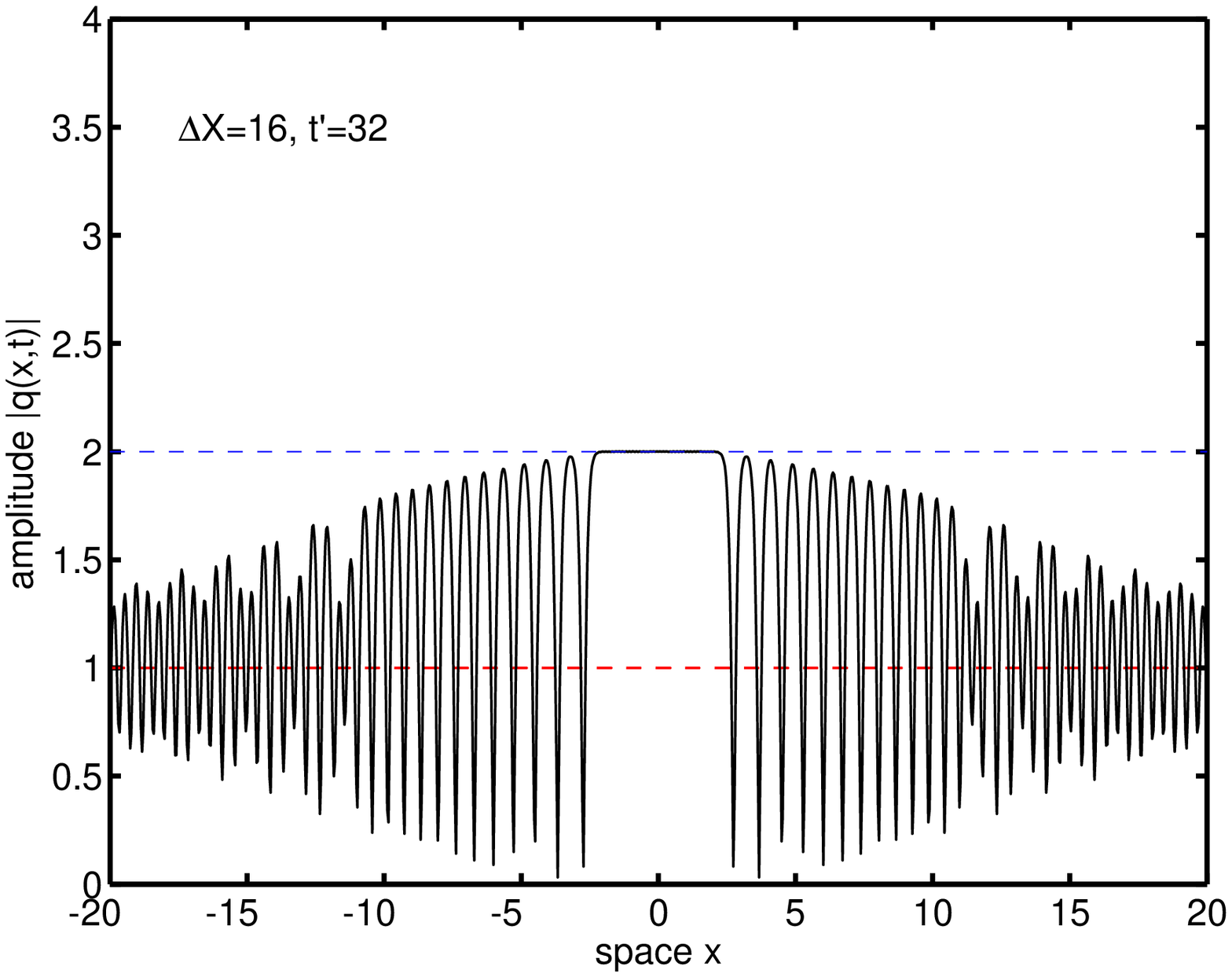}\hss}
\caption{Numerical simulations for different separations 
between jumps, here with a critical arrangement of two frequency jumps 
of amplitude $C_1=C_2=2$:
(a, left)~$\Delta X=2$, 
(b, right)~$\Delta X=16$.
See also Fig.~\ref{f:evolution2jumps}d, which shows the case $\Delta X=5$.
In Fig.~\ref{f:jumpseparation}a the solution is shown at $t'=10$,
in Fig.~\ref{f:jumpseparation}b at $t'=32$.
Increasing the value of $\Delta X$ has two main effects: 
it increases the maximum width of the genus-0 portion, and it simultaneously
lengthens the time scales over which the evolution occurs. See also Fig.~\ref{f:genus2jumps}c.}
\label{f:jumpseparation}
\kern2\medskipamount
\hbox to \textwidth{\hss\epsfxsize\figwidth\epsfbox{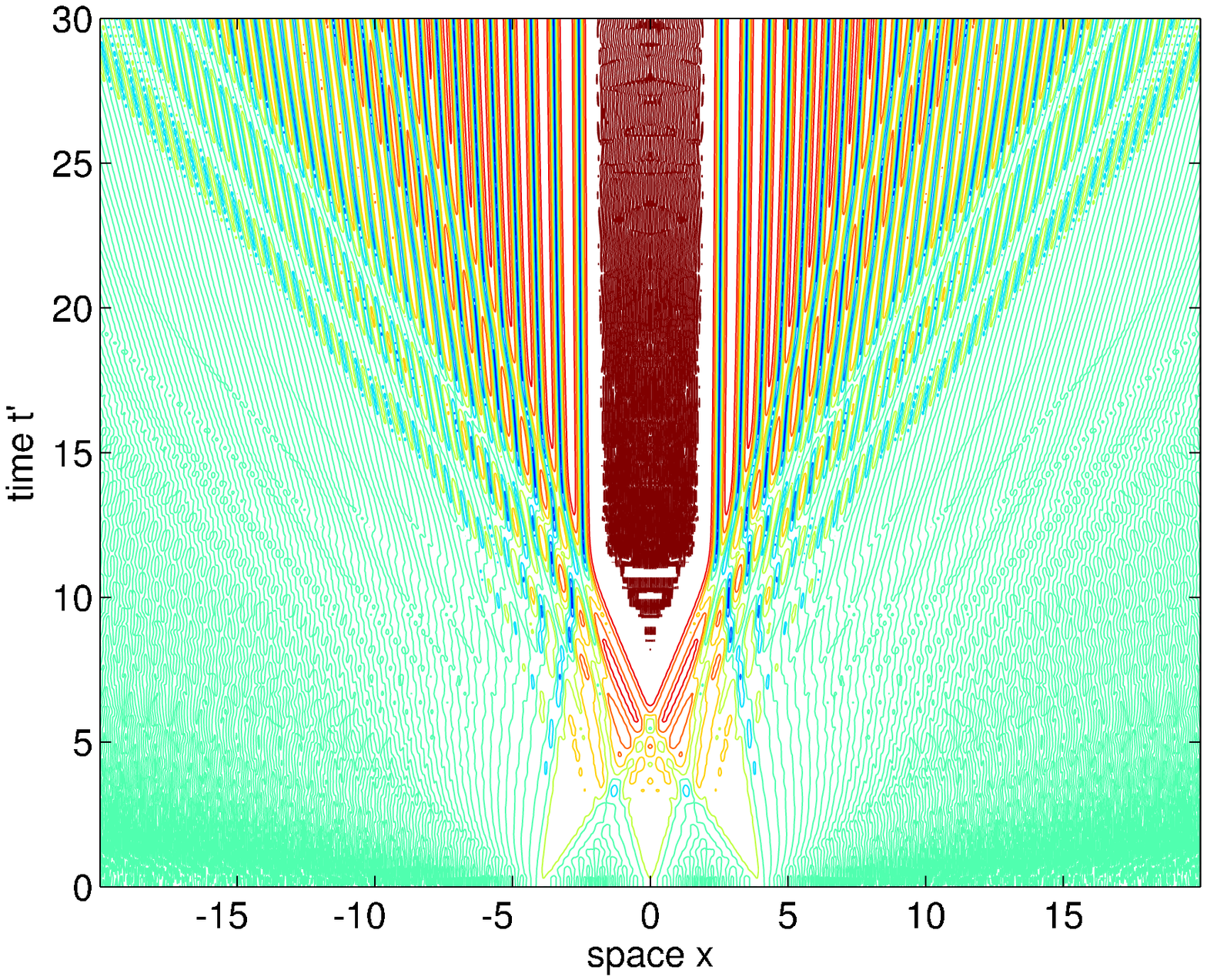}\hss
\epsfxsize\figwidth\epsfbox{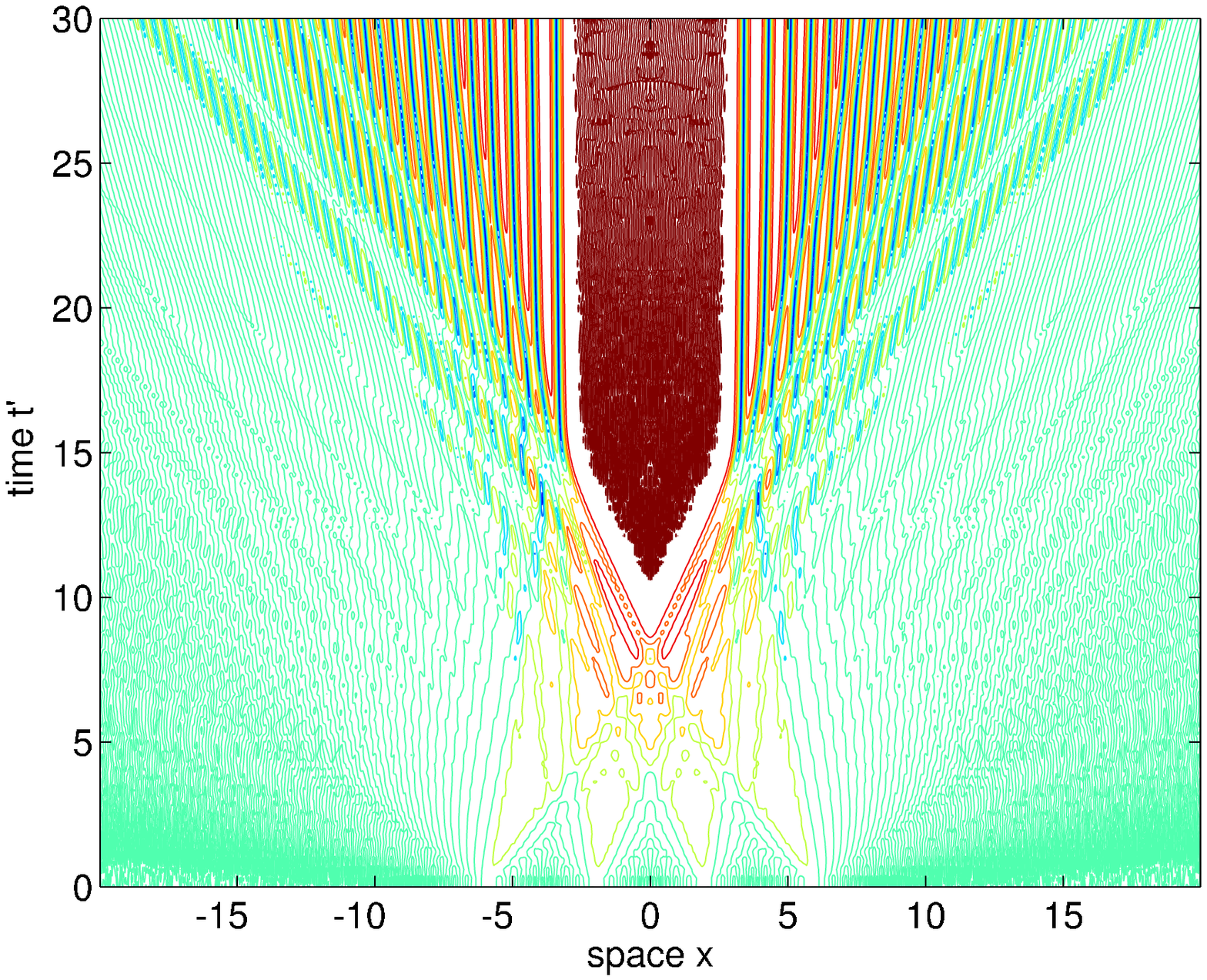}\hss}
\caption{Critical arrangements of more than two jumps: 
(a, left)~$N=3$ and $C=4/3$; 
(b, right)~$N=4$ and $C=1$.
In both cases $q_0=1$ and $\Delta X=4$.}
\label{f:criticalmanyjumps}
\end{figure}

\begin{figure}[b!]
\hbox to \textwidth{\hss\epsfxsize\figwidth\epsfbox{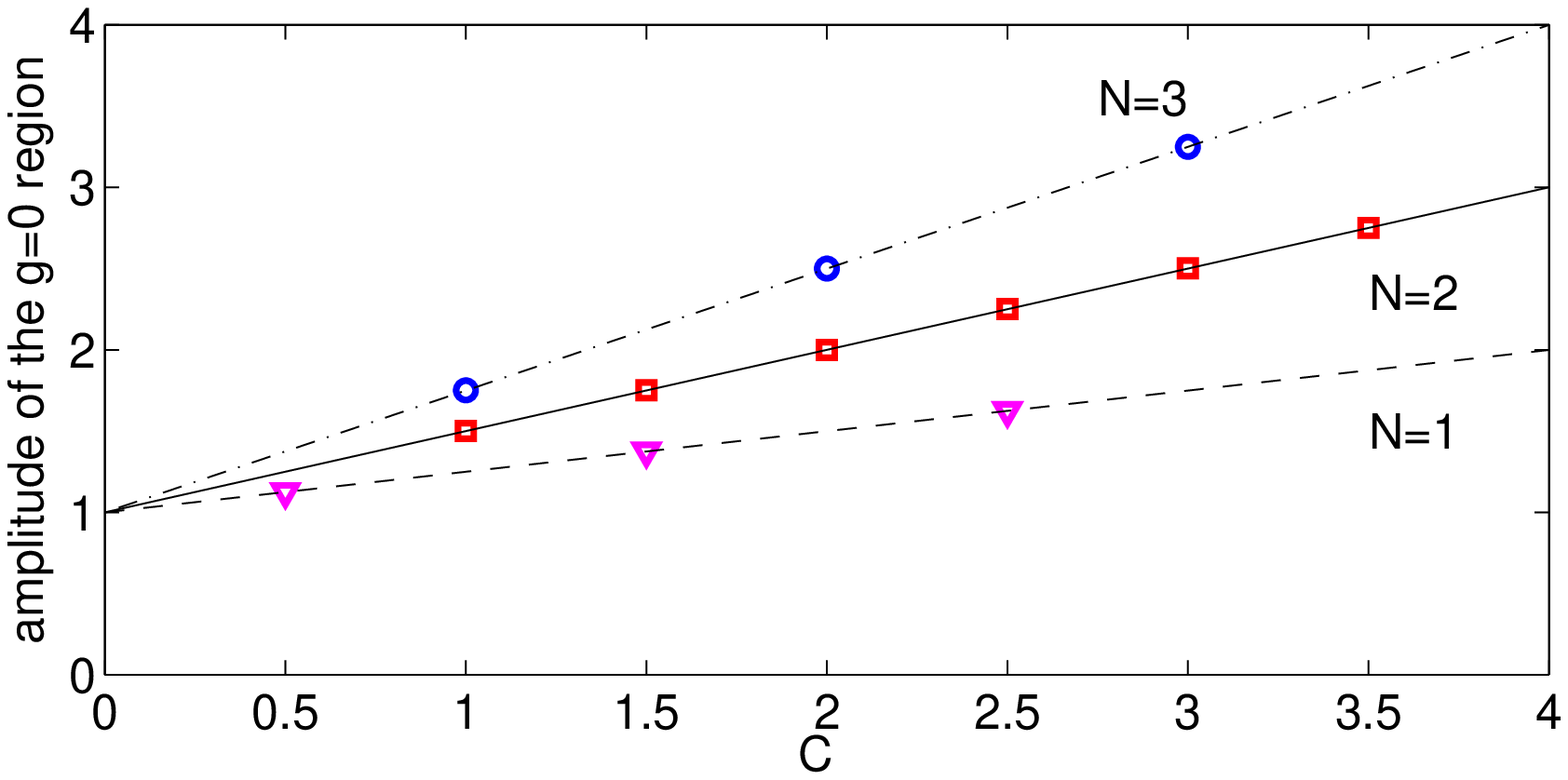}\hss
\epsfxsize\figwidth\epsfbox{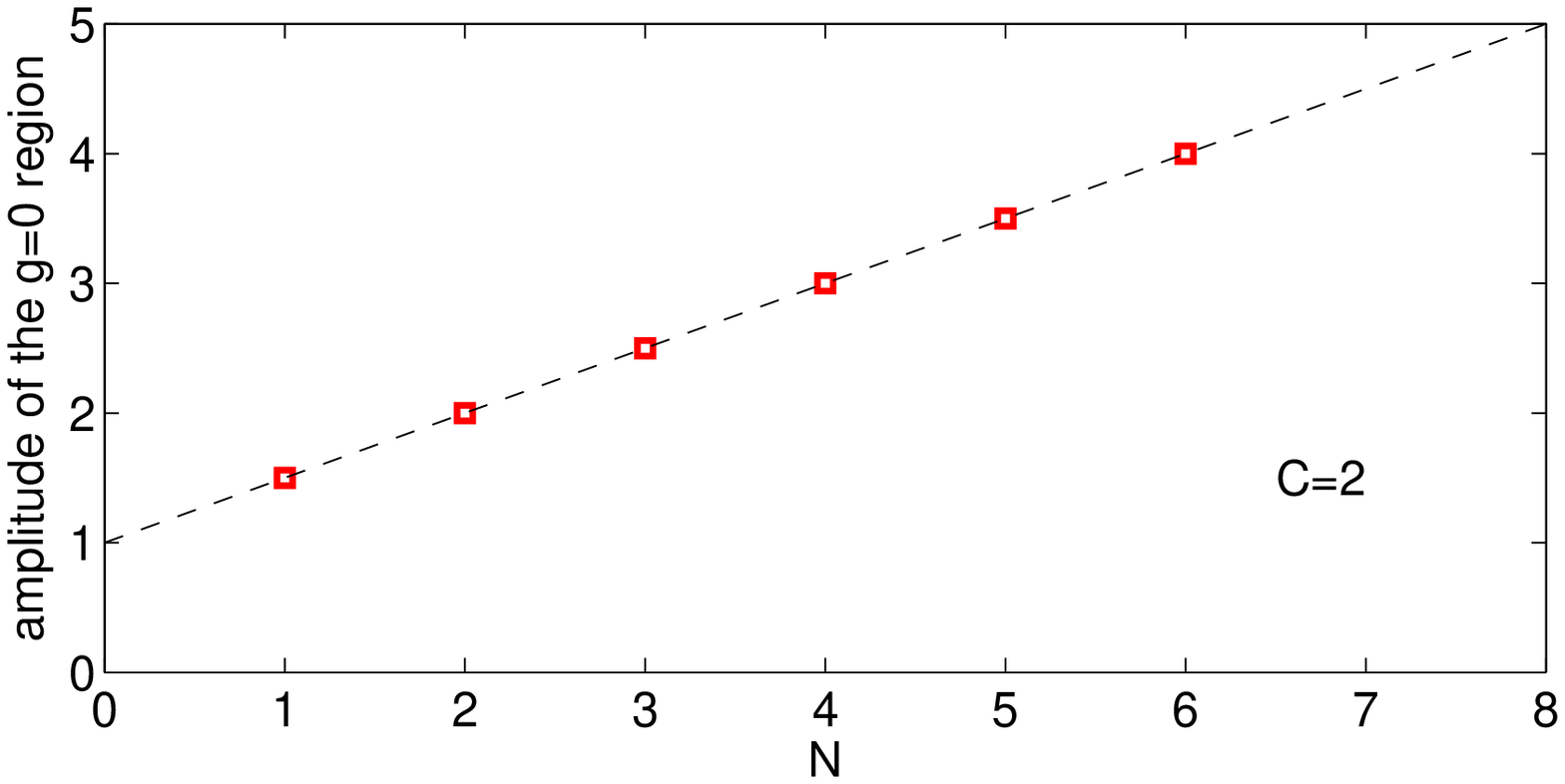}\hss}
\kern-\medskipamount
\caption{(a, left) The numerically calculated amplitude of the 
genus-0 region produced by
an arrangement of $N$ equal-size frequency jumps as a function of 
the indidual jump size~$C$.
Triangles: $N=1$; squares: $N=2$; circles: $N=3$.
For comparison, 
the solid, dashed and dot-dashed lines show the straight lines $y=(N-1)\,C/4$.
(b, right) The amplitude of the genus-0 region produced
by an arrangement of frequency jumps each of size $C=2$, as a function
of the number $N$ of jumps.
Squares: numerical simulations;
dashed line: the line $y=1+N/2$.}
\label{f:g0size}
\kern3\medskipamount
\hbox to \textwidth{\hss\epsfxsize1.0\figwidth\epsfbox{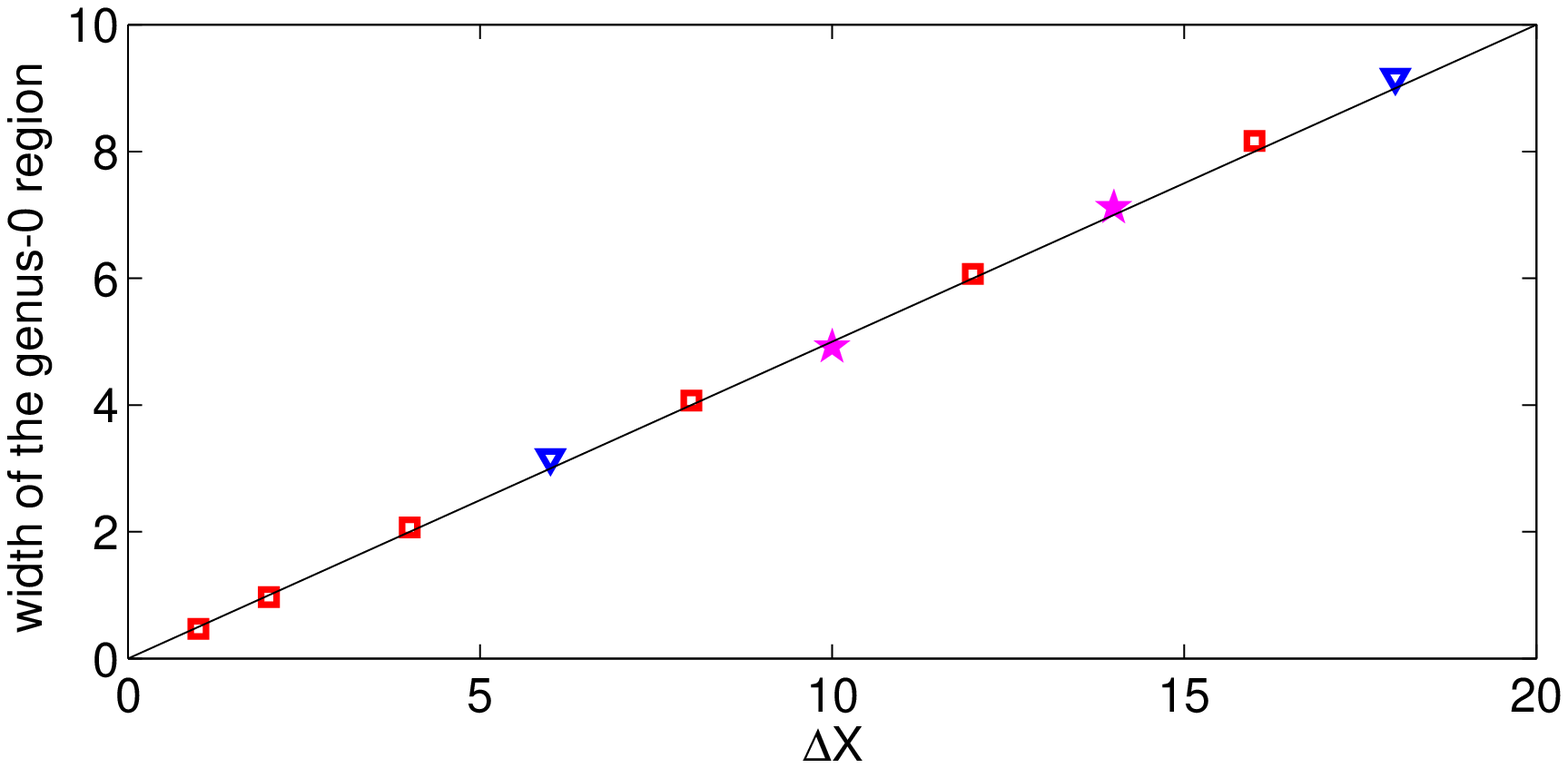}\hss
\epsfxsize1.0\figwidth\epsfbox{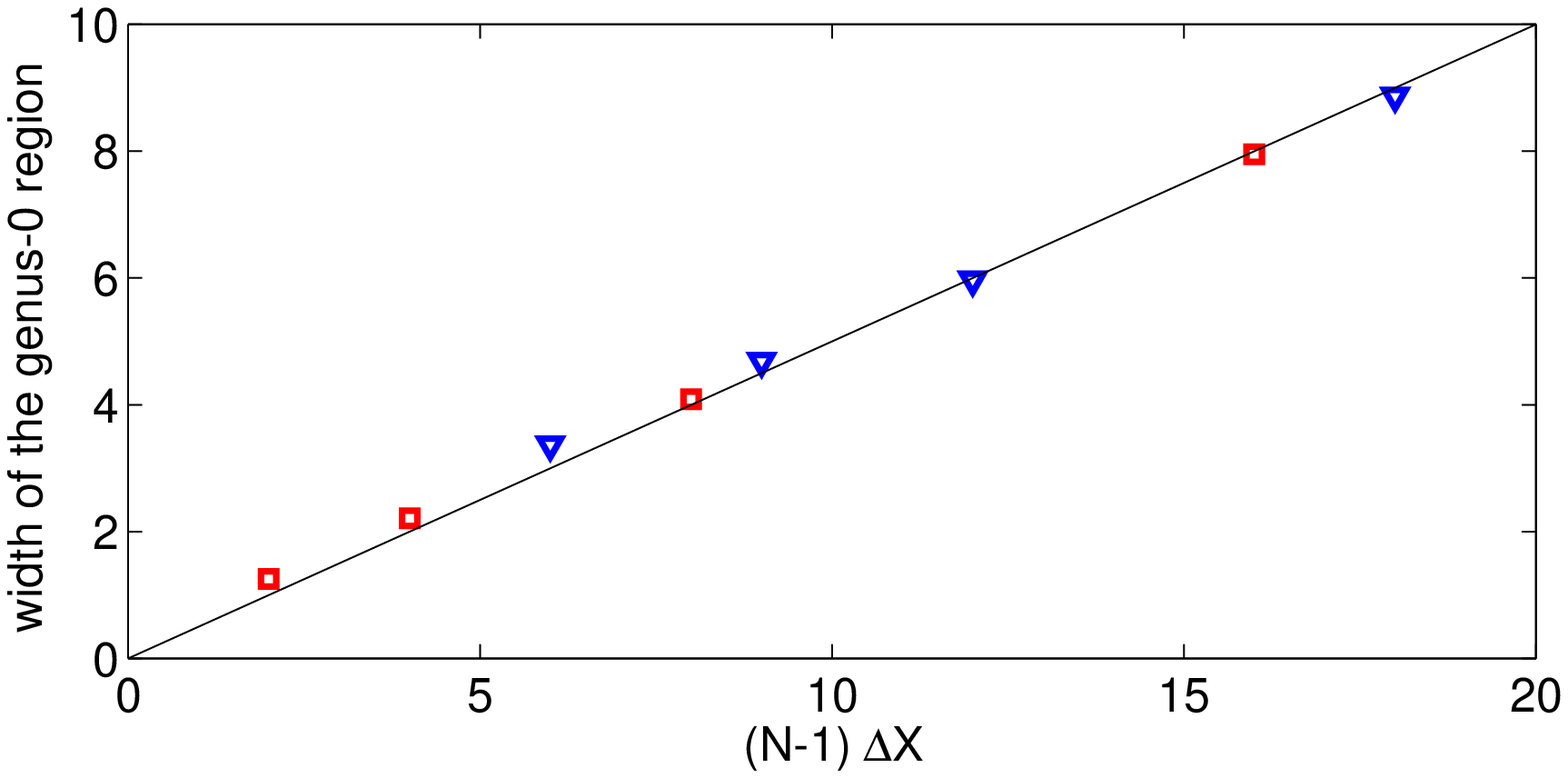}\hss}
\kern-\medskipamount
\caption{Width of the genus-0 region resulting from the interaction of
a critical set of frequency jumps as a function of the spatial separation
between them: (a, left)~$N=2$;
solid line: the straight line $y=\Delta X/2$;
squares: simulations with $q_0=1$ and~$C=2$;
triangles: simulations with $q_0=2$ and~$C=4$;
stars: simulations with $q_0=3$ and $C=6$.
(b, right)~$N=3$ (squares) and $N=4$ (triangles), with $q_0=1$ in both cases;
solid line: the straight line $y= (N-1)\Delta X/2$.
Note that, since the characteristic periodi of the rapid oscillations 
is~$O(\epsilon)$, it is only possible to measure the width 
of the genus-0 region with precision $\Delta x=\pm\epsilon$.}
\label{f:g0width}
\end{figure}

All the numerical results presented are obtained by numerically integrating 
the NLS equation~\eqref{e:NLS} with $\epsilon^2=0.1$
using a fourth-order Fourier split-step method.  
We consider an initial condition $q(x,0)=|q(x,0)|\,\exp[i\varphi(x,0)/\epsilon]$, 
where the initial phase $\varphi(x,0)$ is obtained by integrating
Eq.~\eqref{e:ugeneral}:
\begin{gather}
\varphi(x,0)= \sum_{j=1}^N \big[ - x/2 + (x-X_j)H(x-X_j)\,\,\big]\,C_j\,,
\label{e:phigeneral}
\end{gather}
where the function
$H(x)=1$ if $x\ge0$ and $H(x)=0$ if $x<0$ represents the Heaviside
unit step.
The initial pulse amplitude was taken to be a super-Gaussian, namely 
$|q(x,0)|=q_0\,e^{-(x/\Delta X_0)^{2M}}$ with $M=40$.
The integration step size and the width $\Delta X_0$ of the pulse 
were chosen to be respectively sufficiently small and sufficiently large 
that none of the numerical results described in this work are affected 
by them.
In order to remove the factor $\epsilon$ in front of the time
derivative in Eq.~\eqref{e:NLS}, the numerical simulations were
set-up in terms of the fast scale $t'=t/\epsilon$.
Accordingly, 
all the numerical results will be described in terms of this time variable.
In what follows we describe the behavior of the solutions
for varying values of the parameters $X_j$ and $C_j$ and for 
varying number of frequency jumps.
We first consider the case of two frequency jumps, 
to numerically identify regions of different genus (Fig.~\ref{f:twojumps})
and show a few snapshots of the time evolution (Fig.~\ref{f:evolution2jumps}).
We then proceed to describe three sets of simulations:
The first set (Fig.~\ref{f:jumpsize}) shows the effect of gradually 
increasing the size of the frequency jump. 
The second set (Fig.~\ref{f:numberofjumps}) describes the effect of
increasing the number of jumps.
Finally, the third set (Fig.~\ref{f:jumpseparation})
shows the effect of changing the initial separation between these jumps.
Although numerical experiments were performed with various values of amplitude,
all the figures in this work except Fig.~\ref{f:g0width}a are relative to 
the case $q_0=1$.

\paragraph{Two frequency jumps.}
Figure~\ref{f:twojumps} shows contour lines of $|q(x,t)|$ in the 
$(x,t)$-plane for different value of the jump size $C_1=C_2=2u_0=:C$.
Because of the finite, non-zero value of $\epsilon$,
not all of the features presented in the previous section
are immediately apparent.  
In other words, because of the finite value of $\epsilon$, the
solution of the NLS equation is only approximately described by 
a finite-genus solution in each region.
Moreover, numerically reconstructing the genus of the 
solution is a highly nontrivial problem, which is outside the
scope of this work.
Nonetheless, some regions of genus-0 and genus-1
(whose boundaries have been delimited by dashed lines in Fig.~\ref{f:twojumps})
are recognizable 
(e.g., regions of genus-0 are often flat, and therefore appear white 
in the contour plots in Fig.~\ref{f:twojumps}), and their location 
displays a very good qualitative and quantitative agreement with 
the analytical calculations presented in section~\ref{s:multiphase}.
Note in particular that, 
even though the value of $\epsilon$ is not particularly small,
the value of~$C$ which corresponds to the critical case is 
exactly that which was predicted analytically, namely $C=2$.
Figure~\ref{f:evolution2jumps} shows four snapshots illustrating 
the evolution with time of the critical case shown in 
Fig.~\ref{f:twojumps}c.

It is also interesting to note that the hexagonal patterns visible
in Figs.~\ref{f:twojumps}
(in particular, near the center $x=0$ in 
Figs.~\ref{f:twojumps}a and~\ref{f:twojumps}b) 
show the interaction of two counter propagating genus-1 (periodic) waves, 
which produce, as a result, a genus-2 solution of the NLS equation. 
Note that in the absence of nonlinearity, a superposition of two 
simple waves would produce just a parallelogrammic pattern. 
Thus, the hexagonal pattern is the result of nonlinear phase shifts 
in the interaction. (Note that the hexagonal pattern can also be found 
in genus-2 solutions of the Kadomtsev-Petviashvili equation
describing two-dimensional shallow water waves\cite{segur}.)

\paragraph{Size of the frequency jumps.}
This set of simulations, 
some results of which are shown in Fig.~\ref{f:jumpsize},
is relative to a two-jump initial condition.
We take the two jumps to be equal-size, $C_1=C_2=2u_0=C$,
and to be initially positioned at $X_1=X_2=\pm2.5$.
Looking at the behavior of the solution as $C$ varies,
one can clearly observe that 
the amplitude of the genus-0 portion of the solution increases linearly
with increasing value of~$C$.
More precisely, $|q|_\max=q_0+C/2$, 
in agreement with the analytical calculations
presented in the previous section (cf.\ Corollary \ref{c:genus0amplitude}).
See also Fig.~\ref{f:g0size}a later.
Moreover, in a similar way as in the single-jump case, one can see that
the maximum amplitude of a stable genus-0 region is $|q|_\max=2q_0$.
When $u_0$ is larger than the threshold value $q_0$,
the genus-0 region eventually disappears, as explained 
in section~\ref{s:multiphase}
and as described in Fig.~\ref{f:genus2jumps}b.
Unlike the single-jump case, however, one can obtain larger values of
$\rho$ for limited times, as shown in Figs.~\ref{f:jumpsize}c,d.
Note how, over these shorter times, the amplitude of the
genus-0 region is still given by $|q|_\max=q_0+ C/2$.
Note also, however, that the genus-0 solution closes more rapidly
for increasingly large values of $C$, 
and, above the threshold $C=4$ (i.e., $C=q_0$) does not open at all,
at least for a fixed value of $\Delta X$.
These results are in very good agreement with the analytical calculations
described in the previous sections 
(e.g., see Proposition~\ref{p:genus0region?} and 
Corollaries~\ref{c:genus0amplitude} and~\ref{c:genus0asymptotic?}).

\paragraph{Number of frequency jumps.}
In this set of simulations, 
some of which are shown in Fig.~\ref{f:numberofjumps},
the number $N$ of frequency jumps was varied, while the size of each 
jump and their initial spatial separation are kept fixed respectivaly
at $C=2$ and $\Delta X=5$.
From Fig.~\ref{f:numberofjumps}, it is apparent that the amplitude of 
the genus-0 part also increases linearly with the number of jumps.
More precisely, $|q|_\max=q_0+ NC/4$,
in agreement with the analytical calculations presented in the previous
sections (cf.\ Corollary~\ref{c:genus0amplitude}).
See also Fig.~\ref{f:g0size}b later.
The genus-0 region, however, will eventually close up if the \textit{total} 
jump exceeds the same threshold as in the previous case:
$C_\mathrm{tot}=NC=4q_0$ (cf.\ Corollary~\ref{c:genus0asymptotic?}).
For increasingly larger numbers of jumps, the amplitude of the genus-0
region follows the same law as above, namely $|q|_\max=q_0+NC/4$,
and can temporarily achieve very large values.
The genus-0 region also tends to close up more rapidly, however, and, 
above a certain number of jumps, does not open at all
(at least for a fixed value of $\Delta X$).

\paragraph{Spatial separation between frequency jumps.}
In this set of simulations 
we look at the effect of increasing the spatial separation $\Delta X$ 
between the frequency jumps.
Figure~\ref{f:jumpseparation} is relative to a two-jump initial condition
with $C=2$. 
The consequences of changes in $\Delta X$ are not as obvious
as those of changes in the number of jumps or the jump size.  
One can see, however, that increasing $\Delta X$ has two main effects: 
on one hand, it lenghtens the time scales over which the 
evolution occurs, since the various regions need to travel for
longer distances to come into contact and to interact;
at the same time, it determines the width of the stable
genus-0 region in the critical case.
In fact, numerical results show a remarkable fact, namely that 
the width~$X_\infty$ of the stable genus-0 region depends linearly 
on the initial spatial separation~$\Delta X$ between the jumps.
In the case of two frequency jumps, one simply has $X_\infty=\Delta X/2$.
(Cf.\ Figs.~\ref{f:twojumps}c, \ref{f:evolution2jumps}
and~\ref{f:jumpseparation}.)
A similar relation however holds independently of the number of jumps. 
For a critical arrangement of an arbitrary number $N$ of equal-size,
equally spaced positive jumps, it is 
$X_\infty= (N-1)\,\Delta X/2$,
where $\Delta X$ is the initial separation between consecutive jumps.
(Cf.~Fig.~\ref{f:criticalmanyjumps}, and 
note that $(N-1)\,\Delta X$ is just the distance among the farthest two 
frequency jumps.)
These relations are verified in Figs.~\ref{f:g0width}a,b.

\paragraph{Summary.}
Let us recapitulate for convenience the main results of the 
numerical experiments discussed in the above paragraphs. 
Given~$N$ positive frequency jumps with individual sizes~$C_j$ and 
separations~$\Delta X_j$, we can observe that:
\begin{itemize}
\itemsep0pt
\parsep0pt
\item[(i)]
The value of the asymptotic genus in the central portion of the pulse 
is only determined by the total jump $C_\tot=C_1+\dots+C_N$
(or, equivalently, by the size per jump $C=C_\tot/N$ in the case of 
$N$ jumps of equal size).
More precisely, the solution will be genus-0 iff $C_\tot<q_0$,
in very good agreement with the analytical results 
(see Corollary~\ref{c:genus0asymptotic?}).
Of course, the detailed behavior of the solution for 
finite times depends on the precise values of all the $C_j$ 
and~$\Delta X_j$.
\item[(ii)]
If a genus-0 region is desired which expands forever at the 
center of the pulse, its height is limited by the critical value 
of the cumulative jump~$C_\tot$.
More precisely, $|q|_\max= 2q_0$, 
independently of the number of jumps.
For equal-size jumps, this value is obtained for $C=4q_0/N$, 
in very good agreement with the analytical results
(see Corollary~\ref{c:genus0amplitude}).
\item[(iii)]
It is possible to achieve genus-0 regions of much larger 
amplitudes over limited times by employing a sufficiently
large number of jumps.
More precisely, $|q|_\max=q_0+C_\tot/4$\, 
(i.e., $|q|_\max=q_0+NC/4$ for equal-size jumps),
as shown in Fig.~\ref{f:g0size},
and again in very good agreement with the analytical results
(see Corollary~\ref{c:genus0amplitude}).
\item[(iv)]
For a critical arrangement of an arbitrary number of equal-size,
equally spaced positive frequency jumps, 
the width of the stable genus-0 region
is half of the initial separation between the farthest two frequency jumps.
This fact has not been demonstrated analytically,
but is nonetheless very solidly supported by the numerical results,
as shown in Fig.~\ref{f:g0width}.
\end{itemize}
In the next section we discuss how these results (and in particular
items iii and iv) can be used to generate intense, ultra-short 
optical pulses.

\section{Applications: Generation of intense short optical pulses}
\label{s:opticalpulses}

\begin{figure}[b!]
\hbox to \textwidth{\hss\epsfxsize\figwidth\epsfbox{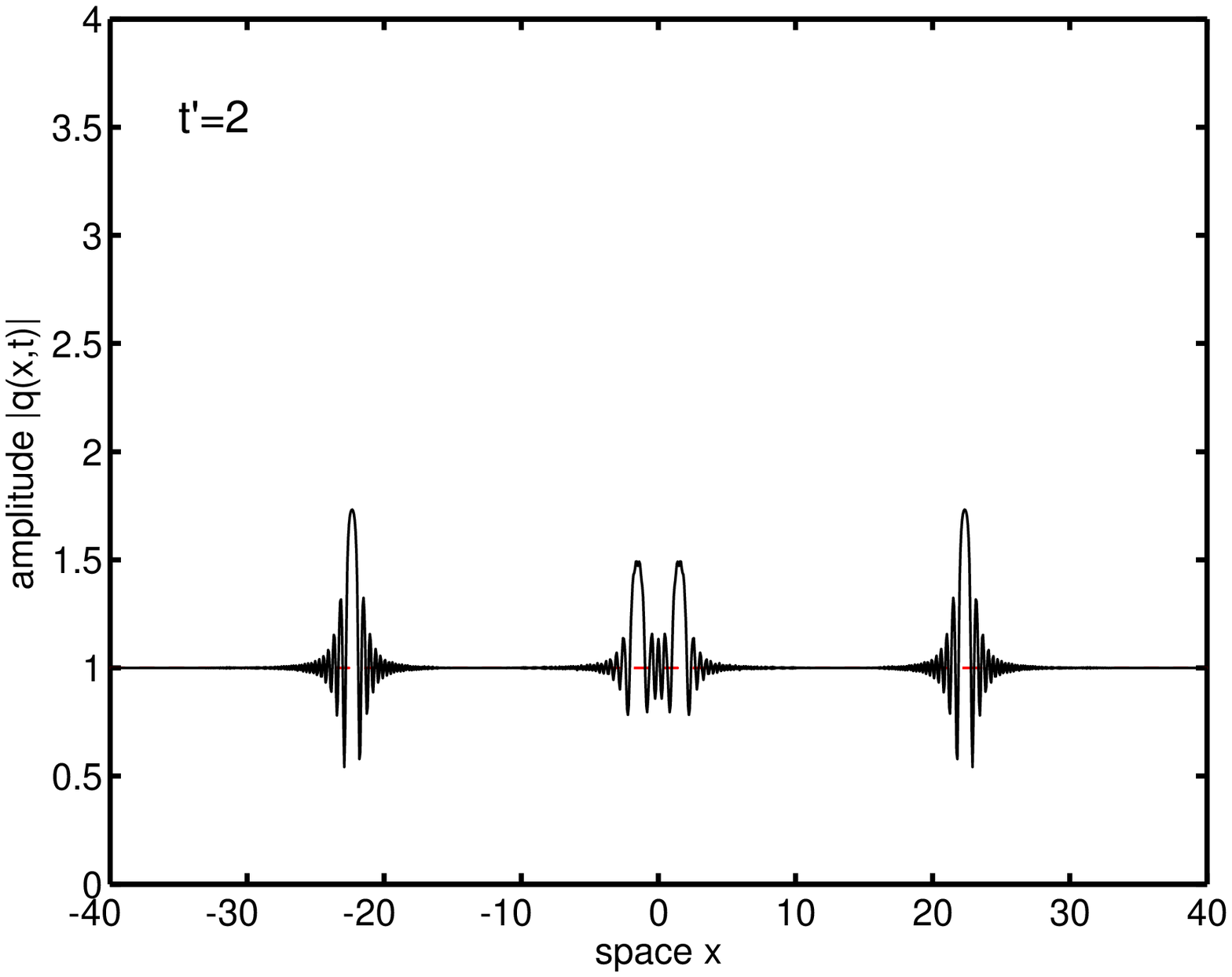}\hss
\epsfxsize\figwidth\epsfbox{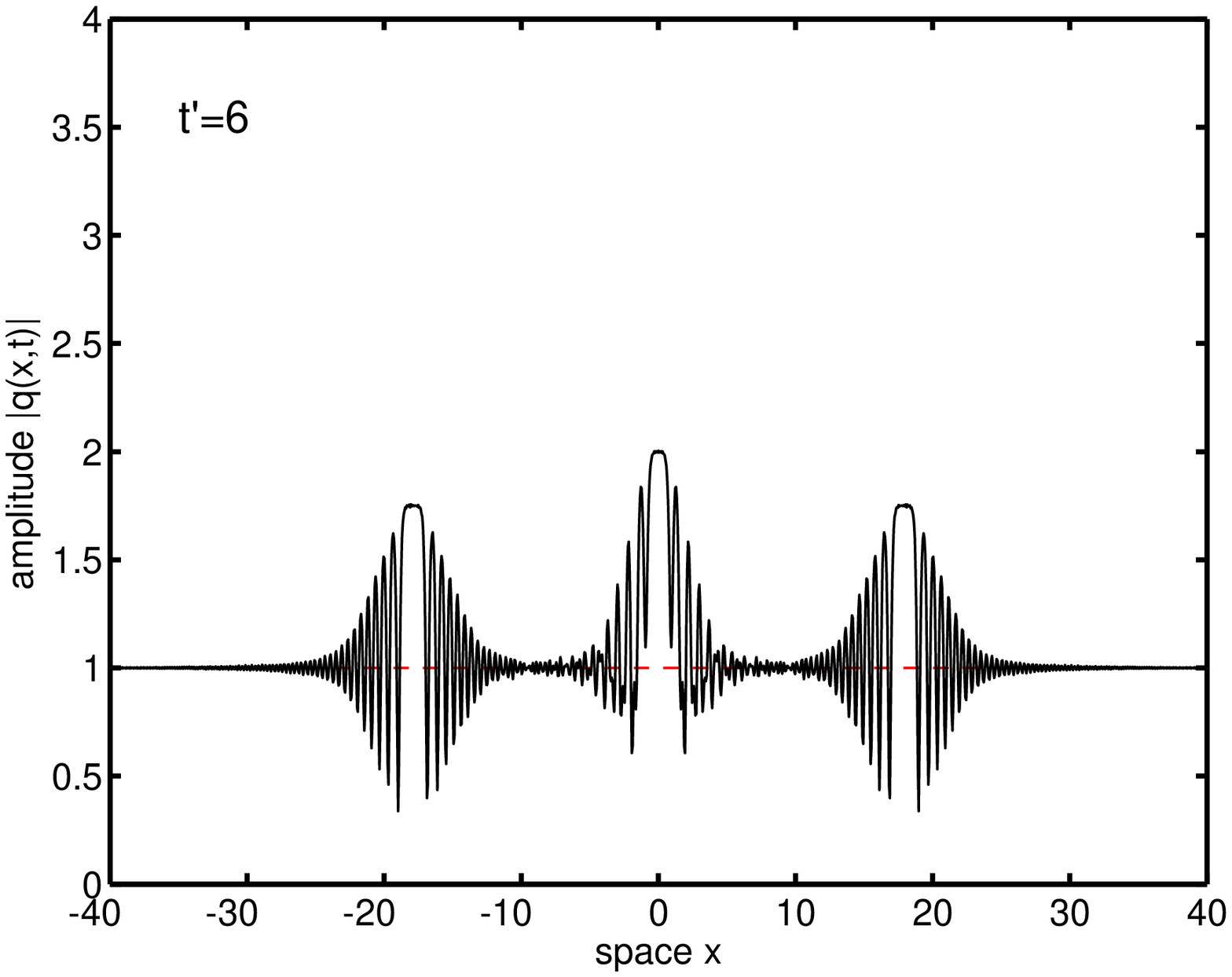}\hss}
\hbox to \textwidth{\hss\epsfxsize\figwidth\epsfbox{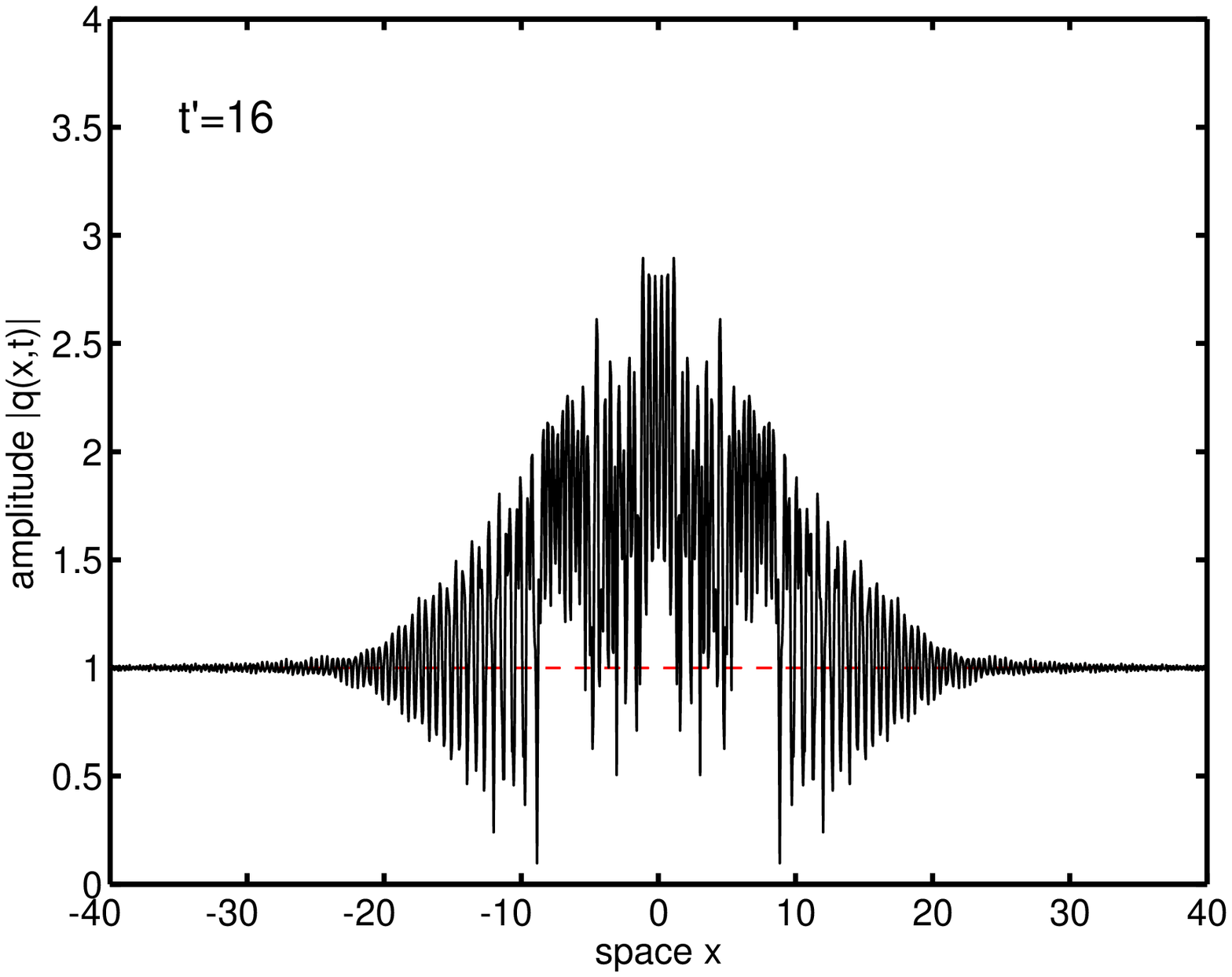}\hss
\epsfxsize\figwidth\epsfbox{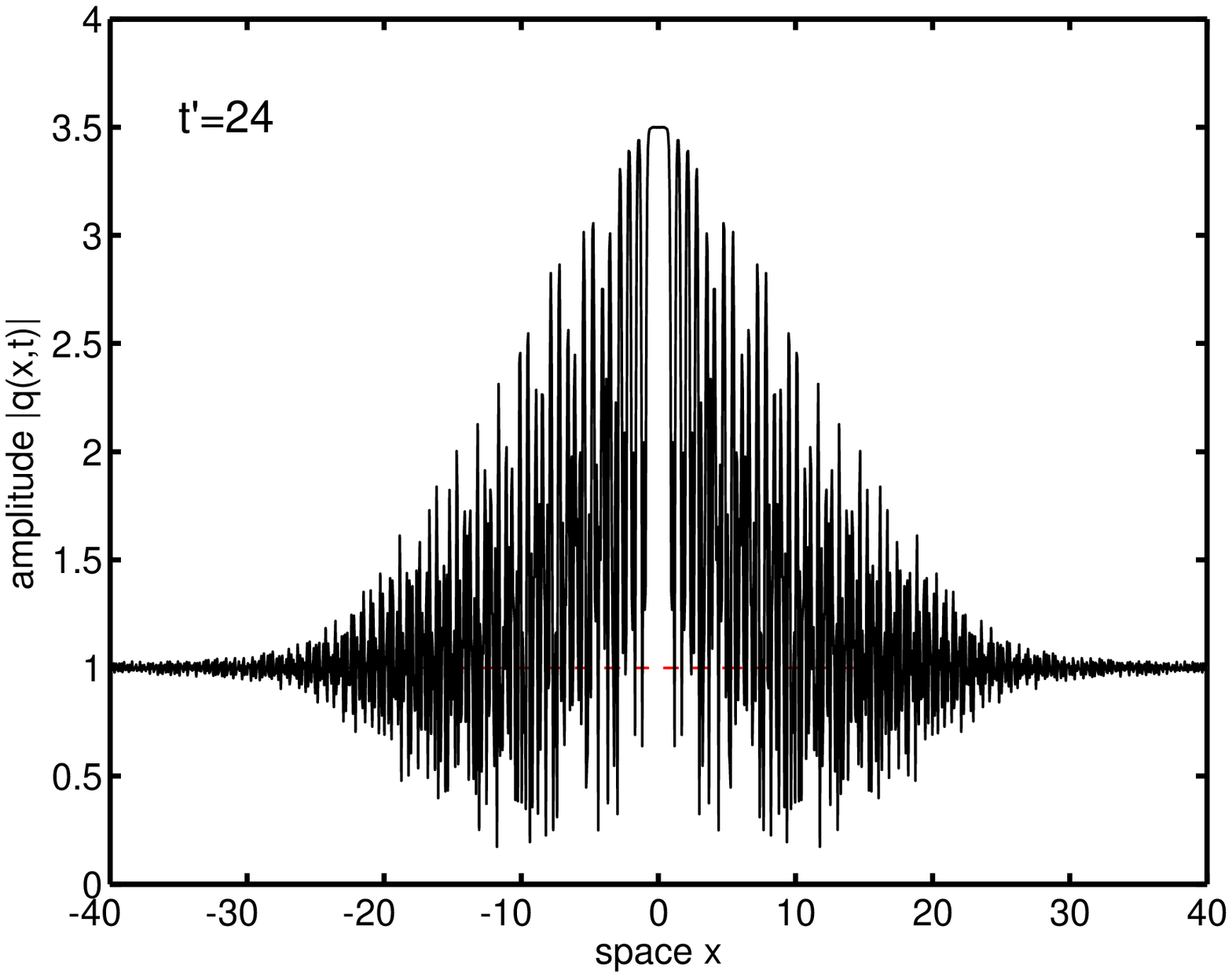}\hss}
\vskip0.25\bigskipamount
\caption{Numerical simulations of the NLS equation with a supercritical
four-jump initial condition.  The frequency jumps, located at $X_4=-X_1=24$
and $X_3=-X_2=2$, have amplitude $C_2=C_3=2$ (so that the interaction
of the two center jump produces a critical genus-0 region in the center) 
and $C_1=C_4=3$, so that the two individual sub-critical jumps open up
a genus-0 region which moves towards the center of the pulse.  The
solution is shown at: 
(a, top left)~$t'=2$, (b, top right)~$t'=6$, 
(c, bottom left)~$t'=16$, (d, bottom right)~$t'=24$.
A contour plot of this solution is shown in Fig.~\ref{f:nlscontours}a.}
\label{f:evolution4jumps}
\end{figure}

\begin{figure}[b!]
\medskip
\hbox to \textwidth{\hss\epsfxsize\figwidth\epsfbox{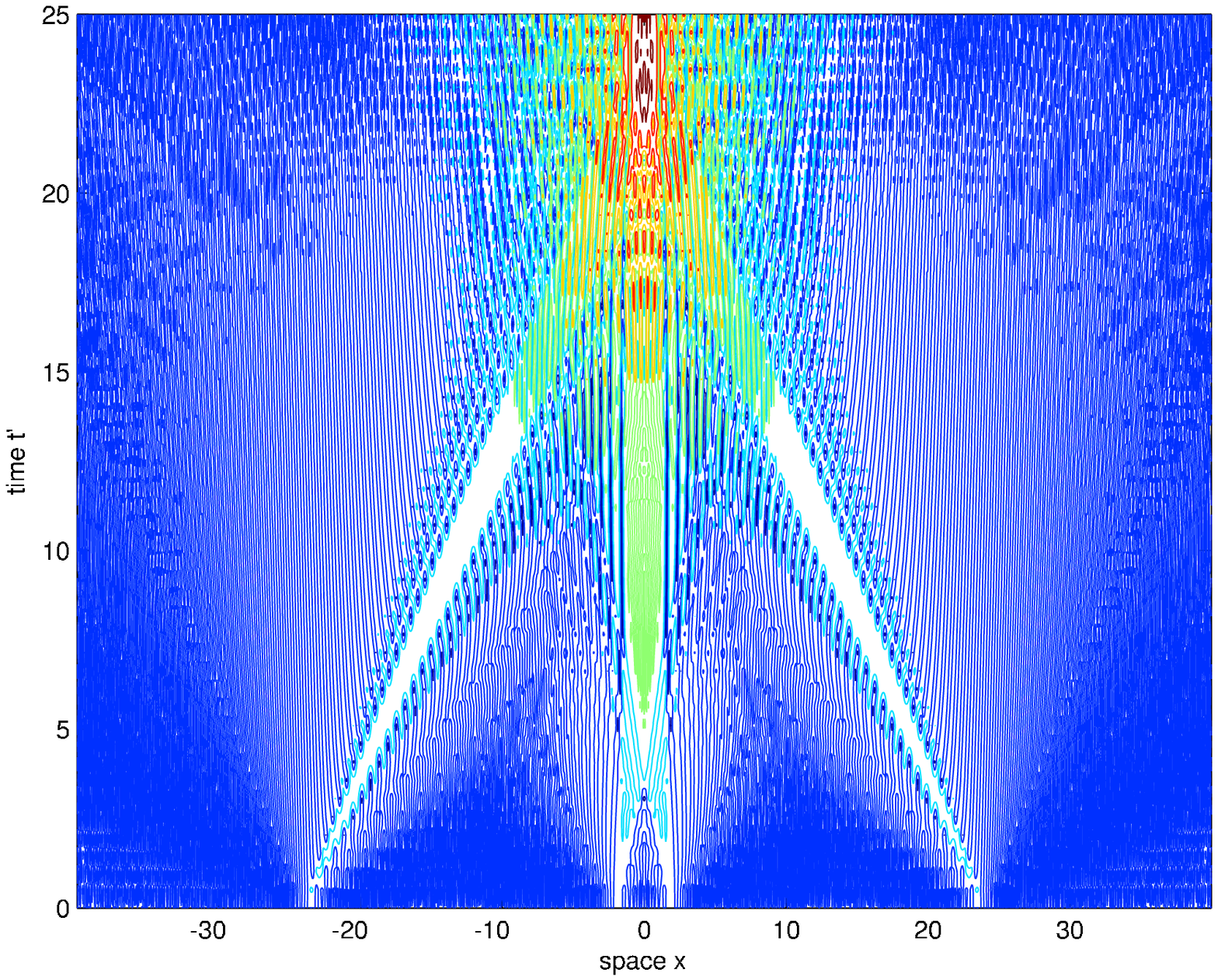}\hss
\epsfxsize\figwidth\epsfbox{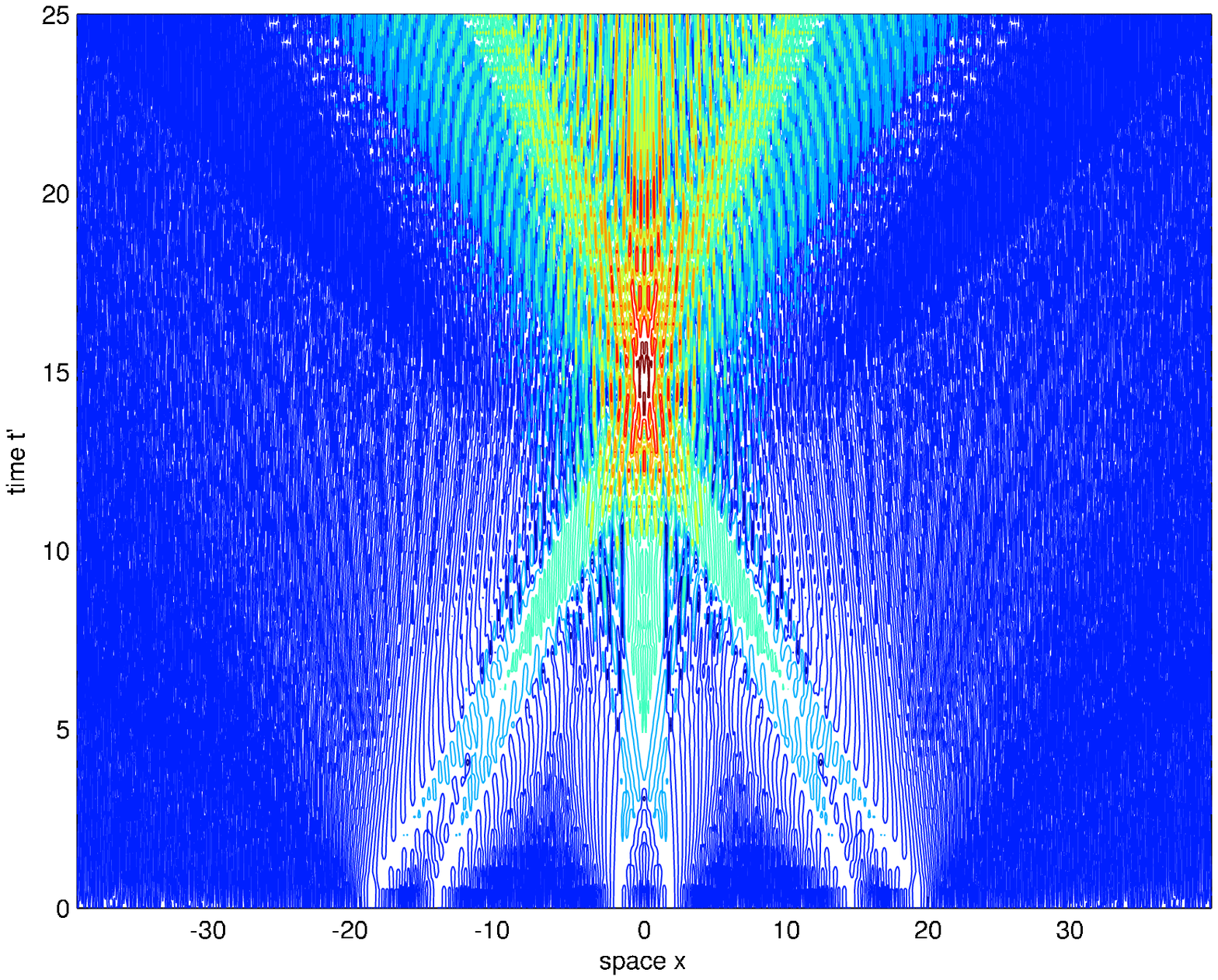}\hss}
\medskip
\hbox to \textwidth{\hss\epsfxsize\figwidth\epsfbox{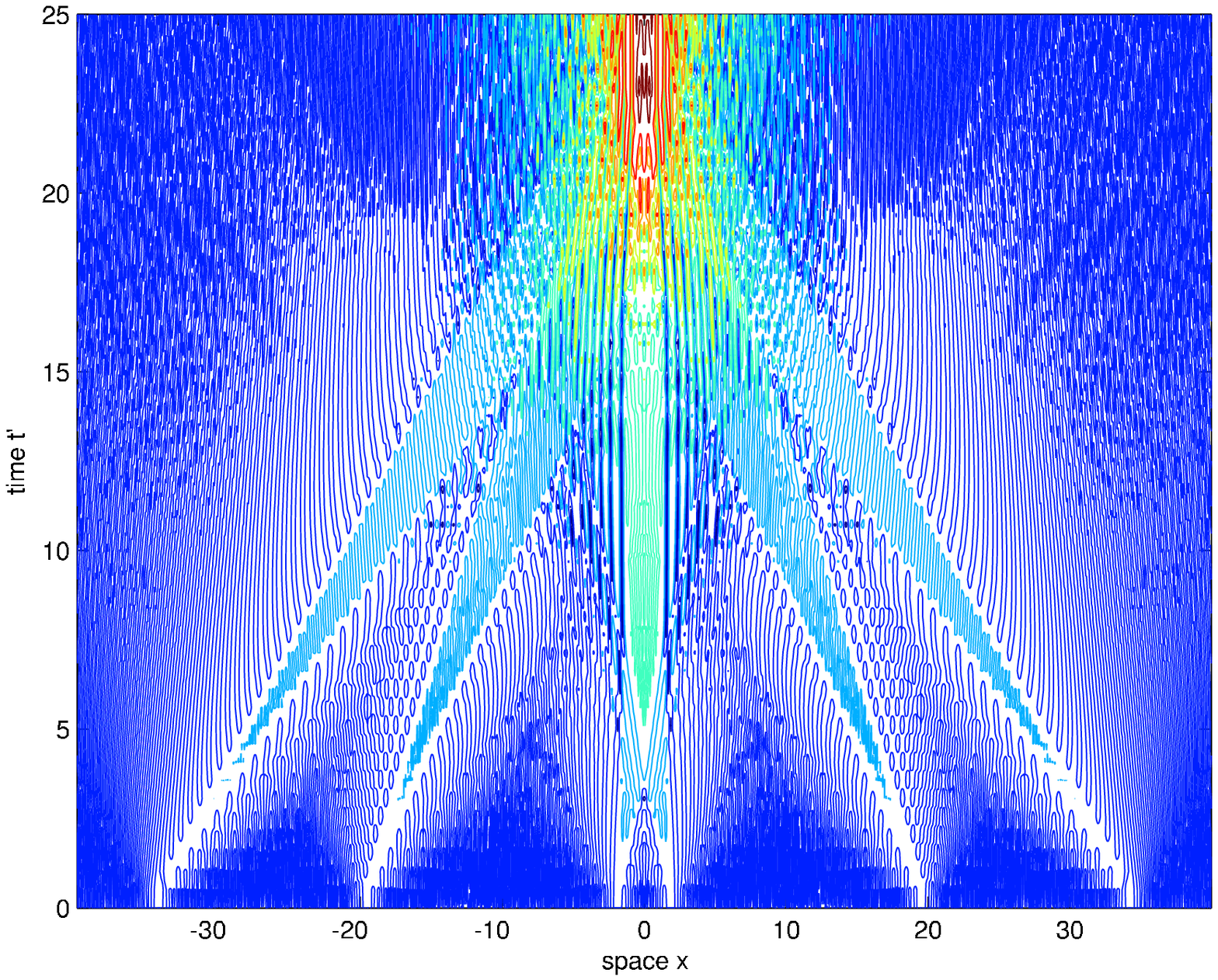}\hss
\epsfxsize\figwidth\epsfbox{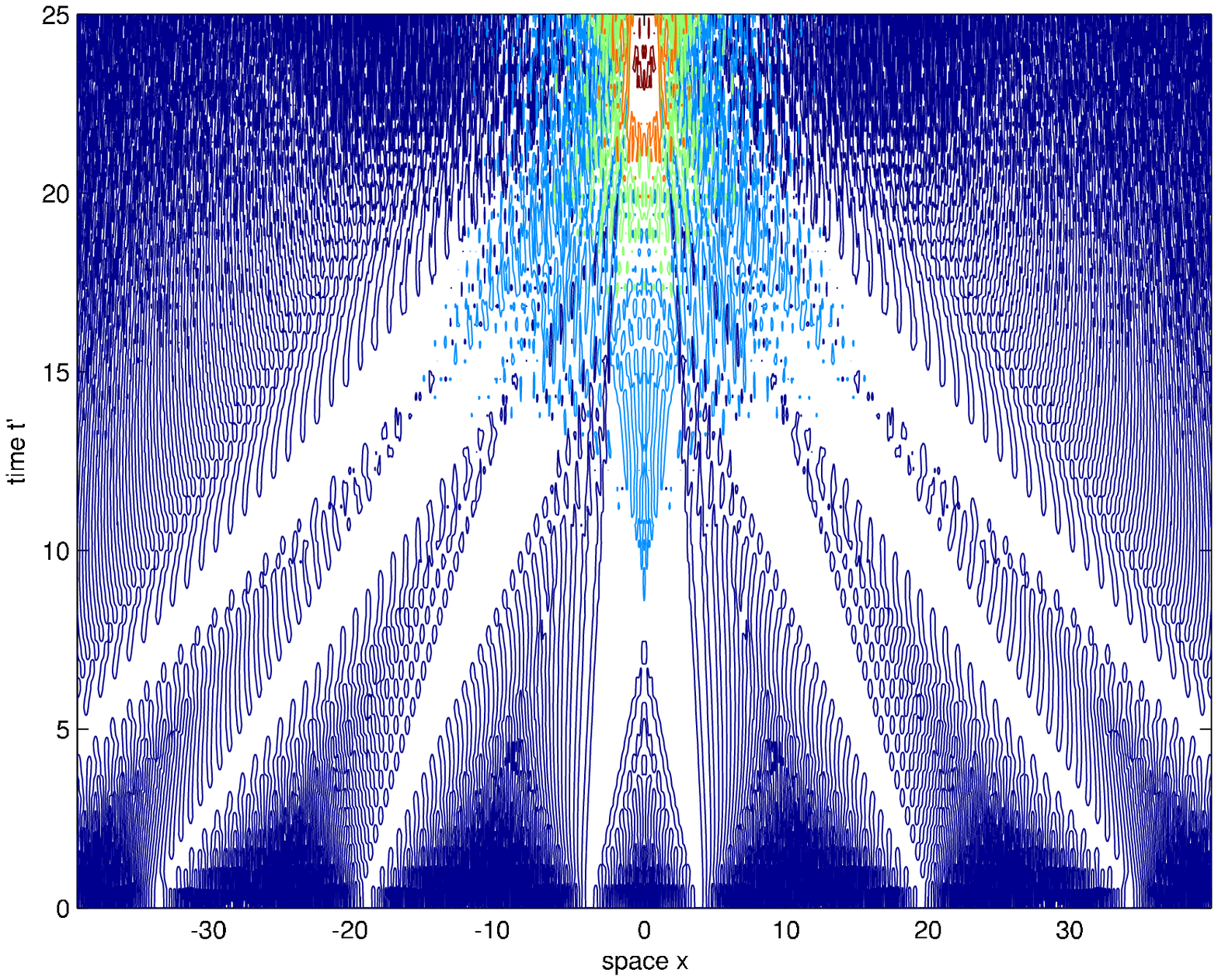}\hss}
\caption{Contour plots of the numerical solution of the NLS equation
with various choices of frequency jumps:
(a, top left)~four jumps, with parameters corresponding to the solution shown
in Figs.~\ref{f:evolution4jumps}a--d;
(b, top right)~six jumps, positioned at $-X_3=X_4=2$,
$-X_2=X_5=15$, $-X_1=X_6=20$ and all with amplitude $C=2$;
(c, bottom left)~six jumps, with same parameters as before except 
$-X_2=X_5=20$ and $-X_1=X_6=35$;
(d, bottom right)~eight jumps, each with amplitude $C=2$ and positioned at
$-X_4=X_5=4$, $-X_3=X_6=20$, $-X_2=X_7=35$, $-X_1=X_8=50$.}
\label{f:nlscontours}
\kern-\medskipamount
\end{figure}

The analytical and numerical results described in the previous sections 
can be used to design
an initial condition that results in a genus-0 region of arbitrarily
high amplitude at any desired distance down the fiber
by employing a sufficiently long continuous wave (CW) pulse at the outset,
as we now describe.
(Recall that, for fiber optics $x$ is the retarded time
and $t$ is the propagation distance along the fiber.)
Of course, the same techniques can also be applied in any physical context
where the dispersionless limit of the NLS equation is relevant.
The basic building block consists of two frequency jumps of 
critical size,
which generate a genus-0 region of constant width, as shown in 
Fig.~\ref{f:twojumps}c.  By including additional frequency jumps
to both sides of this basic pair, one can then temporarily produce 
an overall genus-0 region of higher amplitude.
This basic process is illustrated in Fig.~\ref{f:evolution4jumps},
which shows the time evolution of a four-jump initial condition.
A contour plot of this solution is shown in Fig.~\ref{f:nlscontours}a.
We now discuss some practical issues related to the generation 
of these high-intensity pulses: namely, how to control their amplitude, 
the distance along the fiber at which they are produced, 
their temporal width, and post-processing through
filtering in order to eliminate the high-frequency oscillations.


\paragraph{Amplitude, location and width.}
By appropriately choosing the number, amplitude and location of the
additional frequency jumps, it is possible to precisely specify the
amplitude of the overall genus-0 region as well as to prescribe its
width and the 
distance along the fiber at which it is generated.
This is illustrated in Fig.~\ref{f:nlscontours}, which contains 
contours of four different initial conditions.
Figure~\ref{f:nlscontours}a corresponds to the four-jump 
solution shown in Fig.~\ref{f:evolution4jumps}:
the two central frequency jumps are critical ($C_2=C_3=2$), while the single, 
sub-critical frequency jumps on either side of them ($C_1=C_4=3$) provide 
two expanding genus-0 regions which go to interact with the main one
resulting in an overall genus-0 region of amplitude 
$|q|_\max=q_0+C_\mathrm{tot}/4=7/2$, 
once more in perfect agreement with the analytical prediction.

With similar arrangements of four jumps, 
one can obtain any value of amplitude $|q|_\max<4$, 
at which point the two side jumps become critical, and do not generate 
a genus-0 region anymore.  
This limitation on the pulse amplitude can be overcome, however, 
by considering a larger number of frequency jumps.
Figures~\ref{f:nlscontours}b,c are both relative to a six-jump initial
condition with all jumps having amplitude $C_j=2$, 
so that each pair of adjacent jumps generates a genus-0 region of 
constant width.  These regions then interact to create an overall
genus-0 region of amplitude~4.
In Fig.~\ref{f:nlscontours}b, the jump locations 
$-X_3=X_4=2$, $-X_2=X_5=15$, $-X_1=X_6=20$
were chosen so that the overall genus-0 region arises near $t'=15$.
In Fig.~\ref{f:nlscontours}c, instead, the side jumps
$-X_2=X_5=20$, $-X_1=X_6=35$ were positioned further away,
so that the genus-0 region emerges near $t'=25$ and is maintained
for a longer interval.
Finally, Fig.~\ref{f:nlscontours}d shows an eight-jump initial condition,
with equal-size jumps $C_j=2$, 
generating an overall amplitude $|q|_\max=5$.
Once more, the jump locations were chosen so that this region 
emerges near $t'=25$.
A convenient way to choose the jump locations is to remember that
each frequency jump moves with velocity corresponding to the 
average frequency across the jump.

Finally, we note that the width of the central genus-0 region 
can be easily adjusted by changing the temporal separation between the two 
critical frequency jumps,
by virtue of the linear relation between these two quantities
which was discussed in section~\ref{s:numerics}.

\begin{figure}[b!]
\kern\bigskipamount
\hbox to \textwidth{\hss
\epsfysize0.275\textwidth\epsfbox{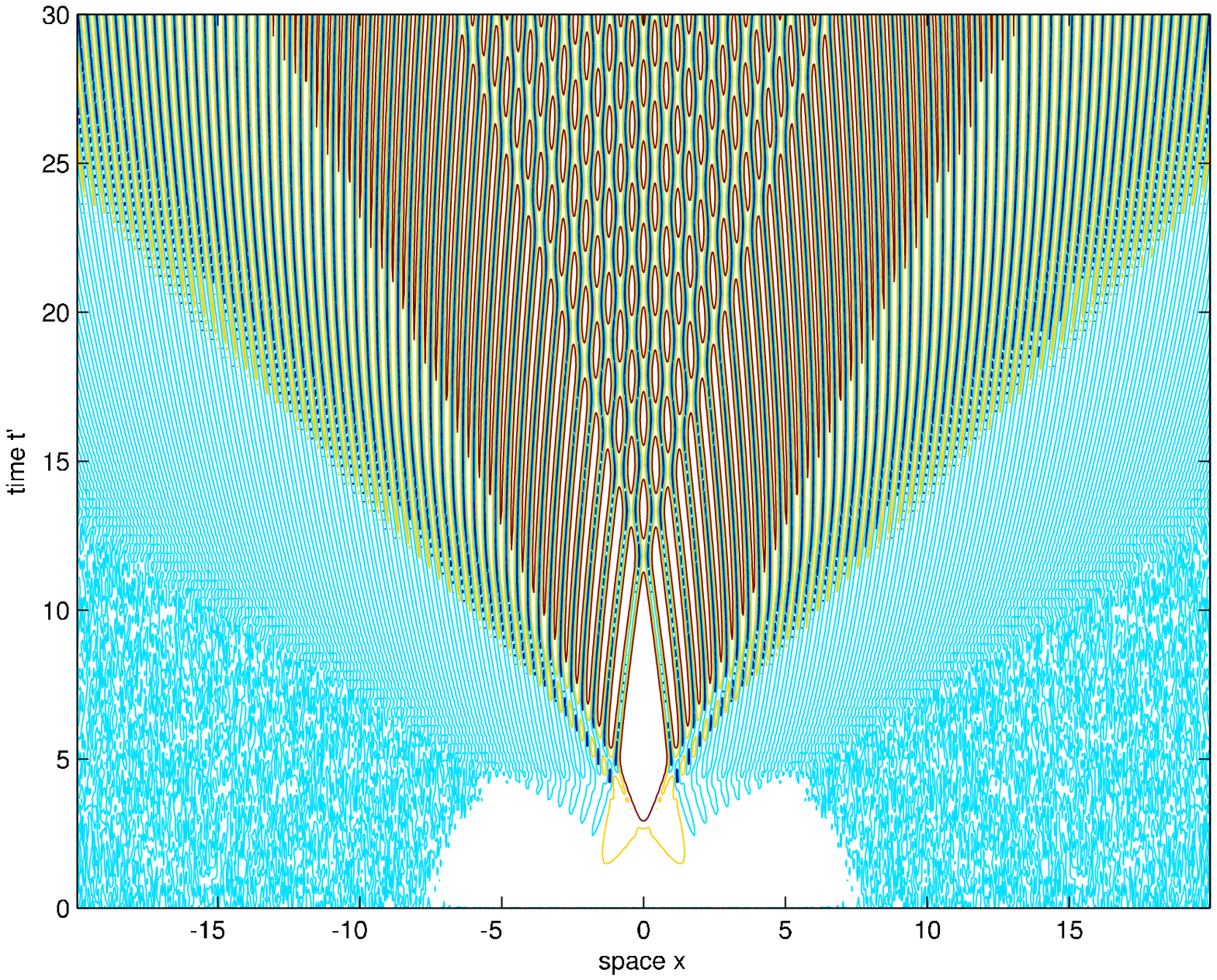}\hss
\epsfysize0.275\textwidth\epsfbox{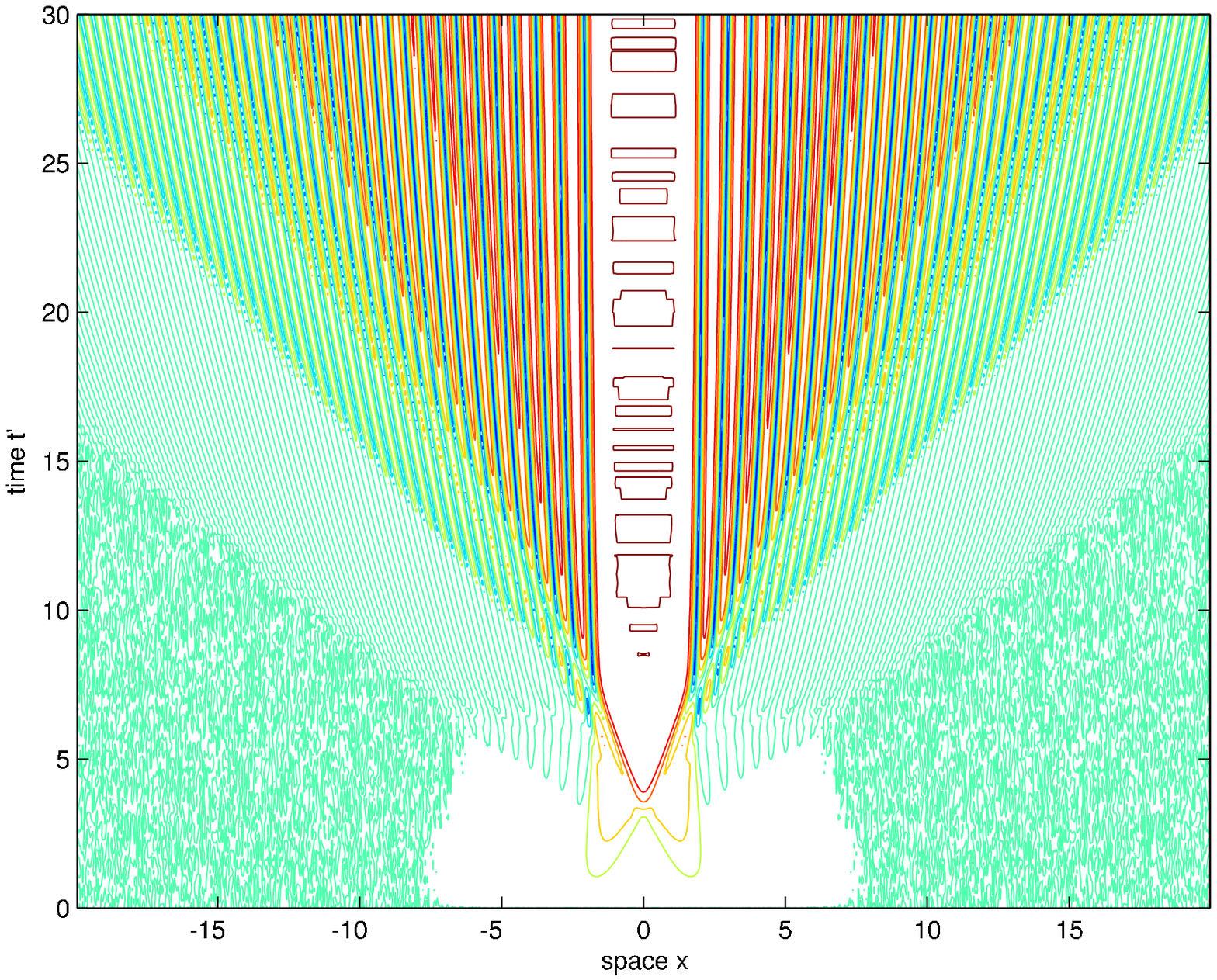}\hss
\epsfysize0.275\textwidth\epsfbox{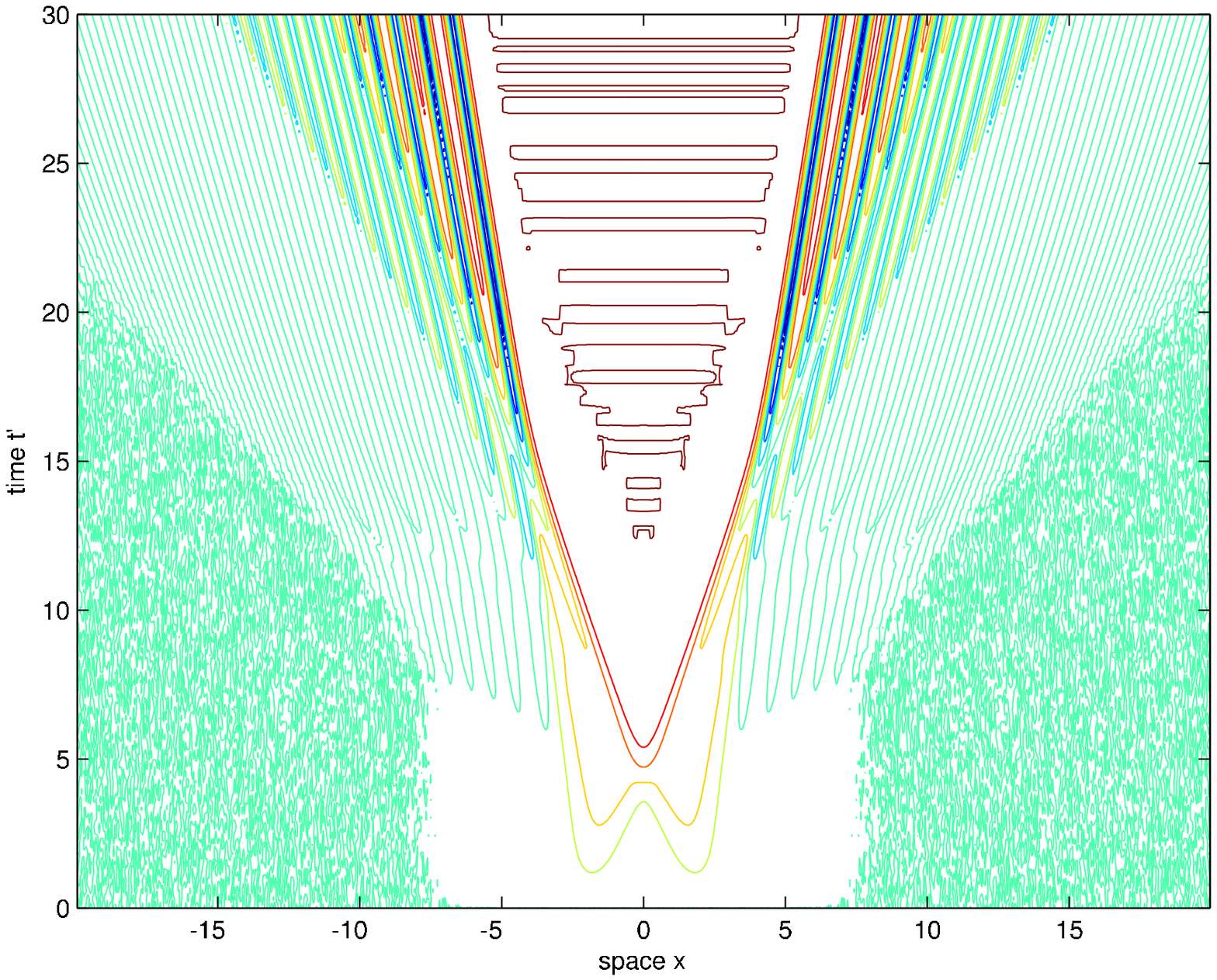}\hss}
\caption{Numerical simulations of NLS with smooth frequency transitions,
with two frequency jumps located at $X_2= -X_1= 2$:
(a, left)~$C=3$; (b, center)~$C=2$, (c, right)~$C=1$.
These figures should be compared to Figs.~\ref{f:twojumps}b--d.}
\label{f:tanh}
\kern-\medskipamount
\end{figure}

\paragraph{Smooth frequency transitions.}
The remarks at the end of section~\ref{s:arbitraryjumps}
about the general stability of the analytical results with respect to 
small changes in the initial conditions also apply with regard to 
the discontinuity in the frequency jumps.
That is, we expect that the general picture described here will be
an approximate description of the pulse behavior even if the 
frequency transitions were continuous,
as would be the case in any experimental setup.
To test this prediction, we performed numerical simulations in which, 
instead of~Eq.~\eqref{e:ugeneral}, 
the initial frequency was chosen according to
\begin{gather}
u(x,0)= \frac12\sum_{j=1}^N C_j\,\tanh(x - X_j)\,,
\label{e:utanh}
\end{gather}
which implies
$\varphi(x,0)= (1/2)\sum\nolimits_{j=1}^N C_j\,\log\big(\cosh(x-X_j)\big)$\,.
The comparison of the numerical results for discontinuous and smooth
frequency transitions shows remarkable agreement between the two cases,
with regard to both the bifurcation diagrams and the the amplitude of
the genus-0 regions.
As an example, in Fig.~\ref{f:tanh} we show numerical simulations 
of two frequency jumps with the same parameter values as in
Figs.~\ref{f:twojumps}b--d. Note also that the hexagonal pattern near the center $x=0$
in Fig.~\ref{f:tanh}a shows a genus-2 solution of the NLS equation (see also
Figs.~\ref{f:twojumps}). The evolution of the Riemann invariants corresponding to
Fig.~\ref{f:tanh}a is given by a smooth-version of Fig.~\ref{f:asymptoticinvariants}.

\paragraph{Filtering.}
The high-frequency oscillations that appear
on each side of the genus-0 region may be undesirable for practical purposes.
These oscillations can be effectively removed 
by appropriately filtering the pulse at the fiber output,
resulting in a smooth, high-amplitude short optical pulse.
For example, post-processing with a Lorentzian filter with frequency response
$F(\omega)=(\sqrt{R}/\pi)/(1+R\omega^2)$ with $R\ll1$
effectively ``wipes out'' the genus-1 and genus-2 oscillations.
A drawback of this method however is that, in the case of a narrow pulse,
the conversion efficiency (i.e., the amount of energy in the
initial CW pulse that is preserved in the final high-amplitude pulse) 
is rather low after filtering.  
A higher conversion efficiency can be obtained either by employing 
an Erbium-doped fiber amplifier 
or by increasing the temporal separation between the two central 
frequency jumps, so as to obtain a wider genus-0 region.

\paragraph{Parameter scaling.}
As a concluding remark we note that, 
even though almost all the figures in this work describe solutions
produced by initial conditions with amplitude $q_0=1$, 
all of these numerical results can easily be related to different values of 
initial amplitude thanks to the scaling symmetry of the NLS equation.  
Namely, if $q(x,t)\to a\, q(x,t)$, the results in this work 
still apply upon rescaling $x\to x'=x/a$, $t\to t'=t/a^2$, 
implying $\rho(x,t)\to \rho'(x',t')=a^2\rho(a x',a^2t')$ and 
$u(t,x)\to u'(x',t')=a\,u(a x',a^2t')$.
Importantly, 
the same scalings can also be used to relate the results to choices 
of spatial or temporal units different from those discussed in 
Appendix~\ref{s:units}.

\section*{Acknowledgements}

We would like to thank W.~L.\ Kath and M.~Fiorentino for providing 
technical specifications regarding appropriate experimental devices,
F.-R. Tian for useful discussions on the Whitham equations
and M. Hoefer for his careful reading of the text.
This work was partially supported by the National Science Foundation under
grant numbers DMS-0404931 and DMS-0506101.

\appendix
\newcounter{eqno}
\section*{Appendix}
\renewcommand\thesubsection{A.\arabic{subsection}}
\renewcommand\theequation{A.\arabic{equation}}
\renewcommand\thefigure{A.\arabic{figure}}
\setcounter{figure}0
\setcounter{equation}0

\subsection{Nondimensionalizations and scalings}
\label{s:units}

It has been known for more than thirty years that the propagation 
of a coherent light pulse in optical fibers is governed by the NLS equation,
which, in physical units, is
\unskip\cite{Agrawal,HK1995}
\begin{equation}
i\partialderiv Ez - \frac12\, k''\,\partialderiv[2]Et + \gamma\,\,|E|^2E=0\,,
\label{e:NLS0}
\end{equation}
where $E(t,z)$ is the slowly varying amplitude of the complex envelope of the
electric field of the pulse,
$t=t_\mathrm{lab}-z/c_g$ is the retarded time 
and $z$ is the propagation distance.
Here, $c_g= 1/k'$ is the group velocity,
$k(\omega)$ the propagation constant,
$k''=d^2k/d\omega^2$ the dispersion coefficient
and $\gamma$ the nonlinear coefficient.
In the following, we will use the typical value 
$\gamma= 3\,\mathrm{W}^{-1}\mathrm{km}^{-1}$.

Introducing the dimensionless variables $q=E/\sqrt{P_*}$,
$T= t/t_*$ and $Z=z/z_\nl$, 
where $P_*$ and $t_*$ are some characteristic power and temporal duration,
and where $z_*=z_\nl=1/(\gamma\,P_*\!)$ is the nonlinear length
(the characteristic distance over which nonlinear effects take place), 
Eq.~\eqref{e:NLS0} becomes
\begin{equation}
i\partialderiv qZ - \frac12\,\d\,\partialderiv[2]qT + |q|^2q= 0\,.
\label{e:nls}
\end{equation}
The dimensionless dispersion coefficient
$\d=\,k''/k''_*=z_\nl/z_\mathrm{disp}$ 
(with $k''_*=t_*^2/z_\nl$) quantifies 
the relative importance of nonlinear and dispersive effects in 
Eq.~\eqref{e:NLS0}.
In this work we are interested in particular to situations where 
$0<\d\ll1$.
This can be achieved in two different ways: constant dispersion 
or dispersion management.
Note that the dimensionless dispersion coefficient $\d$ in
Eq.~\eqref{e:nls} is obtained equivalently as $\d=k''/k_*$ or as
$\d= D/D_*$, with $D= -(2\pi c/\lambda^2)\,k''$,
with $D_*=-(2\pi c/\lambda^2)\,k''_*$, and where 
$D$ is the dispersion coefficient in ps$/$(nm$\cdot$km).
Note $2\pi c/\lambda^2=0.78\,($nm$\cdot$ps$)^{-1}$ at
at $\lambda=1.55\,\mu$m.

\paragraph{Typical units.}

We first consider the case of fibers with constant dispersion. 
A unit power 
$P_*=2$\,W (which can be obtained using commercially available high-power 
Erbium-Doped-Fiber-Amplifier lasers)
implies $z_\nl=0.16$\,km.
Then a characteristic time unit of $t_*=50$\,fs
(which is appropriate in order to describe short optical pulses) 
yields $k''_*= 1.5\,\mathrm{ps}^2/\mathrm{km}$.
Then the dimensionless dispersion coefficient $\d$ in Eq.~\eqref{e:nls} is
$\d= D/D_*$, with 
$D_*=-1.2$\,ps$/$(nm$\cdot$km).
A value of $\d=0.1$ 
(such as the one used in all the simulations presented in this work)
can then be obtained by employing a fiber
with dispersion coefficient $D=-0.12$\,ps$/$(nm$\cdot$km).
Alternatively, the same effects can be observed for lower-amplitude, 
longer pulses over longer propagation distances:
for example, with a unit power $P_*=20$\,mW, and a unit time $t_*=0.5$\,ps,
the unit distance would be $z_*=16$\,km.

Fibers with a larger nonlinear coefficient 
(such as a dispersion-shifted fiber or a photonic-crystal fiber, 
for example) would have
a correspondingly smaller nonlinear length $z_\nl$, which
would make it possible to observe the same nonlinear effects
over shorter distances. 
Also, for comparable values of $t_*$ and $P_*$, a higher nonlinearity 
would result in a larger value of $k''_*$, which means that 
fibers with correspondingly larger dispersion coefficients 
may be used.
Alternatively, a higher nonlinearity would make it possible to achieve 
the same value of~$k''_*$ 
(and hence $\d$) with shorter temporal scales~$t_*$ 
and/or with smaller peak powers~$P_*$.

It should be noted that Eq.~\eqref{e:NLS0} neglects the effect of
fiber loss,
and therefore the peak powers listed above should be taken as the average
of the pulse powers over the whole transmission span.  
This is not a serious issue if the total propagation distance is 
much shorter than the characteristic distances for fiber absorption,
as with the sample parameters chosen earlier
(for standard telecommunication fibers, a power loss coefficient of
0.2\,dB/km is typical), 
or if the effects of damping can be minimized by placing one or
more (Erbium-doped and/or Raman) fiber amplifiers inside the 
transmission span, since it is well-known that the average dynamics
in systems with loss and periodic amplification are still governed
by the NLS Eq.~\eqref{e:NLS} with constant coefficients.\cite{PRL1991v66p161}
We should also note that quasi-lossless propagation of optical pulses
over several hundred kilometers has recently been achieved by making use 
of two bidirectional Raman pumps plus fiber Bragg gratings.\cite{Turitsyn}
Finally, 
note that Eq.~\eqref{e:NLS0} also neglects higher-order dispersion.
As shown in Ref.~\fullcite{SJAM59p2162}, third-order dispersion can affect
pulse behavior.
In order for the results described in this work to apply, therefore,
it is necessary that the fiber dispersion be approximately constant
throughout the bandwidth of the pulse, which means that third-order
dispersion should be small.

\paragraph{Dispersion management.}

If realizing small values of $\d$ in a stable way 
should require an unpractical level of control over fiber dispersion,
the problem might be circumvented by making use of dispersion management.
With dispersion management, a very small value of $\epsilon$ can be achieved 
as an average between opposite the dispersion coefficients
of fibers with alternating signs of dispersion,
even if the local values of dispersion are not individually small.

The propagation of optical pulses in systems with dispersion management 
is still described by the NLS Eq.~\eqref{e:nls}, where however the 
dimensionless dispersion coefficient $\d$ in Eq.~\eqref{e:nls} is replaced 
by a periodic function~$d(z)$.
The specific choice of $d(z)$ is called a dispersion map.
The relative effect of the periodic dispersion variations 
can be quantified by the reduced map strength parameter~$s$
which is defined as\cite{OL23p1668} $4s=\<|\,d(\,\cdot\,)-\d\,|\>\,z_a$,
where now $\d=\<d(\,\cdot\,)\>$
and the average is taken over the period of the dispersion map,~$z_a$.
As long as the strength~$s$ of the dispersion map is small, 
it is possible to average over the rapid variations
of the pulse profile originating in the NLS equation with 
periodic coefficients.
Indeed, it is well-known that, just like with loss/amplification,
the NLS Eq.~\protect\eqref{e:nls}
describes the evolution of the leading-order portion of the
pulse envelope\cite{OL16p1385}
in a system with a moderate amount of 
dispersion management.
That is, the leading order part of the pulse satisfies again
the NLS equation~\eqref{e:nls}, except that the parameter $\d$
now represents the average dispersion.
Therefore, the use of dispersion management might be desirable because it
allows one to more easily obtain small values of effective dispersion
than for a system with constant dispersion.
Because on average the system is still governed by the NLS equation,
the results presented in this work will be preserved
as long as the dispersion map strength~$s$ is small.
If the map strength~$s$ becomes large, however, the leading order dynamics 
are described not by the NLS, but rather by a nonlocal evolution
equation of NLS-type\cite{OL23p1668,OL21p327}.
Determining the behavior of the solutions with small $\d$
for this type of system
is a highly non-trivial task, which is beyond the scope of this work.

\setcounter{eqno}{\value{equation}}
\subsection{Genus-1 solutions of NLS and Whitham's averaging method}
\label{a:elliptic}
\setcounter{equation}{\value{eqno}}

Here we briefly review the construction of genus-1 solutions of the 
NLS equation and the derivation of the genus-1 NLS-Whitham equations,
with the aim of clarifying the connection 
between the Riemann invariants and the spectrum of the Lax operator.
For a detailed construction of higher-genus solutions involving theta
functions, see Refs.~\fullcite{BBEIM1994,GesztesyHolden}.

In order to construct traveling wave solutions of the NLS equation~\eqref{e:NLS},
it is convenient to introduce the fast space and time scales 
$x'= x/\epsilon$ and $t'=t/\epsilon$, and write Eq.~\eqref{e:NLS} as 
$iq_{t'}-\frac12 q_{x'x'}+|q|^2q=0$.
We then look for solutions of the form
\begin{equation}
q(x',t')= e^{i[c x'+ (\frac12 c^2+\lambda )t']}\big(a + f(x'+c t')\big)\,,
\label{e:stationary}
\end{equation}
with $a,c,\lambda\in\Real$ and $f(\,\cdot\,)$ a real function of its argument.
Substituting Eq.~\eqref{e:stationary} into NLS and 
integrating once, one obtains
\begin{equation}
(f')^2= 
f^4 - \sigma_1 f^3 + \sigma_2 f^2 - \sigma_3 f + \sigma_4 
=:\mu(f)\,,
\label{e:uprime}
\end{equation}
where the prime denotes differentiation with respect to~$x'$,
$\sigma_1=-a$, $\sigma_2= 4(3a^2-\lambda+c^2/2)$, 
$\sigma_3= -a\sigma_2$, $\sigma_4=C$ 
and $C$ is the integration constant.
It is convenient to represent the right-hand side of Eq.~\eqref{e:uprime} as 
\begin{equation}
\mu(f)= (f-r_1)(f-r_2)(f-r_3)(f-r_4)\,,
\end{equation}
where the roots $r_1,\dots,r_4$ are related to the coefficients 
in Eq.~\eqref{e:uprime} via the relations
\begin{subequations}
\label{e:g1sigmadef}
\begin{gather}
\sigma_1= r_1+r_2+r_3+r_4\,,\qquad
\sigma_2= r_1r_2 + r_1r_3 + r_1r_4 + r_2r_3 +r_2r_4 + r_3r_4 \,,
\\
\sigma_3= r_1r_2r_3 + r_1r_2r_4 + r_1r_3r_4 + r_2r_3r_4\,,\qquad
\sigma_4= r_1r_2r_3r_4\,.
\end{gather}
\end{subequations}
Integrating formally Eq.~\eqref{e:uprime} one has 
$\int df/\sqrt{\mu(f)}= \pm(x'-x'_0)$.
The integral then defines the solution of NLS in terms of elliptic functions.
Note however that real solutions are confined to those values of $f$ 
such that $\mu(f)>0$.
It is then clear that the values of the four roots $r_1,\dots,r_4$ 
(which will be the Riemann invariants when performing the Whitham averaging
described in the next subsection)
determine the edges of the forbidden regions, or gaps,
which are the values of $f$ for which $\mu(f)<0$.
Several situations can occur depending on the values of 
the parameters in Eq.~\eqref{e:uprime};  
it is relatively easy however to see that bounded solutions only exist 
when all four roots of $\mu(f)$ are real.
In this case, labeling the roots as $r_1<r_2<r_3<r_4$
without loss of generality,
$f(x')$ is given by the elliptic sine:
$f(x')= k\,\mathop{\rm sn}(x',k)$, 
where the elliptic modulus $k\in[0,1]$ 
is given by $k^2= (r_4-r_3)(r_2-r_1)/[(r_4-r_2)(r_3-r_1)]$.
The elliptic parameter\cite{Abramowitz} is $m=k^2$.
The value of $f$ oscillates between $r_2$ and $r_3$,
and the spatial period of the solution is given by $4K(k)$,
where $K(k)$ is the elliptic integral of the first kind.\cite{Abramowitz}
In the limiting case $r_4=r_3=-r_1=-r_2$,
the period of the solution tends to infinity,
and one obtains the dark soliton solutions of NLS: for example, 
when $k=0$ it is $a=0$ and $f(x')= \lambda\,\tanh (\lambda\,x')$.
Note that that the solution of NLS is completely specified up to 
an overall phase in terms of the four roots $r_1,\dots,r_4$.
For a more detailed discussion of single-phase solutions of the NLS equation
in the focusing as well as defocusing regimes, as well as their stability,
see Refs.~\fullcite{PRE68p45601,Kamchatnov}.

It is important to note that, for any finite value of 
the small parameter $\epsilon$, the presence of $\epsilon$
in Eq.~\eqref{e:NLS} can be taken into account by simply replacing~$(x,t)$
by the fast variables $(x'=x/\epsilon, t'=t/\epsilon)$.
This phenomenon is the origin of the high-frequency oscillations in 
the solution of NLS with small dispersion coefficient,
discussed throughout this work.
The nontrivial issue, however, is to understand the qualitative behavior 
of the solution in the weak dispersion limit, as formulated in 
section~\ref{s:nlssmalldisp}.

Let us then briefly outline how to obtain the genus-1 NLS-Whitham equations.
It is well-known that 
the NLS equation admits an infinite number of conservation laws: 
\begin{equation}
\partialderiv {F_n}{t' }=\partialderiv{G_n}{x'}\,,
\label{e:NLSconslaw}
\end{equation}
$n=1,\dots,\infty$.  
For example, the first four conserved densities are
\begin{subequations}
\label{e:densities}
\begin{gather}
F_1= |q|^2\,,\\
F_2=  i (q^*q_{x'}-qq^*_{x'})\,,\\
F_3= |q_{x'}|^2+ |q|^4\,,\\
F_4= i(q^*q_{x'x'x'}-qq^*_{x'x'x'})-3i|q|^2(q^*q_{x'}-qq^*_{x'})\,.
\end{gather}
\end{subequations}
Then we consider small modulations of the genus-1 solution given by
the elliptic function, in the sense that the deformation due to the 
modulation is of order $\epsilon$ over one period of the wave.
Following the standard perturbation method of multiple scales,
we introduce the slow time and space scales $x=\epsilon x'$ and $t=\epsilon t'$. 
We now have two sets of scales, consisting of a set of fast scales 
for the oscillations and a set of slow scales for the modulations. 
With those variables, we have
\[
\frac{\partial}{\partial x'}\to \frac{\partial}{\partial x'}+\epsilon\frac{\partial}{\partial x}\,,
\quad 
\frac{\partial}{\partial t'}\to \frac{\partial}{\partial t'}+\epsilon\frac{\partial}{\partial t}\,.
\]
Then, expanding the conserved quantities as 
$F_n=F^{0}_n+\epsilon F^1_n+\cdots$ and $G_n=G_n^0+\epsilon G^1_n+\cdots$
and taking the average of the conservation laws~\eqref{e:NLSconslaw} 
over the fast scale~$x'$, 
at order $\epsilon$ we obtain the NLS-Whitham equations
in conservation form:
\begin{equation}
\partialderiv {\<F_n^0\>}t= \partialderiv {\<G_n^0\>}x\,.
\label{e:nlswhithamconsform}
\end{equation}
Here $\<F^0_n\>$ and $\<G_n^0\>$ are the average of the leading order 
conserved densities and fluxes associated with the solution 
of the NLS equation, 
and the average of a periodic function 
$f(x',t')=\hat f(\theta)$ 
is given by
\begin{equation}\label{1average}
\def\txint{\mathop{\textstyle\int}\limits}
\<f\>=
  \frac1L \txint_{-L/2}^{L/2} f(x',t')\,dx'
  =\txint_0^1 \hat f(\theta)\,d\theta\,,
\end{equation}
where $L$ is the spatial period and $\theta=(x'+ct')/L$. 
For the genus-1 solutions described above, $L=4K(k)$.
Note that 
the averaged quantities $\langle F^0_n\rangle$ and $\langle G^0_n\rangle$ 
are functions of the spectral parameters $r_1,\dots,r_4$,
Thus the modulations are expressed in terms of the slow motion 
of the spectral parameters~$r_1,\ldots, r_4$.
This also means that we need the first four equations 
from~\eqref{e:nlswhithamconsform}.
By computing the averages, one finds that the variables $r_1,\dots,r_4$ 
are precisely the Riemann invariants for the system.
Then, by writing Eqs.~\eqref{e:nlswhithamconsform} in diagonal form, 
one finally obtains the genus-1 NLS-Whitham equations \eqref{e:nlsWhitham} 
with $g=1$. 
(That Eqs.~\eqref{e:nlswhithamconsform} are diagonalizabile is not obvious, 
but always true for all values of genus; 
see section~\ref{a:NLS-Whitham} for further details.)

\setcounter{eqno}{\value{equation}}
\subsection{The NLS-Whitham equations}
\label{a:NLS-Whitham}
\setcounter{equation}{\value{eqno}}

Here we briefly review some well-known results for the NLS-Whitham equations.
For more details on the finite genus solutions of the NLS equation, 
the averaging process and the NLS-Whitham equations, 
we refer the reader to Refs.~\fullcite{BBEIM1994,SJAM59p2162}
and references therein.

The genus-$g$ NLS-Whitham equations can be obtained by extending
the approach described in section~\ref{a:elliptic}
to hyper-elliptic solutions of NLS and $2g+2$
conservation laws~\cite{OTCMCC1986}.
It is more convenient, however, to take advantage of the integrable
structure of the NLS equation, as described in Ref.~\fullcite{SJAM59p2162}.
Recall that genus-g solutions of NLS are associated to 
the Riemann surface $R:w^2=\mu_g(z)$, where 
$\mu_g(z)= \prod_{k=1}^{2g+2}(z-r_k)$ 
(cf.\ Eq.~\eqref{e:mudef}).
It was shown in Refs.~\fullcite{CPAM1980p739,OTCMCC1986,FAA22p37,TMP71p584}
that the NLS-Whitham equations~\eqref{e:nlswhithamconsform} can be 
elegantly written as 
\begin{equation}
\partialderiv{\omega_1}{t}= \partialderiv{\omega_2}x\,,
\label{a:nlsWhithamAbelian}
\end{equation}
where $\omega_1$ and $\omega_2$ are the meromorphic (Abelian)
differentials of the second kind associated with the Riemann surface~$R$:
\begin{subequations}
\label{a:differentials}
\begin{gather}
\omega_1= \frac12 \bigg[\, 1 + 
  \frac{ z^{g+1} - \frac12\sigma_1z^g + \alpha_1z^{g-1} + \cdots
    + \alpha_{g-1}z + \alpha_g }
    {\sqrt{\mu_g(z)}}
  \,\bigg]\,dz
  = \big(1+O(1/z)\big)\,dz\,,
\\
\omega_2= \frac12 \bigg[\, z + 
  \frac{ z^{g+2} - \frac12\sigma_1 z^{g+1}
    + \frac12\big(\sigma_2-\frac14\sigma_1^2)z^g + \gamma_1z^{g-1} + \cdots
    + \gamma_{g-1}z + \gamma_g }   
    {\sqrt{\mu_g(z)}}
  \,\bigg]\,dz
  = \big(z+O(1/z)\big)\,dz\,,
\end{gather}
\end{subequations}
where $\sigma_1,\dots,\sigma_3$ are the generalization of Eqs.~\eqref{e:g1sigmadef},
\begin{gather}
\sigma_1= \sum_{1\le j\le 2g+2}\r_j\,,\qquad
\sigma_2= \sum_{1\le j<k\le 2g+2}\r_j\r_k\,,\qquad
\sigma_3= \sum_{1\le i<j<k\le 2g+2}\r_i\r_j\r_k\,
\label{a:sigma12}
\end{gather}
($\sigma_3$ will appear later on)
and where the coefficients $\alpha_j$ and $\gamma_j$
are given by the solution of two $g\times g$ linear systems of equations
\begin{subequations}
\label{a:alphagammacoeffs}
\begin{gather}
I_j^{g+1} - \frac12\sigma_1I_j^g + \alpha_1I_j^{g-1} + \cdots
    + \alpha_{g-1}I_j^1 + \alpha_g I_j^0 = 0\,,
\\
I_j^{g+2} - \frac12\sigma_1 I_j^{g+1}
    + \frac12\big(\sigma_2-\frac14\sigma_1^2)I_j^g + \gamma_1I_j^{g-1} + \cdots
    + \gamma_{g-1}I_j^1 + \gamma_g I_j^0 = 0\,,
\end{gather}
\end{subequations}
$j=1,\dots,g$, with the real-valued functions $I_j^k$ given by
\begin{equation}
I_j^k= \frac12 \mathop{\textstyle\int}\limits_{\r_{2j+1}}^{\r_{2j+2}}
  \frac{z^k}{\sqrt{-\mu_g(z)}}\,\, dz\,.
\label{a:Acycles}
\end{equation}
Equations~\eqref{a:alphagammacoeffs} express the requirement that
the A-periods of $\omega_1$, $\omega_2$
be zero: that is, 
$\oint_{a_j}\omega_1=\oint_{a_j}\omega_2=0$, where the $a_j$
are the A-cycles on the Riemann surface~$R$.

Evaluating the residue of $\omega_{1,2}$ at $z=\r_k$ 
in Eq.~\eqref{a:nlsWhithamAbelian}, one obtains 
the genus-$g$ NLS-Whitham equations in Riemann invariant form:
\begin{equation}
\partialderiv[]{\r_k}{t'} = 
  s_k(\r_1,\dots,\r_{2g+2})\, \partialderiv[]{\r_k}x\,,
\label{a:nlsWhitham}
\end{equation}
$k=1,\dots,2g+2$.
The characteristic speeds are given by 
$s_k(\r_1,\dots,\r_{2g+2})= s(\r_k)$,
with
\begin{gather}
s(r)= \frac{ \r^{g+2} - \frac12\sigma_1 \r^{g+1}
    + \frac12\big(\sigma_2-\frac14\sigma_1^2)\r^g + \gamma_1\r^{g-1} + \cdots
    + \gamma_{g-1}\r + \gamma_g }
  { \r^{g+1} - \frac12\sigma_1\r^g + \alpha_1\r^{g-1} + \cdots
    + \alpha_{g-1}\r + \alpha_g }\,.
\label{a:invariantspeed}
\end{gather}
Since the NLS-Whitham equations~\eqref{a:nlsWhitham} describe the
evolution of the Riemann invariants~$\r_1,\dots,\r_{2g+2}$ with
respect to the slow time and space variables, and since the
Riemann invariants are also the branch points of the Riemann surface
corresponding to finite-genus solutions of the NLS equation,
it follows that the NLS-Whitham equations describe the slow 
modulation of finite-genus solutions of the NLS equation.

Note that the abelian differentials in Eqs.~\eqref{a:differentials}
can also be expanded in terms of the conservation laws of the NLS equation:
\begin{gather}
\omega_1\sim \bigg( 1 + \sum_{n=1}^\infty \frac{n\<F^0_n\>}{z^{n+1}} \bigg)\,dz\,,
\qquad
\omega_2\sim \bigg( z + \sum_{n=1}^\infty \frac{n\<G^0_n\>}{z^{n+1}} \bigg)\,dz\,,
\end{gather}
where $\<F^0_n\>$ and $\<G^0_n\>$ are the conserved densities 
and fluxes of the NLS equation, averaged over the fast phases.
The average of a quasi-periodic function of $g$ phases 
$f(x',t')= \hat f(\theta_1,\ldots,\theta_g)$, 
with $\theta_k=(x'+c_k t')/L_k$ and $L_k$ being the period for 
the phase~$\theta_k$, is given by
\[
\def\txint{\mathop{\textstyle\int}\limits}
\<f\> =
  \lim_{L\to\infty}\frac1L\txint_{-L/2}^{L/2}f(x',t')\,dx'=
  \txint_0^1\cdots\txint_0^1\hat f(\theta_1,\ldots,\theta_g)
    \,d\theta_1\cdots d\theta_g\,,
\]
where we have used the ergodic assumption, 
i.e., that the $g$ phases produce a dense orbit on the
Jacobian variety of the Riemann surface\cite{CPAM1980p739}.
In this framework Eqs.~\eqref{a:nlsWhithamAbelian} then 
yield Eqs.~\eqref{e:nlswhithamconsform}
for $n=1,\dots,2g+2$.
For example, the first few densities and fluxes are
\begin{subequations}
\let\f= \txtfrac
\label{a:densities}
\begin{align*}
&\<F^0_1\>= \<\rho\>= -\f14\sigma_2 + \f1{16}\sigma_1^2 + \f12\alpha_1\,,
\\
&\<F^0_2\>= \<\rho u\>=
   -\f14\sigma_3-\f18\sigma_1\sigma_2+\f1{32}\sigma_1^3
   +\f18\sigma_1\alpha_1+\f14\alpha_2\,,
\\
&\<G^0_1\>= \<F_2^0\>=
   -\f14\sigma_3-\f18\sigma_1\sigma_2+\f1{32}\sigma_1^3+\f12 \gamma_1\,,
\\
&\<G^0_2\>= \<\rho u^2 + \f12 \rho^2\>=
   -\f18 \sigma_4 + \f18\sigma_1\sigma_3+\f1{32}\sigma_2^2
   -\f5{64}\sigma_1^2\sigma_2 + \f9{512}\sigma_1^4
   + \f18\sigma_1\gamma_1 + \f14\gamma_2\,,
\end{align*}
\end{subequations}
and Eq.~\eqref{e:dam} is obtained from Eqs.~\eqref{e:nlswhithamconsform} when $g=0$.

Upon evaluating the integrals in Eq.~\eqref{a:Acycles}
and taking the appropriate limits as $\epsilon\to0^+$,
Eq.~\eqref{a:invariantspeed} can be used to calculate the value of
all the speeds listed in sections~\ref{s:singlejump} 
and~\ref{s:multiphase}, as we show in the next subsection.
It should be noted that the general solution of the NLS-Whitham 
equations~\eqref{a:nlsWhitham} can be written in terms of the
Hodograph transformation
\begin{equation}
x - s_k(r_1,\dots,r_{2g+2})\, t = w_k(r_1,\dots,r_{2g+2})\,,
\label{a:hodograph}
\end{equation}
$k=1,\dots,2g+2$, where the functions $w_k$ are expressed in terms
of the initial data via hyperelliptic integrals.
When inverting Eqs.~\eqref{a:hodograph} to find the solution $q(x,t)$,
the locations where the Jacobian of the transformation vanishes determine
the boundary between regions of different genus\cite{CPAM52p655,PLA277p115}.
This information was used  in Ref.~\fullcite{CPAM52p655} 
to determine the location between regions of genus-0 and genus-1 
in some specific situations,
and the method was used in Ref.~\fullcite{CPAM55p1569}
to determine the generic evolution of an arbitrary initial datum for the
Korteweg-de~Vries equation.
It should be noted, however, that the method cannot be easily applied
for the situations considered in the present work,
since the transformation $x\to q(x,0)$ is not invertible
in the case of piecewise constant initial data, which are
the ones of interest here.

\setcounter{eqno}{\value{equation}}
\subsection{Calculation of some characteristic speeds}
\label{a:characteristicspeeds}
\setcounter{equation}{\value{eqno}}

We now briefly describe some calculations regarding the characteristic speeds
of the regularized Riemann invariants for the NLS-Whitham equations.
The results of these calculations and similar others were summarized 
in sections \ref{s:singlejump} and~\ref{s:multiphase}. 
Since the calculations are rather tedious however,
we will limit ourselves to present two examples.
For more details on similar types of calculations we also refer
the reader to Refs.~\fullcite{SJAM52p909,SJAM59p2162}.
When performing these calculations, it is useful to note that 
the characteristic speed in Eq.~\eqref{a:invariantspeed}
is the ratio of two meromorphic differentials
which becomes the ratio of two polynomials
(the denominator of the differentials).
In the case of step initial data, some common factors then arise, 
and the resulting cancellations produce a finite result.

\paragraph{Genus-1 calculations: Equations~(\ref{e:s1jump_i}) 
  and~(\ref{e:g0outerspeed}).}
Our first aim is to calculate $s_3^-$ in case~(i) of section~\ref{s:singlejump}, 
as shown in Fig.~\ref{f:riemann1jump}a.
This is done by evaluating $s_3(\r_1,\dots,\r_4)= s(\r_3)$,
with $s(r)$ given by Eq.~\eqref{a:invariantspeed} with
$r_4=-r_1= u_0+2q_0$, $r_2= -u_0+2q_0$ and $r_3=r_4-\epsilon$, 
and then taking the limit $\epsilon\to0^+$.
In general, 
Eqs.~\eqref{a:alphagammacoeffs} yield the coefficients 
$\alpha_1$ and $\gamma_1$ for the $g=1$ case as
\begin{equation}
\alpha_1=
  - \frac{ I_1^2 - \frac12 \sigma_1 I_1^1 }{I_1^0}\,,
\qquad
\gamma_1= - \frac{ I_1^3 - \frac12 \sigma_1 I_1^2 
  + \frac12 \big( \sigma_2 - \frac14 \sigma_1^2 \big) I_1^1 }
      {I_1^0}\,.
\label{a:g1coeffs}
\end{equation}
In our case, Eqs.~\eqref{a:sigma12} also give
$\sigma_1=r_2+r_4=4q_0+\epsilon$ and 
$\sigma_2= r_2r_4-r_3^2=-2u_0(u_0+2q_0)+\epsilon(u_0-2q_0)$. 
In general,
upon making the substitution $z= r_3 + \Delta\,\sin^2\theta$
with $\Delta= r_4-r_3$.
the integrals $I_1^0,\dots,I_1^3$ take the convenient form:
\begin{equation}
I_1^k= \mathop{\textstyle\int}\limits_0^{\pi/2}
  \frac{(r_3+\Delta\,\sin^2\theta)^k}
    {\sqrt{\big( r_3 - r_1 + \Delta\,\sin^2\theta \big) 
       \big( r_3 - r_2 - \Delta\,\sin^2\theta \big) }}\,\,
    d\theta\,.
\end{equation}
In our case, expanding the power in the numerator and substituting
the explicit value of the invariants, the integrals can also be
written as:
\begin{gather}
I_1^k= \sum_{m=0}^k  \binom km \,(u_0+2q_0)^{k-m}\, \epsilon^m\,\,
  \mathop{\textstyle\int}\limits_0^{\pi/2}
  \frac{\sin^{2m}\theta}
    {\sqrt{\big( 2(u_0+2q_0) - \epsilon\, \cos^2\theta \big) 
       \big( 2u_0 - \epsilon\, \cos^2\theta \big) }}\,\,
    d\theta\,.
\label{a:g1integrals}
\end{gather}
Then, evaluating the integrals in Eq.~\eqref{a:g1integrals}, 
substituting the results and Eq.~\eqref{a:g1coeffs} 
into Eq.~\eqref{a:invariantspeed} and taking the limit $\epsilon\to0^+$,
one then obtains the desired result, namely Eq.~\eqref{e:s1jump_i}.
Note that, even though $\lim_{\epsilon\to0^+}I_1^0=\infty$,
the final result for the characteristic speed is finite.

We should emphasize that, by rescaling $u_0\to2u_0$, 
the above calculation also provides the speed $s_\mathrm{outer}$ in
Eq.~\eqref{e:g0outerspeed} of section~\ref{s:multiphase}.
More precisely, 
upon neglecting degenerate branches in Figs.~\ref{f:riemann2jumps}a--c,
the values of the Riemann invariants that determine $s_3^-$ 
in Fig.~\ref{f:riemann1jump}a
are exactly those of the Riemann invariants that determine 
$s_5^-$ in case~(i) (cf.\ Fig.~\ref{f:riemann2jumps}a)
and $s_9^-$ in cases~(ii) and~(iii) (cf.\ Figs.~\ref{f:riemann2jumps}b,c)
of section~\ref{s:multiphase} upon replacing $u_0\to2u_0$.

\paragraph{Degenerate genus-2 calculations: Equations~(\ref{e:s1jump_ii})
  and~(\ref{p:g0inner}).}
We now want to calculate $s_5^+$ in case~(ii) of section~\ref{s:singlejump}, 
as shown in Fig.~\ref{f:riemann1jump}b.
This is done by evaluating $s_5(\r_1,\dots,\r_6)= s(\r_5)$,
with $s(r)$ still given by Eq.~\eqref{a:invariantspeed} with $g=2$
and with
$r_6=-r_1= u_0+2q_0$, $r_4= -r_3= -u_0+2q_0$, 
$r_5 =-r_2= r_4+\epsilon$, 
and then taking again the limit $\epsilon\to0^+$.
In general, 
Eqs.~\eqref{a:alphagammacoeffs} yield the coefficients 
$\alpha_1$ $\alpha_2$, $\gamma_1$ and $\gamma_2$ 
for the $g=2$ case as:
\begin{subequations}
\label{a:g2coeffs}
\begin{gather}
\alpha_{j+1}= (-1)^{j+1}\frac1{[1,0]}
  \bigg([3,j]-\frac12\sigma_1\,[2,j]\bigg)\,,
\\
\gamma_{j+1}= (-1)^{j+1}\frac1{[1,0]}
  \bigg([4,j]-\frac12\sigma_1\,[3,j] + \frac12\bigg(\sigma_2-\frac14\sigma_1^2\bigg)[2,j]\bigg)\,,
\\
\noalign{\noindent $j=0,1$, where}
[i,j]= \det\begin{pmatrix} I_1^i & I_1^j \\ I_2^i & I_2^j \end{pmatrix}\,.
\end{gather}
\end{subequations}
In our case, thanks to the symmetry of the initial datum
for the invariants, Eqs.~\eqref{a:sigma12} give 
$\sigma_1=0$ 
and 
$\sigma_2=-r_4^2-r_5^2-r_6^2=(-3u_0^2+4u_0q_0-12q_0^2)+2(u_0-2q_0)\epsilon-\epsilon^2$.
Upon making the substitutions $z= r_3 + (r_4 - r_3)\sin^2\theta$
on $I_1^k$ and $z= r_5 + (r_6 - r_5)\sin^2\theta$ on $I_2^k$
and using the reflective symmetry of the initial datum,
the integrals $I_{1,2}^0,\dots,I_{1,2}^4$ take the form:
\begin{subequations}
\label{a:g2integrals}
\begin{gather}
I_1^k= (-1)^kr_4^k \,\,\mathop{\textstyle\int}\limits_0^{\pi/2}
  \frac{\cos^{2k}2\theta}
    { \sqrt{\big( r_5^2 - r_4^2 \cos^2 2\theta\big) 
       \big( r_6^2 - r_4^2 \cos^2 2\theta\big) } }\,\,
    d\theta\,,
\\
I_2^k= \mathop{\textstyle\int}\limits_0^{\pi/2}
  \frac{\big( r_5 + \Delta\,\sin^2\theta  \big)^k}
    { \sqrt{\big( r_5 - r_4 + \Delta\,\sin^2\theta  \big) 
      \big( r_4 + r_5 + \Delta\,\sin^2\theta  \big) 
      \big( 2\,r_5 + \Delta\,\sin^2\theta  \big) 
      \big( r_5 + r_6 + \Delta\,\sin^2\theta  \big) } }\,\,
    d\theta\,,
\end{gather}
\end{subequations}
where $\Delta= r_6-r_5$.
Then, evaluating the above integrals, 
substituting Eq.~\eqref{a:g2integrals} and Eq.~\eqref{a:g2coeffs} 
into Eq.~\eqref{a:invariantspeed} and taking the limit $\epsilon\to0^+$,
one then obtains the desired result, namely Eq.~\eqref{e:s1jump_ii}.
Again, note that even though some of the integrals are divergent in the 
limit $\epsilon\to0^+$, the result for the characteristic speeds is finite.

Again, we emphasize that, 
by rescaling $q_0\to q_0+\frac12 u_0$ and $u_0\to2u_0$,
the above calculation also provides both of the speeds 
$\smash{s_\mathrm{inner}^{(1)}}$ and $\smash{s_\mathrm{inner}^{(2)}}$ 
in Eq.~\eqref{e:g0innerspeed}
of section~\ref{s:multiphase}.
More precisely, it is easy to see that,
upon neglecting degenerate branches in Figs.~\ref{f:riemann2jumps}b--d,
the values of the Riemann invariants that determine $s_5^+$ 
in Fig.~\ref{f:riemann1jump}b
are the same as those that determine $s_9^+$ in Fig.~\ref{f:riemann2jumps}b--c
$q_0\to q_0+\frac12 u_0$.
Similarly, upon neglecting degenerate branches Figs.~\ref{f:riemann2jumps}b--d
one sees that 
the values that determine $s_5^+$ in Fig.~\ref{f:riemann1jump}b
also coincide with those that determine $s_7^+$ in Fig.~\ref{f:riemann2jumps}b
and $s_5^+$ in Fig.~\ref{f:riemann2jumps}c upon $u_0\to2u_0$.

\def\@biblabel#1{#1.}

\end{document}